\documentclass{scrreprt}

\usepackage[a4paper,total={6in, 8in}]{geometry}
\usepackage[utf8]{inputenc}
\usepackage{authblk}
\usepackage{booktabs}
\usepackage{color}
\usepackage{comment}
\usepackage{enumitem}
\usepackage{framed}
\usepackage{graphicx}
\usepackage{hhline}
\usepackage[hyperindex,breaklinks]{hyperref}
\usepackage{longtable}
\usepackage{multirow}
\usepackage{pdflscape}
\usepackage{pgfplots}
\usepackage{pifont}
\usepackage{placeins}
\usepackage{tablefootnote}
\usepackage{tabularx}
\usepackage[skip=10pt plus1pt, indent=0pt]{parskip}
\usepackage{fancyhdr}
\usetikzlibrary{calc}
\usepackage{tablefootnote}

\providecommand{\keywords}[1]{\small \textbf{\textit{Keywords---}} #1}

\definecolor{shadecolor}{rgb}{0.9,0.9,0.9}

\newcommand{\takeaway}[1]{\hfill \break \begin{shaded} \textbf{Takeaways:} #1 \end{shaded}}

\newcommand{\NewPageCustom}{\newpage}

\title{Soccer on Social Media}
\subtitle{Comparative Analysis of Social Media Strategies Used by \\ European Soccer Leagues and Teams}

\author[1,2]{Mehdi Houshmand Sarkhoosh}
\author[1,2]{Sayed Mohammad Majidi Dorcheh}
\author[1,3]{Sushant Gautam}
\author[2,3]{Cise Midoglu}
\author[2,3]{Saeed Shafiee Sabet}
\author[1,2,3]{P{\aa}l Halvorsen}
\affil[1]{Oslo Metropolitan University (OsloMet), Norway}
\affil[2]{Forzasys AS, Norway}
\affil[3]{Simula Metropolitan Center for Digital Engineering (SimulaMet), Norway }

\date{}

\begin{document}

\pagenumbering{roman}
\maketitle

\section*{Abstract}

In the era of digitalization, social media has become an integral part of our lives, serving as a significant hub for individuals and businesses to share information, communicate, and engage. 
This is also the case for professional sports, where leagues, clubs and players are using social media to reach out to their fans. In this respect, a huge amount of time is spent curating multimedia content for various social media platforms and their target users. With the emergence of Artificial Intelligence (AI), AI-based tools for automating content generation and enhancing user experiences on social media have become widely popular. However, to effectively utilize such tools, it is imperative to comprehend the demographics and preferences of users on different platforms, understand how content providers post information in these channels, and how different types of multimedia are consumed by audiences. This report presents an analysis of social media platforms, in terms of demographics, supported multimedia modalities, and distinct features and specifications for different modalities, followed by a comparative case study of select European soccer leagues and teams, in terms of their social media practices. Through this analysis, we demonstrate that social media, while being very important for and widely used by supporters from all ages, also requires a fine-tuned effort on the part of soccer professionals, in order to elevate fan experiences and foster engagement.

\keywords{soccer/association football, multimedia, social media strategy, followers, visitor engagement, content cadence, website traffic}

\newpage

\setcounter{tocdepth}{3}
\tableofcontents
\newpage
\listoffigures
\newpage
\listoftables
\newpage

\pagenumbering{arabic}
\normalsize

\chapter{Introduction}\label{section:introduction}

In today's digital age, social media has become an indispensable aspect of our daily lives. People are spending a significant amount of time curating multimedia content for various platforms, targeting diverse user groups. The influence of social media continues to expand, as highlighted by 2023's statistics~\cite{DIGITAL_2023_STATSHOT_REPORT}. A remarkable 60\% of the global population, corresponding to 4.80 billion individuals, are actively engaged on social media platforms. This engagement has seen a surge with an addition of 150 million new users in the past year alone. On average, users dedicate 2 hours and 24 minutes daily to their social media activities. However, the reach of these platforms varies globally. While North America and certain European regions report high engagement rates of around 74-84\%, areas like Middle Africa lag behind at just 7\%. Eastern Asia, Southern America, and other parts of Europe also showcase strong engagement, with rates between 72-84\%. Conversely, regions like Southern Asia and Western Africa have lower engagement, at 41\% and 13\% respectively. As these platforms continue to evolve to cater to varied consumer preferences, it's evident that businesses must stay updated and strategically harness the power of social media in their marketing strategies.

Given the vast and growing influence of social media, there's an undeniable need for businesses and content creators to optimize their approach. The sheer volume of users and the time they spend on these platforms underscore the potential reach and impact of well-curated content. This is where the integration of technology, particularly Artificial Intelligence (AI), becomes crucial. 
With the rise of AI, multimedia content generation processes can be (partly) automated, offering a myriad of benefits. Automation can lead to faster content production, reduced human error, and the ability to tailor content more precisely to target demographics. We have earlier developed such tools for the soccer domain, which automatically detect events~\cite{Nergård2020, Nergård2021-Automated, Nergård2021-3D}, clip events~\cite{Valand2021-Automated,Valand2021-AI-Based}, select thumbnails~\cite{Husa2022-Automatic,Husa2022-HOST-ATS} and summarize games~\cite{gautam2023,gautam2022_summarization,gautam2022_soccer}. The next step is then to prepare the content for a particular distribution channel. In this respect, social media has become a hub for AI-based automation tools to improve user experience. However, to effectively leverage AI tools, it is important to understand the demographics and preferences of users on each platform. This understanding will not only enhance user engagement but also ensure that the content resonates with the intended audience, further emphasizing the importance of our study.

In the first part of this report, we provide an analysis of $7$ social media platforms, namely TikTok~\cite{TikTokHelp2023}, Instagram~\cite{InstagramHelp2023}, Facebook~\cite{FacebookHelp2023},X~\cite{TwitterHelp2023}\footnote{The platform is referred to as "Twitter" in the rest of this report.}, Snapchat~\cite{SnapchatHelp2023}, LinkedIn~\cite{LinkedInHelp2023}, and YouTube~\cite{YouTubeHelp2023}, in terms of demographics (e.g., age and gender distribution), as well as supported modalities and multimedia specifications. Our analysis will help identify the most popular social media platforms for different age groups and genders, and understand the unique features of each platform. This knowledge can be used to curate highlights including video, image, and text, which are optimized for each platform, develop personalized content, and enhance user engagement, ultimately leading to better social media experiences. 

In the second part of this report, we present specific case studies. First, we explore the social media presence of $5$ prominent European soccer leagues, namely the English Premier League (England)~\cite{PremierLeague2023}, Bundesliga (Germany)~\cite{Bundesliga2023}, La Liga (Spain)~\cite{LaLiga2023}, Serie A (Italy)~\cite{SerieA2023}, and Ligue 1 (France)~\cite{Ligue1UberEats2023}, and compare their activities to the social media presence of two smaller Europeean leagues from Scandinavia, namely Allsvenskan (Sweden)~\cite{Allsvenskan2023} and Eliteserien (Norway)~\cite{Eliteserien2023}. Next, we compare the social media activities of some of the most prominent teams in the aforementioned leagues with those of Norwegian soccer teams. Through this analysis, we hope to provide insights into the differences and similarities in the social media practices of different soccer leagues and teams, and how they utilize social media platforms to engage with fans and promote their respective organizations.

All information presented in this document is based on data retrieved between 22.06.2023 and 22.07.2023. 
Table~\ref{tab:terminology} presents an overview of terminology used throughout the report. 

\begin{table}[ht]

    \scriptsize
    \centering
    \caption{Key terms and concepts related to social media and digital engagement.}
    \label{tab:terminology}

    
        \begin{tabular}{|p{1.5cm}|p{2cm}||p{10cm}|}
        \hline
       
        \textbf{Category} 
        & \textbf{Term} 
        & \textbf{Description} 
        \\ \hline
        
        \multirow{9}{1.5cm}{Social Media Concepts}
        
        & Social Media Platform 
        & Digital platforms and applications that enable users to create, share, or exchange information, ideas, pictures, and videos in virtual communities and networks. 
        \\ \hhline{~--}
        
        & Content 
        & Any form of media, such as text, image, video, and audio, created and shared on social media platforms to convey a message or information. 
        \\ \hhline{~--}
        
        & Hashtag 
        & A word or phrase preceded by a hash sign (\#) used on social media platforms to identify messages on a specific topic. They help categorize content and make it discoverable. 
        \\ \hhline{~--}
        
        & Trend 
        & Pattern or topic that has become popular on social media within a specific timeframe. They often reflect current events, cultural moments, or viral content. 
        \\ \hline
        
        \multirow{14}{1.5cm}{Engagement Metrics}
        
        & Engagement 
        & The interaction between users and content on social media platforms. It can be measured through likes, comments, shares, and other forms of interaction. 
        \\ \hhline{~--}
        
        & Engagement Rate 
        & A metric that measures the level of engagement a piece of content receives relative to its reach or views. It's often calculated as the total engagement divided by total followers (or impressions) multiplied by 100. 
        \\ \hhline{~--}
        
        & Like
        & A common form of engagement where users show appreciation or agreement with content by clicking a "like" button. 
        \\ \hhline{~--}
        
        & View 
        & The number of times a piece of content, especially a video, has been watched by users. 
        \\ \hhline{~--}
    
        & CTR 
        & The ratio of users who click on a specific link to the number of total users who view a page, email, or advertisement. 
        \\ \hhline{~--}
    
        & Impression 
        & The number of times content is displayed on a user's screen. \\ \hhline{~--}
        
        & Reach 
        & The total number of unique users who have seen a particular piece of content or advertisement on social media. 
        \\ \hline
        
        \multirow{7}{1.5cm}{User Dynamics}
        
        & Gen Z 
        & The demographic cohort succeeding Millennials. Typically, researchers and commentators use birth years from the mid-to-late 1990s to the early 2010s to define Gen Z. 
        \\ \hhline{~--}
        
        & Follower 
        & A user who subscribes to another user's profile on social media to see their content in their feed. 
        \\ \hhline{~--}
        
        & Interaction 
        & Actions taken by users in response to content on social media. This includes liking, commenting, sharing, and more. 
        \\ \hline
    
        \multirow{3}{1.5cm}{Multimedia Content}
        
        & Multimedia Modality 
        & Different forms of content used for interaction in communication channels. In the context of social media platforms, this refers to various media types such as text, image, video, and audio. 
        \\ \hline
    
        \multirow{9}{1.5cm}{Traffic Sources} 
        
        & Traffic Source 
        & The origin or medium through which users find and visit a website or platform. 
        \\ \hhline{~--}
        
        & Direct 
        & Visitors who arrive at a website by entering the URL directly into their browser. 
        \\ \hhline{~--}
    
        & Referral 
        & Visitors who arrive at a website from external websites that link to it. 
        \\ \hhline{~--}
        
        & Search 
        & Visitors who arrive at a website after using a search engine. 
        \\ \hhline{~--}
        
        & Social 
        & Visitors who arrive at a website from social media platforms. 
        \\ \hhline{~--}
        
        & Mail 
        & Visitors who arrive at a website from email campaigns or marketing efforts. 
        \\ \hhline{~--}
        
        & Display 
        & Visitors who arrive at a website after clicking an advertisement displayed on another website or platform. 
        \\ \hline

        \end{tabular}
        
    
\end{table}
\FloatBarrier

\clearpage
\chapter{Related Work}\label{section:related-work}

\section{Social Media Strategies in the Context of Sports}

Previous research on social media strategies in the context of sports have shed light on the significance of social media as a powerful tool for sports organizations. These studies have examined various aspects ranging from fan engagement to brand image and marketing effectiveness, and provide a comprehensive understanding of the role of social media in the context of sports. 

McCarthy et al.~\cite{McCarthy2022} found that English soccer clubs employ diverse strategies on platforms such as Facebook, Twitter, and Instagram, including behind-the-scenes content and contests, to engage fans. Obradović et al.~\cite{Obradovic2019} highlighted the use of social media by Premier League soccer clubs for communication, brand promotion, and revenue generation. 
Yun et al.~\cite{Yun2020} investigated the drivers of soccer fan loyalty in the context of the Australian A-League. They found that fan engagement, team brand image, and cumulative fan satisfaction significantly influenced fan loyalty. This study highlighted the importance of various factors in shaping attitudinal and behavioral loyalty among soccer fans, with enduring involvement with the team playing a moderating role. Trail et al.~\cite{Trail2017} conducted a longitudinal study focusing on team-fan role identity and its impact on self-reported attendance behavior and future intentions in college sporting events. Their research emphasized the role of identity in predicting attendance behavior and support for the team over time, shedding light on the importance of fan identity in the context of sports. 

Alduayji~\cite{Alduayji2019} examined online co-creation behavior in a sports context, shedding light on the nature of online interactions among sports fans in the digital space. Winand et al.~\cite{Winand2019} conducted a content analysis of FIFA's Twitter account, examining how international sport federations use social media to engage with their followers. This research explored various strategies employed by sport organizations to connect with their audiences through social media. Eddy et al.~\cite{Eddy2021} explored engagement with sport sponsor activations on Twitter, shedding light on how the structure and content of sponsored messages impact follower engagement. Naraine et al.~\cite{Naraine2019} analyzed user engagement within the Twitter community of professional sport organizations. Their research revealed demographic and temporal insights into the engagement patterns of Twitter users following professional sports teams.

\section{AI-Based Content Generation and Enhancement}

To engage audiences and stand out in the crowded digital landscape, content creators continually seek innovative ways to create and deliver compelling and personalized content. Recent advancements in multimedia research have the potential to revolutionize social media content creation and enhancement technologies. 

Darbinyan~\cite{Darbinyan2023} discusses how AI is playing a game-changing role in many aspects of our lives, especially the social media landscape with automatic content moderation and personalized recommendations, as well as how AI is reshaping the way we interact and connect online. Belsky~\cite{Belsky2023} discusses how generative AI models are changing the world of content creation, with impacts on marketing, software, design, entertainment, and interpersonal communications. It also explores how these models can be used to automatically generate content such as articles, blog posts, or social media posts. Kapoor~\cite{Kapoor2018} discusses the findings of 132 papers on social media and social networking published between 1997 and 2017. It examines the behavioral side of social media, investigates the aspect of reviews and recommendations, and studies its integration for organizational purposes. Recent guide by Sprout~\cite{sproutsocial2021} outlines the social media behaviors and expectations of each generation, and what businesses should consider when targeting each generation. It highlights the unique preferences, behaviors, and expectations of different group on social platforms. Wang et al.~\cite{Wang2022} introduce a Decision Boundary Customizer module and a Mini-History mechanism for generating user-adaptive highlight classifiers. This research can be leveraged to enhance personalized content recommendations on social media platforms. By analyzing user preferences and viewing history, content creators can tailor their recommendations to individual users, increasing user engagement and satisfaction.
Hou et al.~\cite{Hou2022} discuss the application of deep learning techniques in logo detection. This technology can benefit social media content creators by automating brand monitoring. Content containing specific logos can be tracked and analyzed, allowing brands to assess their online presence and measure the effectiveness of marketing campaigns. Sosnovik et al.~\cite{Sosnovik2023} propose a contrastive learning approach for video summarization. Social media platforms often require condensed video content that conveys a story efficiently. Contrastive video summarization can automate the process of generating concise and engaging video summaries, saving time for content creators and offering users more digestible content. 

Social media content often involves diverse topics. Xiong et al.~\cite{Xiong2023} introduce a method for topic-adaptive video highlight detection. This line of research can assist content creators in identifying and highlighting relevant moments in videos based on text inputs, ensuring that content remains relevant and engaging to the intended audience. Gautam et al.~\cite{gautam2022,gautam2022,gautam2022} use such metadata inputs to enhance an end-to-end automated soccer broadcast  pipeline and automatically generate textual game summaries.

Smart cropping and object detection technologies can also significantly impact social media content creation~\cite{apostolidis2021}. These technologies automatically identify and crop relevant portions of images or videos for various display sizes (aspect ratios), making content more visually appealing and accessible across different devices and platforms. In this context, semantic segmentation and inpainting have also been used to retarget legacy video frames to wider aspect ratios, while preserving the original proportions of important objects, thereby enhancing the visual naturality of the retargeted content~\cite{Jin_2023}. Filtering content based on object detection can improve content moderation and user experience on social media. Content creators can use efficient and near real-time object detection algorithms to identify and filter out inappropriate or undesirable content, ensuring that their platforms remain safe and user-friendly~\cite{mallmann2020}.

\section{Multimodal Datasets and Applications}

Researchers have been exploring the potential of multimodal datasets curated from social media for various applications. DepressionNet~\cite{Zogan2021} is a novel framework proposed to detect depression by summarizing user-generated content from Twitter. Lu et al.~\cite{Lu2021} focus on fast and effective identification of pandemic-related rumors on social media, highlighting the need for few-shot learning techniques due to the scarcity of early-stage instances. Gautam et al.~\cite{gautam2023,midoglu2022} discuss the opportunities and challenges of leveraging multimodal data to enhance AI understanding, particularly in the soccer domain. These studies collectively emphasize the significance of multimodal datasets in enhancing AI models for diverse applications in the social media landscape.

\takeaway{\begin{itemize}
    \item Recent advances in deep learning, multimedia analysis methods, and multimodal datasets have the potential to facilitate and enhance social media content creation in various ways.
    \item These advancements can empower content creators to deliver personalized, engaging, and visually appealing content to their audiences.
    \item They also improve content moderation and brand monitoring capabilities.
    \item As technology continues to evolve, it is likely that further innovations will continue to shape the landscape of social media content creation and enhancement.
\end{itemize}}

\clearpage
\chapter{Overview of Social Media Platforms}\label{section:social-media-platforms}

Social media platforms have become an integral part of modern communication, each offering unique features tailored to different user needs. 

\begin{itemize}
    \item \textbf{TikTok}: Primarily known for its short-form videos, appeals largely to younger audiences with its viral trends and music-driven content.

    \item \textbf{Instagram}: With its emphasis on visual aesthetics, has become the hub for influencers, brand partnerships, and visual storytelling.

    \item \textbf{Facebook}: The pioneer in the social networking realm, offers a comprehensive suite of features from news feeds to marketplaces, although it has recently seen a shift in its demographic as younger users migrate to newer platforms.

    \item \textbf{Twitter}: Stands out for its real-time updates and direct engagement, making it a favorite among professionals, journalists, and public figures.

    \item \textbf{Snapchat}: With its ephemeral content and augmented reality lenses, remains a favorite among Gen Z, offering a unique storytelling format.

    \item \textbf{LinkedIn}: Serves as the go-to platform for professionals, providing resources for networking, industry news, and career development.

    \item \textbf{YouTube}: The premier platform for video sharing and consumption. With its diverse content, it caters to a global audience, enabling creators to build careers, shape culture, and inform the masses. It stands unique for its depth, reach, and transformative impact on digital storytelling.
\end{itemize}

As these platforms continue to evolve in response to technological advancements and changing user behaviors, they remain pivotal in shaping the way we connect, share, and engage with the world around us~\cite{ACCC2023}.

\NewPageCustom
\section{Demographic Analysis}\label{section:demographic-analysis}

In this section, we will explore the user demographics of popular social media platforms, in terms of age and gender distribution. From the younger, predominantly Gen Z audience of TikTok to the more professional and career-oriented user base of LinkedIn, each social media platform has its unique demographic makeup and usage patterns. Analyzing the demographics of users on social media platforms makes it possible to gain valuable insights into the characteristics of each platform's core user base and tailor content and engagement strategies accordingly. 

\begin{table}[ht]

    \centering
    \scriptsize
    \caption[{Popularity of different social media platforms.}]{Popularity of different social media platforms in terms of the total number of users, per platform.}
    \label{tab:platforms-popularity}
    

        \begin{tabular}{|l|c||c|c|c|c|c|c|}
            \hline
            
            & \textbf{Population} 
            & \textbf{TikTok} 
            & \textbf{Instagram} 
            & \textbf{Facebook} 
            & \textbf{Twitter} 
            & \textbf{SnapChat} 
            & \textbf{LinkedIn} 
            \\ \hline  
            
            Norway 
            & 5.4 million 
            & 1.57 million 
            & 2.95 million 
            & 3.30 million  
            & N/A
            & 3.45 million 
            & 2.3 million 
            \\ \hline
            
            World
            & 7.8 billion 
            & 2 billion 
            & 1.2 billion 
            & 3.71 billion 
            & 330 million 
            & 428.4 million 
            & 740 million 
            \\ \hline
            
        \end{tabular}
        
   
\end{table}
\FloatBarrier

\begin{table}[ht]

    \centering
    \scriptsize
    \caption[Demographics of different social media platforms.]{Demographics of different social media platforms in terms of age and gender distribution in percentage, per platform.}
    \label{tab:platforms-demographic-analysis}
    

        \begin{tabular}{|c|c||c|c|c|c|c|c|}
            \hline
            \textbf{Age} 
            & \textbf{Gender} 
            & \textbf{TikTok} 
            & \textbf{Instagram} 
            & \textbf{Facebook} 
            & \textbf{Twitter} 
            & \textbf{SnapChat} 
            & \textbf{LinkedIn} 
            \\ \hline  
            
            \multirow{2}{*}{13-17}            
            & Female 
            & 13.25 
            & 4 
            & 2.1
            & 2.8 
            & N/A
            & N/A
            \\ \cline{2-8} 
            & Male 
            & 10.2 
            & 4.9 
            & 3.7
            & 3.7 
            & N/A 
            & N/A 
            \\ \hline
            
            \multirow{2}{*}{18-24} 
            & Female 
            & 12.2 
            & 13.4 
            & 8.9 
            & 8.4 
            & 23 
            & 8.7 
            \\ \cline{2-8} 
            & Male 
            & 9.1 
            & 16.8 
            & 12.6
            & 10.6 
            & 20.1 
            & 11.6 
            \\ \hline           
            
            \multirow{2}{*}{25-34} 
            & Female 
            & 11.3 
            & 14.1 
            & 12.3
            & 12.5 
            & 13.7 
            & 25.2 
            \\ \cline{2-8} 
            & Male 
            & 8.3 
            & 17.1 
            & 17.6 
            & 17.4 
            & 12.03 
            & 33.8 
            \\ \hline 
            
            \multirow{2}{*}{35-44}             
            & Female 
            & 10.57 
            & 8.1 
            & 8.5 
            & 7.02 
            & 6.3 
            & 7.5 
            \\ \cline{2-8} 
            & Male 
            & 8.2 
            & 7.7 
            & 10.9
            & 9.6 
            & 5.5 
            & 10.12 
            \\ \hline
            
            \multirow{2}{*}{45-54} 
            & Female 
            & 5.27 
            & 3.5 
            & 5.5 
            & 5.4 
            & N/A 
            & 1.2 
            \\ \cline{2-8} 
            & Male 
            & 3.7 
            & 4.3 
            & 6.1
            & 7.84 
            & N/A 
            & 1.6 
            \\ \hline 
            
            \multirow{2}{*}{55-64} 
            & Female 
            & 3.1 
            & 2.2 
            & 3.8 
            & 6.4 
            & N/A 
            & N/A 
            \\ \cline{2-8} 
            & Male 
            & 3.2 
            & 1.5 
            & 3.5 
            & 6.84 
            & N/A 
            & N/A 
            \\ \hline         

            \multirow{2}{*}{65+} 
            & Female 
            & N/A 
            & 1.2 
            & 3 
            & 5.4 
            & N/A 
            & N/A
            \\ \cline{2-8}             
            & Male 
            & N/A 
            & 0.9 
            & 2.6 
            & 8.84 
            & N/A 
            & N/A 
            \\ \hline
            
        \end{tabular}
        

\end{table}
\FloatBarrier

In Table~\ref{tab:platforms-popularity}, we provide the total number of users on each platform in Norway and the world, and in Table~\ref{tab:platforms-demographic-analysis}, we provide the age and gender distribution in percentage on each social media platform.\footnote{(Table~\ref{tab:platforms-demographic-analysis}) For instance, in the TikTok column under the age group 13-17 and gender Female, the value 13.25 means that 13.25\% of TikTok's user base is comprised of females aged 13-17.} Popularity and user demographics information presented in Tables~\ref{tab:platforms-demographic-analysis} and ~\ref{tab:platforms-popularity} have been derived from~\cite{Statista2023}. 

The tables show that Facebook is the most popular social media platform in the world, with a staggering 3.71 billion active users as of 2021. Although it decreased to 3.03 billion by 2023~\cite{statistaFacebook2023}, it still remains the most popular. However, when we look specifically at Norway, Snapchat emerges as the most popular platform, with 3.45 million users. This suggests that while some platforms may dominate globally, local trends and preferences can still have a significant impact on the popularity of social media platforms in specific regions. 

The age and gender distribution of users on each platform can also reveal important insights. For example, while Facebook has a diverse age range of users, with the majority of its users aged between 18 and 44, TikTok is most popular among younger users aged 18-24. LinkedIn, on the other hand, has a more mature user base, with the majority of its users aged between 25 and 44. 

\takeaway{\begin{itemize}
    \item Social media platforms, each with unique features and demographics, play pivotal roles in modern communication.
    \item TikTok appeals to younger audiences with its short-form videos.
    \item LinkedIn offers professional networking capabilities.
    \item The understanding of each platform's specifications, multimedia modalities, and user demographics is essential for effective content creation and engagement.
\end{itemize}}

\NewPageCustom
\section{Multimedia Modalities}\label{section:multimedia-modalities}

Social media platforms offer different modalities for users to interact and engage with content. Each platform has its own unique features and user base, which influence the types of modalities available. Table~\ref{tab:platforms-modalities} summarizes the different modalities supported and promoted by popular social media platforms.

\begin{table}[ht]

    \scriptsize
    \centering
    \caption[Multimedia modalities supported by different social media platforms.]{Multimedia modalities supported by different social media platforms.}
    \label{tab:platforms-modalities}


        \begin{tabular}{|l|l||c|c|c|c|c|c|}
        \cline{3-8}
        
        \multicolumn{2}{l|}{}
        & \textbf{TikTok} 
        & \textbf{Instagram} 
        & \textbf{Facebook} 
        & \textbf{Twitter} 
        & \textbf{SnapChat} 
        & \textbf{LinkedIn} 
        \\ \hline

        \multicolumn{2}{|l||}{\textbf{Video}}
        & \ding{52} 
        & \ding{52}
        & \ding{52}
        & \ding{52}
        & \ding{52} 
        & \ding{52}
        \\ \hline
        
        \multicolumn{2}{|l||}{\textbf{Image}}
        & \ding{54}
        & \ding{52} 
        & \ding{52} 
        & \ding{52}
        & \ding{52}
        & \ding{52} 
        \\ \hline
        
        \multicolumn{2}{|l||}{\textbf{Audio}}
        & \ding{54}
        & \ding{54}
        & \ding{54}
        & \ding{52}
        & \ding{54}
        & \ding{54}
        \\ \hline
        
        \multirow{3}{*}{\textbf{Text}} 
        & Short
        & \ding{52} 
        & \ding{52} 
        & \ding{52} 
        & \ding{52} 
        & \ding{52}
        & \ding{52} 
        \\ \cline{2-8}
        
        & Overlay\tablefootnote{Text overlay on top of the video.}
        & \ding{52} 
        & \ding{52}
        & \ding{52} 
        & \ding{52}
        & \ding{52} 
        & \ding{52}
        \\ \cline{2-8}
        
        & Standalone\tablefootnote{Standalone text above or below content.}
        & \ding{52}
        & \ding{52}
        & \ding{52}
        & \ding{52}
        & \ding{52}
        & \ding{52}
        \\ \hline
        
        \multirow{3}{*}{\textbf{Captions}} 
        & Hashtags
        & \ding{52}
        & \ding{52}
        & \ding{52}
        & \ding{52}
        & \ding{52}
        & \ding{52}
        \\ \cline{2-8}
        
        & Mentions
        & \ding{52}
        & \ding{52}
        & \ding{52}
        & \ding{52}
        & \ding{52}
        & \ding{52}
        \\ \cline{2-8}
        
        & Links 
        & \ding{52}
        & \ding{52}
        & \ding{52}
        & \ding{52}
        & \ding{52}
        & \ding{52}
        \\ \hline
          
        \multirow{3}{1.9cm}{\textbf{Post Description}}
        & Hashtags
        & \ding{52}
        & \ding{52}
        & \ding{52}
        & \ding{52}
        & \ding{52}
        & \ding{52}
        \\ \cline{2-8}
        
        & Mentions
        & \ding{52}
        & \ding{52}
        & \ding{52}
        & \ding{52}
        & \ding{52}
        & \ding{52}
        \\ \cline{2-8}
        
        & Links
        & \ding{52}
        & \ding{52}
        & \ding{52}
        & \ding{52}
        & \ding{52}
        & \ding{52}
        \\ \hline
        
        \end{tabular}
        
    
\end{table}
\FloatBarrier

\textbf{TikTok}, a short video-sharing platform, intertwines visuals with sound to offer a rich, multimodal experience. The "For You" page blends algorithm-driven selections and hashtags, offering content tailored to user preferences. Interaction is also multimodal, encompassing likes, comments, shares, and the duet feature, where users co-create videos, adding layers of collaborative expression.

\textbf{Instagram} primarily leverages visual modalities, with photos and videos at its core. Yet, it extends its multimodality through captions, comments, and the temporal nature of Instagram Stories. The platform's explore page blends user interests with visual content, showcasing a rich tapestry of images, videos, and user interactions.

\textbf{Facebook} offers a comprehensive multimodal experience, integrating text updates, photos, videos, reactions, and emojis. The platform fosters community building through groups, where shared interests come alive through a blend of textual, visual, and video content.

\textbf{Twitter}, while anchored in its text-based microblogging nature, has evolved to integrate photos, videos, and gifs, creating a more varied, multimodal communication landscape. From short tweets to visual media, Twitter encapsulates global conversations in diverse modalities.

\textbf{Snapchat}, with its ephemeral visuals, integrates the transient nature of content with augmented reality filters and lenses. Its stories feature, coupled with direct interactions, combines various modes of communication, from visual to textual, offering a transient yet impactful interaction platform.

\textbf{LinkedIn}, catering to professionals, marries textual updates with videos, articles, and images. The LinkedIn Pulse extends its multimodal nature, allowing users to craft articles, blending text with images and videos, to share industry insights, innovations, and narratives.

\takeaway{\begin{itemize}
    \item Social media platforms cater to diverse user interactions, as each platform has unique niches and engagement styles.    
    \item TikTok is video-centric with features such as the "For You" page and duets.
    \item Instagram emphasizes photo and video sharing, including Stories.
    \item Facebook is a connection hub with varied newsfeeds and group features.
    \item Twitter is a microblogging platform known for concise posts and trending topics.
    \item Snapchat offers ephemeral content with augmented reality enhancements.
    \item LinkedIn focuses on professional networking and industry insights.  
\end{itemize}}

\NewPageCustom
\section{Video Specifications}\label{section:video-specifications}

Each social media platform has its own video specifications that content creators need to keep in mind when uploading videos. These specifications can include requirements for video format, resolution, aspect ratio, length, file size, etc. Adhering to these specifications can help ensure that videos display properly on different devices, and provide the best viewing experience for users. Table~\ref{tab:platforms-video-specifications} summarizes the video specifications for popular social media platforms.

\textbf{TikTok} supports a variety of video formats, including MP4 and MOV. The recommended resolution for TikTok videos is 1080x1920 (vertical) or 1920x1080 (horizontal), with an aspect ratio of 9:16 or 16:9. TikTok videos can be up to 60 seconds long, but many popular videos are shorter. The maximum file size for videos on TikTok is 287.6 MB.

\textbf{Instagram} supports a variety of video formats, including MP4 and MOV. The recommended resolution for Instagram videos is 1080x1080 (square), 1080x1350 (vertical), or 1080x608 (horizontal), with an aspect ratio of 1:1, 4:5, or 16:9. Instagram videos can be up to 60 seconds long in the feed and up to 15 seconds long in Instagram Stories. The maximum file size for videos on Instagram is 4 GB.

\textbf{Facebook} supports a variety of video formats, including MP4 and MOV. The recommended resolution for Facebook videos is 1080x1080 (square), 1080x1350 (vertical), or 1080x608 (horizontal), with an aspect ratio of 1:1, 4:5, or 16:9. Facebook videos can be up to 240 minutes long, but shorter videos tend to perform better. The maximum file size for videos on Facebook is 4 GB.

\textbf{Twitter} supports video uploads in MP4 and MOV formats for mobile apps, and only MP4 for the web. The recommended resolution for Twitter videos is 1280x720 (landscape and portrait), with an aspect ratio of 1:1 for square videos, 16:9 for landscape videos, and 9:16 for portrait videos. Videos on Twitter can be up to 140 seconds long. The maximum file size for videos uploaded from the Twitter app is 512MB, while for the Twitter web client, it is 1GB.

\textbf{Snapchat} supports MP4 and MOV video formats. The recommended resolution for Snapchat videos is 1080x1920 (vertical), with an aspect ratio of 9:16. Snapchat videos can be up to 60 seconds long, but most are shorter. The maximum file size for videos on Snapchat is 32 MB.

\textbf{LinkedIn} supports a variety of video formats, including MP4 and MOV. The recommended resolution for LinkedIn videos is 360p, 480p, 720p, or 1080p, with an aspect ratio of 1:2.4 to 2.4:1. LinkedIn videos can be up to 10 minutes long, but shorter videos tend to perform better. The maximum file size for videos on LinkedIn is 5 GB.

\takeaway{\begin{itemize}
    \item Each social media platform has video specifications uniquely tailored for the best video display, content creators should adapt to these specifications for optimal user experience and engagement.
    \item TikTok and Snapchat primarily use 9:16 vertical videos. 
    \item Instagram, Facebook, and Twitter have varied ratios, including square and horizontal. 
    \item Video lengths vary from 15-second stories on Instagram to 240 minutes on Facebook. 
    \item File size limits range from 32 MB on Snapchat to 5 GB on LinkedIn. 
\end{itemize}}

\newgeometry{margin=3cm}
\begin{landscape}

    \scriptsize
    \centering
    
    \begin{longtable}{l|l||c|c|c|c|c|c|c|c|c|c|c|c|c|c|}
    \caption[Video specifications for different social media platforms.]{Video specifications for different social media platforms~\cite{Hootsuite2022,FacebookHelp2023,Neoreach2022}.}
    \label{tab:platforms-video-specifications} 
    \\ \cline{3-16}

    \multicolumn{2}{c|}{}
    & \multicolumn{2}{c|}{\textbf{TikTok}} 
    & \multicolumn{3}{c|}{\textbf{Instagram}} 
    & \multicolumn{4}{c|}{\textbf{Facebook}} 
    & \textbf{Twitter} 
    & \multicolumn{2}{c|}{\textbf{Snapchat}} 
    & \multicolumn{2}{c|}{\textbf{LinkedIn}} 
    \\ \cline{3-16}

    \multicolumn{2}{c|}{}
    & \multicolumn{1}{p{1cm}|}{\textbf{Story}} 
    & \multicolumn{1}{p{1cm}|}{\textbf{Post}} 
    & \multicolumn{1}{p{1cm}|}{\textbf{Reel}} 
    & \multicolumn{1}{p{1cm}|}{\textbf{Feed}} 
    & \multicolumn{1}{p{1cm}|}{\textbf{Story}} 
    & \multicolumn{1}{p{1cm}|}{\textbf{Feed}} 
    & \multicolumn{1}{p{1cm}|}{\textbf{Video Ad}} 
    & \multicolumn{1}{p{1cm}|}{\textbf{In-stream}} 
    & \multicolumn{1}{p{1cm}|}{\textbf{Story}} 
    & \multicolumn{1}{p{1cm}|}{\textbf{Video}} 
    & \multicolumn{1}{p{1.55cm}|}{\textbf{Video-Spotlight}} 
    & \multicolumn{1}{p{1cm}|}{\textbf{Story}} 
    & \multicolumn{1}{p{1cm}|}{\textbf{Native Video}} 
    & \multicolumn{1}{p{1cm}|}{\textbf{Video Ad}} 
    \\ \hline \hline

    \multicolumn{2}{|p{2.75cm}||}{\textbf{Life Span}} 
    & \multicolumn{1}{p{1cm}|}{24h} 
    & \multicolumn{1}{p{1cm}|}{Instant decay} 
    & \multicolumn{1}{p{1cm}|}{14+d} 
    & \multicolumn{1}{p{1cm}|}{48h} 
    & \multicolumn{1}{p{1cm}|}{24h}
    & \multicolumn{1}{p{1cm}|}{6h} 
    & \multicolumn{1}{c|}{N/A}
    & \multicolumn{1}{c|}{N/A}
    & \multicolumn{1}{p{1cm}|}{24h}
    & \multicolumn{1}{p{1cm}|}{15-18 min}
    & \multicolumn{1}{p{1cm}|}{Instant decay} 
    & \multicolumn{1}{p{1cm}|}{24h}
    & \multicolumn{1}{p{1cm}|}{24h} 
    & \multicolumn{1}{c|}{N/A}
    \\ \hline  
    
    \multicolumn{1}{|l|}{\multirow{2}{1.5cm}{\textbf{File Size}}}
    & \multicolumn{1}{|p{1.25cm}||}{Minimum} 
    & \multicolumn{2}{c|}{N/A}
    & \multicolumn{3}{c|}{N/A} 
    & \multicolumn{1}{c|}{N/A} 
    & \multicolumn{1}{c|}{N/A} 
    & \multicolumn{1}{c|}{N/A} 
    & \multicolumn{1}{c|}{N/A} 
    & \multicolumn{1}{c|}{N/A}
    & \multicolumn{1}{c|}{N/A} 
    & \multicolumn{1}{c|}{N/A} 
    & \multicolumn{1}{p{1cm}|}{75KB}
    & \multicolumn{1}{p{1cm}|}{75KB}
    \\ \cline{2-16} 

    \multicolumn{1}{|l|}{} 
    & \multicolumn{1}{|p{1.25cm}||}{Maximum} 
    & \multicolumn{2}{p{2cm}|}{Android:72MB Apple:287MB} 
    & \multicolumn{3}{p{3cm}|}{$\leq$10min:650MB $\leq$60min:3.6GB} 
    & \multicolumn{1}{p{1cm}|}{4GB} 
    & \multicolumn{1}{p{1cm}|}{4GB} 
    & \multicolumn{1}{c|}{N/A} 
    & \multicolumn{1}{c|}{N/A} 
    & \multicolumn{1}{p{1cm}|}{512MB} 
    & \multicolumn{1}{p{1.55cm}|}{1GB} 
    & \multicolumn{1}{c|}{N/A} 
    & \multicolumn{1}{p{1cm}|}{5GB}
    & \multicolumn{1}{p{1cm}|}{200MB}
    \\ \hline 

    \multicolumn{1}{|l|}{\multirow{2}{1.5cm}{\textbf{Length}}}
    & \multicolumn{1}{|p{1.25cm}||}{Minimum} 
    & \multicolumn{1}{c|}{N/A} 
    & \multicolumn{1}{c|}{N/A} 
    & \multicolumn{1}{c|}{N/A} 
    & \multicolumn{1}{p{1cm}|}{10min} 
    & \multicolumn{1}{p{1cm}|}{15s} 
    & \multicolumn{1}{p{1cm}|}{240min} 
    & \multicolumn{1}{p{1cm}|}{240min} 
    & \multicolumn{1}{p{1cm}|}{5s} 
    & \multicolumn{1}{p{1cm}|}{1s} 
    & \multicolumn{1}{p{1cm}|}{140s} 
    & \multicolumn{1}{p{1.55cm}|}{3s} 
    & \multicolumn{1}{p{1cm}|}{60s} 
    & \multicolumn{1}{p{1cm}|}{3s}
    & \multicolumn{1}{p{1cm}|}{3s} 
    \\ \cline{2-16} 

    \multicolumn{1}{|l|}{} 
    & \multicolumn{1}{|p{1.25cm}||}{Maximum} 
    & \multicolumn{1}{p{1cm}|}{3min} 
    & \multicolumn{1}{p{1cm}|}{10min} 
    & \multicolumn{1}{p{1cm}|}{15min} 
    & \multicolumn{1}{p{1cm}|}{60min} 
    & \multicolumn{1}{p{1cm}|}{15s} 
    & \multicolumn{1}{p{1cm}|}{240min} 
    & \multicolumn{1}{p{1cm}|}{240min} 
    & \multicolumn{1}{p{1cm}|}{120s} 
    & \multicolumn{1}{p{1cm}|}{120s} 
    & \multicolumn{1}{p{1cm}|}{140s} 
    & \multicolumn{1}{p{1.55cm}|}{180s} 
    & \multicolumn{1}{p{1cm}|}{60s} 
    & \multicolumn{1}{p{1cm}|}{10min} 
    & \multicolumn{1}{p{1cm}|}{30min}  
    \\ \hline
     
    \multicolumn{1}{|l|}{\multirow{22}{1.5cm}{\textbf{File Format}}}
    & \texttt{MP4}
    & \multicolumn{2}{c|}{\ding{52}} 
    & \multicolumn{2}{c|}{\ding{52}} 
    & \multicolumn{1}{c|}{\ding{52}} 
    & \multicolumn{1}{c|}{\ding{52}} 
    & \multicolumn{1}{c|}{\ding{52}} 
    & \multicolumn{1}{c|}{\ding{52}}
    & \multicolumn{1}{c|}{\ding{52}} 
    & \multicolumn{1}{c|}{\ding{52}\footnote{Web.}}
    & \multicolumn{1}{c|}{\ding{52}} 
    & \multicolumn{1}{c|}{\ding{52}} 
    & \multicolumn{1}{c|}{\ding{52}}
    & \multicolumn{1}{c|}{\ding{52}} 
    \\ \cline{2-16}

    \multicolumn{1}{|l|}{}
    & \texttt{MOV}
    & \multicolumn{2}{c|}{\ding{52}} 
    & \multicolumn{2}{c|}{\ding{52}} 
    & \multicolumn{1}{c|}{\ding{52}} 
    & \multicolumn{1}{c|}{\ding{52}} 
    & \multicolumn{1}{c|}{\ding{52}} 
    & \multicolumn{1}{c|}{\ding{52}}
    & \multicolumn{1}{c|}{\ding{52}} 
    & \multicolumn{1}{c|}{\ding{52}\footnote{Mobile.}}
    & \multicolumn{1}{c|}{\ding{52}} 
    & \multicolumn{1}{c|}{\ding{52}} 
    & \multicolumn{1}{c|}{\ding{54}}
    & \multicolumn{1}{c|}{\ding{54}}
    \\ \cline{2-16}
    
    \multicolumn{1}{|l|}{}
    & \texttt{GIF}
    & \multicolumn{2}{c|}{\ding{54}}
    & \multicolumn{2}{c|}{\ding{54}} 
    & \multicolumn{1}{c|}{\ding{52}} 
    & \multicolumn{1}{c|}{\ding{54}}
    & \multicolumn{1}{c|}{\ding{54}} 
    & \multicolumn{1}{c|}{\ding{54}}
    & \multicolumn{1}{c|}{\ding{54}} 
    & \multicolumn{1}{c|}{\ding{54}}
    & \multicolumn{1}{c|}{\ding{54}} 
    & \multicolumn{1}{c|}{\ding{54}} 
    & \multicolumn{1}{c|}{\ding{54}} 
    & \multicolumn{1}{c|}{\ding{54}}
    \\ \cline{2-16}

    \multicolumn{1}{|l|}{}
    & \texttt{AVI}
    & \multicolumn{2}{c|}{\ding{54}} 
    & \multicolumn{2}{c|}{\ding{54}}
    & \multicolumn{1}{c|}{\ding{54}} 
    & \multicolumn{1}{c|}{\ding{54}} 
    & \multicolumn{1}{c|}{\ding{54}}
    & \multicolumn{1}{c|}{\ding{52}}
    & \multicolumn{1}{c|}{\ding{52}} 
    & \multicolumn{1}{c|}{\ding{54}}
    & \multicolumn{1}{c|}{\ding{54}} 
    & \multicolumn{1}{c|}{\ding{54}}
    & \multicolumn{1}{c|}{\ding{54}}
    & \multicolumn{1}{c|}{\ding{54}}
    \\ \cline{2-16}
    
    \multicolumn{1}{|l|}{}
    & \texttt{3G2}
    & \multicolumn{2}{c|}{\ding{54}} 
    & \multicolumn{2}{c|}{\ding{54}} 
    & \multicolumn{1}{c|}{\ding{54}} 
    & \multicolumn{1}{c|}{\ding{54}} 
    & \multicolumn{1}{c|}{\ding{54}} 
    & \multicolumn{1}{c|}{\ding{52}}
    & \multicolumn{1}{c|}{\ding{54}} 
    & \multicolumn{1}{c|}{\ding{54}}
    & \multicolumn{1}{c|}{\ding{54}}
    & \multicolumn{1}{c|}{\ding{54}}
    & \multicolumn{1}{c|}{\ding{54}}
    & \multicolumn{1}{c|}{\ding{54}} 
    \\ \cline{2-16}
    
    \multicolumn{1}{|l|}{}
    & \texttt{3GP}
    & \multicolumn{2}{c|}{\ding{54}}
    & \multicolumn{2}{c|}{\ding{54}} 
    & \multicolumn{1}{c|}{\ding{54}}
    & \multicolumn{1}{c|}{\ding{54}}
    & \multicolumn{1}{c|}{\ding{54}}
    & \multicolumn{1}{c|}{\ding{52}}
    & \multicolumn{1}{c|}{\ding{54}}
    & \multicolumn{1}{c|}{\ding{54}}
    & \multicolumn{1}{c|}{\ding{54}}
    & \multicolumn{1}{c|}{\ding{54}} 
    & \multicolumn{1}{c|}{\ding{54}}
    & \multicolumn{1}{c|}{\ding{54}} 
    \\ \cline{2-16}
    
    \multicolumn{1}{|l|}{}
    & \texttt{ASF}
    & \multicolumn{2}{c|}{\ding{54}}
    & \multicolumn{2}{c|}{\ding{54}}
    & \multicolumn{1}{c|}{\ding{54}} 
    & \multicolumn{1}{c|}{\ding{54}} 
    & \multicolumn{1}{c|}{\ding{54}}
    & \multicolumn{1}{c|}{\ding{52}}
    & \multicolumn{1}{c|}{\ding{54}}
    & \multicolumn{1}{c|}{\ding{54}}
    & \multicolumn{1}{c|}{\ding{54}}
    & \multicolumn{1}{c|}{\ding{54}}
    & \multicolumn{1}{c|}{\ding{52}}
    & \multicolumn{1}{c|}{\ding{54}}
    \\ \cline{2-16}

    \multicolumn{1}{|l|}{}
    & \texttt{M4V}
    & \multicolumn{2}{c|}{\ding{54}} 
    & \multicolumn{2}{c|}{\ding{54}} 
    & \multicolumn{1}{c|}{\ding{54}} 
    & \multicolumn{1}{c|}{\ding{54}}
    & \multicolumn{1}{c|}{\ding{54}} 
    & \multicolumn{1}{c|}{\ding{52}}
    & \multicolumn{1}{c|}{\ding{54}} 
    & \multicolumn{1}{c|}{\ding{54}}
    & \multicolumn{1}{c|}{\ding{54}}
    & \multicolumn{1}{c|}{\ding{54}}
    & \multicolumn{1}{c|}{\ding{54}}
    & \multicolumn{1}{c|}{\ding{54}}
    \\ \cline{2-16}

    \multicolumn{1}{|l|}{}
    & \texttt{MKV}
    & \multicolumn{2}{c|}{\ding{54}} 
    & \multicolumn{2}{c|}{\ding{54}} 
    & \multicolumn{1}{c|}{\ding{54}}
    & \multicolumn{1}{c|}{\ding{54}}
    & \multicolumn{1}{c|}{\ding{54}} 
    & \multicolumn{1}{c|}{\ding{52}}
    & \multicolumn{1}{c|}{\ding{54}}
    & \multicolumn{1}{c|}{\ding{54}} 
    & \multicolumn{1}{c|}{\ding{54}} 
    & \multicolumn{1}{c|}{\ding{54}} 
    & \multicolumn{1}{c|}{\ding{52}}
    & \multicolumn{1}{c|}{\ding{54}}
    \\ \cline{2-16}
    
    \multicolumn{1}{|l|}{}
    & \texttt{MPE}
    & \multicolumn{2}{c|}{\ding{54}} 
    & \multicolumn{2}{c|}{\ding{54}}
    & \multicolumn{1}{c|}{\ding{54}}
    & \multicolumn{1}{c|}{\ding{54}} 
    & \multicolumn{1}{c|}{\ding{54}} 
    & \multicolumn{1}{c|}{\ding{52}}
    & \multicolumn{1}{c|}{\ding{54}} 
    & \multicolumn{1}{c|}{\ding{54}}
    & \multicolumn{1}{c|}{\ding{54}} 
    & \multicolumn{1}{c|}{\ding{54}}
    & \multicolumn{1}{c|}{\ding{54}}
    & \multicolumn{1}{c|}{\ding{54}} 
    \\ \cline{2-16}

    \multicolumn{1}{|l|}{}
    & \texttt{MPV}
    & \multicolumn{2}{c|}{\ding{54}} 
    & \multicolumn{2}{c|}{\ding{54}}  
    & \multicolumn{1}{c|}{\ding{54}}  
    & \multicolumn{1}{c|}{\ding{54}} 
    & \multicolumn{1}{c|}{\ding{54}}  
    & \multicolumn{1}{c|}{\ding{52}} 
    & \multicolumn{1}{c|}{\ding{54}} 
    & \multicolumn{1}{c|}{\ding{54}} 
    & \multicolumn{1}{c|}{\ding{54}} 
    & \multicolumn{1}{c|}{\ding{54}}  
    & \multicolumn{1}{c|}{\ding{54}} 
    & \multicolumn{1}{c|}{\ding{54}} 
    \\ \cline{2-16}
    
    \multicolumn{1}{|l|}{}
    & \texttt{WEBM}
    & \multicolumn{2}{c|}{\ding{54}} 
    & \multicolumn{2}{c|}{\ding{54}} 
    & \multicolumn{1}{c|}{\ding{54}} 
    & \multicolumn{1}{c|}{\ding{54}} 
    & \multicolumn{1}{c|}{\ding{54}} 
    & \multicolumn{1}{c|}{\ding{52}} 
    & \multicolumn{1}{c|}{\ding{54}} 
    & \multicolumn{1}{c|}{\ding{54}} 
    & \multicolumn{1}{c|}{\ding{54}} 
    & \multicolumn{1}{c|}{\ding{54}} 
    & \multicolumn{1}{c|}{\ding{52}} 
    & \multicolumn{1}{c|}{\ding{54}} 
    \\ \cline{2-16}

    \multicolumn{1}{|l|}{}
    & \texttt{WMV}
    & \multicolumn{2}{c|}{\ding{54}}  
    & \multicolumn{2}{c|}{\ding{54}} 
    & \multicolumn{1}{c|}{\ding{54}} 
    & \multicolumn{1}{c|}{\ding{54}} 
    & \multicolumn{1}{c|}{\ding{54}} 
    & \multicolumn{1}{c|}{\ding{52}} 
    & \multicolumn{1}{c|}{\ding{54}} 
    & \multicolumn{1}{c|}{\ding{54}} 
    & \multicolumn{1}{c|}{\ding{54}} 
    & \multicolumn{1}{c|}{\ding{54}} 
    & \multicolumn{1}{c|}{\ding{54}} 
    & \multicolumn{1}{c|}{\ding{54}} 
    \\ \cline{2-16}
    
    \multicolumn{1}{|l|}{}
    & \texttt{ASF}
    & \multicolumn{2}{c|}{\ding{54}} 
    & \multicolumn{2}{c|}{\ding{54}} 
    & \multicolumn{1}{c|}{\ding{54}} 
    & \multicolumn{1}{c|}{\ding{54}} 
    & \multicolumn{1}{c|}{\ding{54}} 
    & \multicolumn{1}{c|}{\ding{54}} 
    & \multicolumn{1}{c|}{\ding{54}} 
    & \multicolumn{1}{c|}{\ding{54}} 
    & \multicolumn{1}{c|}{\ding{54}} 
    & \multicolumn{1}{c|}{\ding{54}} 
    & \multicolumn{1}{c|}{\ding{52}} 
    & \multicolumn{1}{c|}{\ding{54}} 
    \\ \cline{2-16}

    \multicolumn{1}{|l|}{}
    & \texttt{FLV}
    & \multicolumn{2}{c|}{\ding{54}} 
    & \multicolumn{2}{c|}{\ding{54}} 
    & \multicolumn{1}{c|}{\ding{54}}  
    & \multicolumn{1}{c|}{\ding{54}} 
    & \multicolumn{1}{c|}{\ding{54}} 
    & \multicolumn{1}{c|}{\ding{54}} 
    & \multicolumn{1}{c|}{\ding{54}} 
    & \multicolumn{1}{c|}{\ding{54}} 
    & \multicolumn{1}{c|}{\ding{54}} 
    & \multicolumn{1}{c|}{\ding{54}} 
    & \multicolumn{1}{c|}{\ding{52}} 
    & \multicolumn{1}{c|}{\ding{54}} 
    \\ \cline{2-16}
    
    \multicolumn{1}{|l|}{}
    & \texttt{MPEG-1}
    & \multicolumn{2}{c|}{\ding{54}} 
    & \multicolumn{2}{c|}{\ding{54}}
    & \multicolumn{1}{c|}{\ding{54}} 
    & \multicolumn{1}{c|}{\ding{54}}
    & \multicolumn{1}{c|}{\ding{54}}
    & \multicolumn{1}{c|}{\ding{54}}
    & \multicolumn{1}{c|}{\ding{54}}
    & \multicolumn{1}{c|}{\ding{54}}
    & \multicolumn{1}{c|}{\ding{54}}
    & \multicolumn{1}{c|}{\ding{54}}
    & \multicolumn{1}{c|}{\ding{52}}
    & \multicolumn{1}{c|}{\ding{54}} 
    \\ \cline{2-16}

    \multicolumn{1}{|l|}{}
    & \texttt{MPEG-4}
    & \multicolumn{2}{c|}{\ding{54}} 
    & \multicolumn{2}{c|}{\ding{54}} 
    & \multicolumn{1}{c|}{\ding{54}} 
    & \multicolumn{1}{c|}{\ding{54}} 
    & \multicolumn{1}{c|}{\ding{54}} 
    & \multicolumn{1}{c|}{\ding{54}} 
    & \multicolumn{1}{c|}{\ding{54}} 
    & \multicolumn{1}{c|}{\ding{54}} 
    & \multicolumn{1}{c|}{\ding{54}} 
    & \multicolumn{1}{c|}{\ding{54}} 
    & \multicolumn{1}{c|}{\ding{52}} 
    & \multicolumn{1}{c|}{\ding{54}} 
    \\ \cline{2-16}
    
    \multicolumn{1}{|l|}{}
    & \texttt{AVC}
    & \multicolumn{2}{c|}{\ding{54}} 
    & \multicolumn{2}{c|}{\ding{54}} 
    & \multicolumn{1}{c|}{\ding{54}} 
    & \multicolumn{1}{c|}{\ding{54}} 
    & \multicolumn{1}{c|}{\ding{54}} 
    & \multicolumn{1}{c|}{\ding{54}} 
    & \multicolumn{1}{c|}{\ding{54}} 
    & \multicolumn{1}{c|}{\ding{54}} 
    & \multicolumn{1}{c|}{\ding{54}} 
    & \multicolumn{1}{c|}{\ding{54}} 
    & \multicolumn{1}{c|}{\ding{52}} 
    & \multicolumn{1}{c|}{\ding{54}} 
    \\ \cline{2-16}

    \multicolumn{1}{|l|}{}
    & \texttt{VP8}
    & \multicolumn{2}{c|}{\ding{54}} 
    & \multicolumn{2}{c|}{\ding{54}} 
    & \multicolumn{1}{c|}{\ding{54}} 
    & \multicolumn{1}{c|}{\ding{54}} 
    & \multicolumn{1}{c|}{\ding{54}} 
    & \multicolumn{1}{c|}{\ding{54}} 
    & \multicolumn{1}{c|}{\ding{54}} 
    & \multicolumn{1}{c|}{\ding{54}} 
    & \multicolumn{1}{c|}{\ding{54}} 
    & \multicolumn{1}{c|}{\ding{54}} 
    & \multicolumn{1}{c|}{\ding{52}} 
    & \multicolumn{1}{c|}{\ding{54}} 
    \\ \cline{2-16}
    
    \multicolumn{1}{|l|}{}
    & \texttt{VP9}
    & \multicolumn{2}{c|}{\ding{54}} 
    & \multicolumn{2}{c|}{\ding{54}} 
    & \multicolumn{1}{c|}{\ding{54}} 
    & \multicolumn{1}{c|}{\ding{54}} 
    & \multicolumn{1}{c|}{\ding{54}} 
    & \multicolumn{1}{c|}{\ding{54}} 
    & \multicolumn{1}{c|}{\ding{54}} 
    & \multicolumn{1}{c|}{\ding{54}} 
    & \multicolumn{1}{c|}{\ding{54}} 
    & \multicolumn{1}{c|}{\ding{54}} 
    & \multicolumn{1}{c|}{\ding{52}} 
    & \multicolumn{1}{c|}{\ding{54}} 
    \\ \cline{2-16}

    \multicolumn{1}{|l|}{}
    & \texttt{MWV2}
    & \multicolumn{2}{c|}{\ding{54}} 
    & \multicolumn{2}{c|}{\ding{54}} 
    & \multicolumn{1}{c|}{\ding{54}} 
    & \multicolumn{1}{c|}{\ding{54}} 
    & \multicolumn{1}{c|}{\ding{54}} 
    & \multicolumn{1}{c|}{\ding{54}} 
    & \multicolumn{1}{c|}{\ding{54}} 
    & \multicolumn{1}{c|}{\ding{54}} 
    & \multicolumn{1}{c|}{\ding{54}} 
    & \multicolumn{1}{c|}{\ding{54}} 
    & \multicolumn{1}{c|}{\ding{52}} 
    & \multicolumn{1}{c|}{\ding{54}} 
    \\ \cline{2-16}
    
    \multicolumn{1}{|l|}{}
    & \texttt{MWV3}
    & \multicolumn{2}{c|}{\ding{54}} 
    & \multicolumn{2}{c|}{\ding{54}} 
    & \multicolumn{1}{c|}{\ding{54}} 
    & \multicolumn{1}{c|}{\ding{54}} 
    & \multicolumn{1}{c|}{\ding{54}} 
    & \multicolumn{1}{c|}{\ding{54}} 
    & \multicolumn{1}{c|}{\ding{54}} 
    & \multicolumn{1}{c|}{\ding{54}} 
    & \multicolumn{1}{c|}{\ding{54}} 
    & \multicolumn{1}{c|}{\ding{54}} 
    & \multicolumn{1}{c|}{\ding{52}} 
    & \multicolumn{1}{c|}{\ding{54}} 
    \\ \hline
    
    \multicolumn{1}{|l|}{\multirow{6}{1.5cm}{\textbf{Captions}}}
    & \multicolumn{1}{|p{1.25cm}||}{Maximum Length}
    & \multicolumn{2}{p{2cm}|}{150} 
    & \multicolumn{1}{p{1cm}|}{2200} 
    & \multicolumn{1}{p{1cm}|}{2200} 
    & \multicolumn{1}{p{1cm}|}{N/A} 
    & \multicolumn{1}{p{1cm}|}{33,000} 
    & \multicolumn{1}{c|}{N/A}  
    & \multicolumn{1}{c|}{N/A}  
    & \multicolumn{1}{c|}{N/A} 
    & \multicolumn{1}{p{1cm}|}{280} 
    & \multicolumn{1}{p{1cm}|}{25} 
    & \multicolumn{1}{p{1cm}|}{55} 
    & \multicolumn{2}{p{2cm}|}{3000} 
    \\ \cline{2-16}

    \multicolumn{1}{|l|}{}
    & \multicolumn{1}{|p{1.25cm}||}{Truncate Length} 
    & \multicolumn{2}{p{2cm}|}{70} 
    & \multicolumn{1}{p{1cm}|}{125} 
    & \multicolumn{1}{p{1cm}|}{125} 
    & \multicolumn{1}{c|}{N/A} 
    & \multicolumn{3}{c|}{Custom\footnote{Default value for the truncate length is 63 characters, but it can be adjusted to a preferred value.}} 
    & \multicolumn{1}{c|}{N/A} 
    & \multicolumn{1}{p{1cm}|}{23} 
    & \multicolumn{1}{c|}{N/A} 
    & \multicolumn{1}{c|}{N/A} 
    & \multicolumn{2}{p{2cm}|}{200} 
    \\ \cline{2-16}

    \multicolumn{1}{|l|}{}
    & \multicolumn{1}{|p{1.25cm}||}{Auto Generate} 
    & \multicolumn{2}{c|}{\ding{52}} 
    & \multicolumn{3}{c|}{\ding{52}} 
    & \multicolumn{3}{c|}{\ding{52}} 
    & \multicolumn{1}{c|}{N/A} 
    & \multicolumn{1}{c|}{\ding{52}} 
    & \multicolumn{1}{c|}{\ding{52}} 
    & \multicolumn{1}{c|}{N/A} 
    & \multicolumn{1}{c|}{\ding{52}} 
    & \multicolumn{1}{c|}{N/A} 
    \\ \hline

    \multicolumn{1}{|p{1.5cm}|}{\textbf{Framerate}} 
    & FPS
    & \multicolumn{2}{c|}{30} 
    & \multicolumn{3}{c|}{$\geq$30} 
    & \multicolumn{4}{c|}{$\leq$30} 
    & \multicolumn{1}{c|}{30/60}
    & \multicolumn{2}{c|}{30} 
    & \multicolumn{1}{c|}{10-60} 
    & \multicolumn{1}{c|}{$\leq$30} 
    \\ \hline
    
    \multicolumn{1}{|p{1.75cm}|}{\textbf{Codec (V)}} 
    & \texttt{H.264}
    & \multicolumn{2}{c|}{\ding{52}} 
    & \multicolumn{3}{c|}{\ding{52}\footnote{Progressive scan: displays frames in sequence. High Profile: advanced H.264 setting. 2 B frames: two bidirectional frames in sequence. Closed GOP: self-contained video sequence. GOP at half frame rate: picture group length is half of fps. CABAC: H.264 entropy coding. Chroma 4:2:0: color compression method.}}
    & \multicolumn{4}{c|}{\ding{52}} 
    & \multicolumn{1}{c|}{\ding{52}} 
    & \multicolumn{2}{c|}{\ding{52}} 
    & \multicolumn{2}{c|}{\ding{52}} 
    \\ \hline

    \multicolumn{1}{|p{1.75cm}|}{\textbf{Codec (A)}} 
    & \texttt{AAC}
    & \multicolumn{2}{c|}{\ding{52}}
    & \multicolumn{3}{c|}{\ding{52}\footnote{Channels: Stereo (2 channels), Stereo+5.1 (surround with 6 channels). Sample rate: 96Khz (high-definition) or 48Khz (standard)}} 
    & \multicolumn{4}{c|}{\ding{52}\footnote{Stereo AAC: audio encoding with 2 channels. Bitrate: 128Kbps (standard quality).}} 
    & \multicolumn{1}{c|}{\ding{52}}
    & \multicolumn{2}{c|}{\ding{52}} 
    & \multicolumn{2}{c|}{\ding{52}} 
    \\ \hline
    
    \multicolumn{2}{|p{2cm}||}{\textbf{Maximum Resolution}}
    & \multicolumn{2}{p{2cm}|}{1080 x 1920} 
    & \multicolumn{1}{p{1cm}|}{1080 x 1920} 
    & \multicolumn{1}{p{1cm}|}{1080 x 1080} 
    & \multicolumn{1}{p{1cm}|}{1080 x 1920} 
    & \multicolumn{1}{p{1cm}|}{3840 x 2160} 
    & \multicolumn{1}{p{1cm}|}{3840 x 2160} 
    & \multicolumn{1}{c|}{N/A} 
    & \multicolumn{1}{p{1cm}|}{3840 x 2160} 
    & \multicolumn{1}{p{1cm}|}{1080 x 1920} 
    & \multicolumn{1}{p{1.55cm}|}{1080 x 1920} 
    & \multicolumn{1}{p{1cm}|}{1080 x 1920} 
    & \multicolumn{1}{p{1cm}|}{4096 x 2304} 
    & \multicolumn{1}{p{1cm}|}{1536 x 1920} 
    \\ \hline
    
    \multicolumn{2}{|p{2.75cm}||}{\textbf{Recommended Aspect Ratio}} 
    & \multicolumn{2}{p{2cm}|}{9:16} 
    & \multicolumn{1}{p{1cm}|}{9:16} 
    & \multicolumn{1}{p{1cm}|}{1:1} 
    & \multicolumn{1}{p{1cm}|}{9:16} 
    & \multicolumn{1}{p{1cm}|}{4:5} 
    & \multicolumn{1}{p{1cm}|}{16:9} 
    & \multicolumn{1}{p{1cm}|}{16:9} 
    & \multicolumn{1}{p{1cm}|}{9:16} 
    & \multicolumn{1}{c|}{N/A} 
    & \multicolumn{2}{p{2cm}|}{9:16} 
    & \multicolumn{1}{p{1cm}|}{1:2.4 - 2.4:1} 
    & \multicolumn{1}{c|}{N/A} 
    \\ \hline
    
    \multicolumn{1}{|l|}{\multirow{6}{1.5cm}{\textbf{Aspect Ratio}}}
    & 9:16
    & \multicolumn{2}{c|}{\ding{52}}  
    & \multicolumn{3}{c|}{\ding{52}} 
    & \multicolumn{4}{c|}{\ding{52}\footnote{Vertical 4:5, full portrait 9:16.}}
    & \multicolumn{1}{c|}{\ding{52}} 
    & \multicolumn{2}{c|}{\ding{52}} 
    & \multicolumn{1}{c|}{\ding{52}\footnote{Vertical videos are cropped into a square in the feed.}} 
    & \multicolumn{1}{c|}{\ding{54}}
    \\ \cline{2-16}

    \multicolumn{1}{|l|}{}
    & 16:9
    & \multicolumn{2}{c|}{\ding{52}}  
    & \multicolumn{3}{c|}{\ding{52}} 
    & \multicolumn{4}{c|}{\ding{52}}
    & \multicolumn{1}{c|}{\ding{52}} 
    & \multicolumn{2}{c|}{\ding{52}} 
    & \multicolumn{2}{c|}{\ding{52}} 
    \\ \cline{2-16}

    \multicolumn{1}{|l|}{}
    & 4:5
    & \multicolumn{2}{c|}{\ding{54}}  
    & \multicolumn{3}{c|}{\ding{52}} 
    & \multicolumn{4}{c|}{\ding{52}}
    & \multicolumn{1}{c|}{\ding{54}} 
    & \multicolumn{2}{c|}{\ding{54}}
    & \multicolumn{2}{c|}{\ding{54}} 
    \\ \cline{2-16}
    
    \multicolumn{1}{|l|}{}
    & 1:2.4
    & \multicolumn{2}{c|}{\ding{54}} 
    & \multicolumn{3}{c|}{\ding{54}} 
    & \multicolumn{4}{c|}{\ding{52}}
    & \multicolumn{1}{c|}{\ding{54}}
    & \multicolumn{2}{c|}{\ding{52}} 
    & \multicolumn{2}{c|}{\ding{54}} 
    \\ \cline{2-16}
    
    \multicolumn{1}{|l|}{}
    & 2.4:1
    & \multicolumn{2}{c|}{\ding{54}} 
    & \multicolumn{3}{c|}{\ding{54}} 
    & \multicolumn{4}{c|}{\ding{52}}
    & \multicolumn{1}{c|}{\ding{54}} 
    & \multicolumn{2}{c|}{\ding{54}}
    & \multicolumn{2}{c|}{\ding{54}} 
    \\ \cline{2-16}
    
    \multicolumn{1}{|l|}{}
    & 1:1
    & \multicolumn{2}{c|}{\ding{54}}  
    & \multicolumn{3}{c|}{\ding{54}} 
    & \multicolumn{4}{c|}{\ding{52}}
    & \multicolumn{1}{c|}{\ding{54}} 
    & \multicolumn{2}{c|}{\ding{54}} 
    & \multicolumn{2}{c|}{\ding{54}} 
    \\ \hline

    \multicolumn{1}{|l|}{\multirow{5}{1.5cm}{\textbf{Thumbnail}}} 
    & \multicolumn{1}{p{1.25cm}||}{Manual\footnote{Manual thumbnail selection.}} 
    & \multicolumn{2}{c|}{\ding{54}} 
    & \multicolumn{1}{c|}{\ding{52}} 
    & \multicolumn{1}{c|}{\ding{52}} 
    & \multicolumn{1}{c|}{\ding{54}} 
    & \multicolumn{4}{c|}{\ding{52}} 
    & \multicolumn{1}{c|}{\ding{52}} 
    & \multicolumn{2}{c|}{\ding{54}} 
    & \multicolumn{2}{c|}{\ding{52}} 
    \\ \cline{2-16}

    \multicolumn{1}{|l|}{}
    & \texttt{PNG} 
    & \multicolumn{2}{c|}{N/A} 
    & \multicolumn{1}{c|}{\ding{52}}
    & \multicolumn{1}{c|}{\ding{52}} 
    & \multicolumn{1}{c|}{N/A} 
    & \multicolumn{3}{c|}{\ding{52}} 
    & \multicolumn{1}{c|}{N/A}
    & \multicolumn{1}{c|}{\ding{52}}
    & \multicolumn{1}{c|}{\ding{52}} 
    & \multicolumn{1}{c|}{N/A} 
    & \multicolumn{2}{c|}{\ding{52}} 
    \\ \cline{2-16}
    
    \multicolumn{1}{|l|}{}
    & \texttt{JPG} 
    & \multicolumn{2}{c|}{N/A} 
    & \multicolumn{1}{c|}{\ding{52}}
    & \multicolumn{1}{c|}{\ding{52}} 
    & \multicolumn{1}{c|}{N/A} 
    & \multicolumn{3}{c|}{\ding{52}} 
    & \multicolumn{1}{c|}{N/A}
    & \multicolumn{1}{c|}{\ding{52}}
    & \multicolumn{1}{c|}{\ding{52}} 
    & \multicolumn{1}{c|}{N/A} 
    & \multicolumn{2}{c|}{\ding{52}} 
    \\ \cline{2-16}

    \multicolumn{1}{|l|}{}
    & Best Size
    & \multicolumn{2}{p{2cm}|}{1080 x 1920} 
    & \multicolumn{1}{p{1cm}|}{1080 x 1080} 
    & \multicolumn{1}{p{1cm}|}{1080 x 1080} 
    & \multicolumn{1}{c|}{N/A} 
    & \multicolumn{3}{p{3cm}|}{1200 x 628} 
    & \multicolumn{1}{c|}{N/A} 
    & \multicolumn{1}{p{1cm}|}{1200 x 1200} 
    & \multicolumn{1}{p{1cm}|}{1080 x 1920} 
    & \multicolumn{1}{c|}{N/A} 
    & \multicolumn{2}{p{2cm}|}{1200 x 628} 
    \\ \hline

    \multicolumn{2}{|p{2.75cm}||}{\textbf{Overlay\footnote{Additional overlay sticker including music, hashtags, locations, etc.}}} 
    & \multicolumn{2}{c|}{\ding{52}} 
    & \multicolumn{3}{c|}{\ding{52}} 
    & \multicolumn{4}{c|}{\ding{52}} 
    & \multicolumn{1}{c|}{\ding{52}} 
    & \multicolumn{2}{c|}{\ding{52}} 
    & \multicolumn{2}{c|}{\ding{52}} 
    \\ \hline
    
    \multicolumn{2}{|p{2.75cm}||}{\textbf{Pinned Feature}} 
    & \multicolumn{2}{c|}{\ding{52}} 
    & \multicolumn{1}{c|}{\ding{52}} 
    & \multicolumn{1}{c|}{\ding{52}} 
    & \multicolumn{1}{c|}{\ding{54}} 
    & \multicolumn{1}{c|}{\ding{52}} 
    & \multicolumn{1}{c|}{\ding{52}} 
    & \multicolumn{1}{c|}{N/A} 
    & \multicolumn{1}{c|}{\ding{54}} 
    & \multicolumn{1}{c|}{\ding{52}} 
    & \multicolumn{2}{c|}{\ding{54}} 
    & \multicolumn{1}{c|}{\ding{52}} 
    & \multicolumn{1}{c|}{\ding{54}} 
    \\ \hline

    \multicolumn{2}{|p{2.75cm}||}{\textbf{Create Polls}} 
    & \multicolumn{2}{c|}{\ding{52}} 
    & \multicolumn{1}{c|}{\ding{54}} 
    & \multicolumn{1}{c|}{\ding{54}} 
    & \multicolumn{1}{c|}{\ding{52}} 
    & \multicolumn{1}{c|}{\ding{52}} 
    & \multicolumn{1}{c|}{\ding{54}} 
    & \multicolumn{1}{c|}{\ding{52}} 
    & \multicolumn{1}{c|}{\ding{52}} 
    & \multicolumn{1}{c|}{\ding{52}} 
    & \multicolumn{1}{c|}{\ding{54}} 
    & \multicolumn{1}{c|}{\ding{52}} 
    & \multicolumn{1}{c|}{\ding{54}} 
    & \multicolumn{1}{c|}{\ding{52}} 
    \\ \hline

    \multicolumn{2}{|p{2.75cm}||}{\textbf{Carousel}} 
    & \multicolumn{2}{c|}{\ding{54}} 
    & \multicolumn{1}{c|}{\ding{52}} 
    & \multicolumn{1}{c|}{\ding{52}} 
    & \multicolumn{1}{c|}{\ding{54}} 
    & \multicolumn{1}{c|}{\ding{52}} 
    & \multicolumn{1}{c|}{\ding{52}} 
    & \multicolumn{1}{c|}{N/A} 
    & \multicolumn{1}{c|}{\ding{52}}
    & \multicolumn{1}{c|}{\ding{52}}
    & \multicolumn{1}{c|}{\ding{54}} 
    & \multicolumn{1}{c|}{\ding{54}} 
    & \multicolumn{1}{c|}{\ding{52}\footnote{Only on Desktop.}} 
    & \multicolumn{1}{c|}{\ding{54}} 
    \\ \hline

\end{longtable}
\end{landscape}
\restoregeometry

\NewPageCustom
\section{Image Specifications}\label{section:image-specifications}

Each social media platform has its own image specifications that content creators need to keep in mind when uploading images. These specifications can include requirements for image format, resolution, aspect ratio, file size, etc. Adhering to these specifications can help ensure that images display properly on different devices, and provide the best viewing experience for users. Table~\ref{tab:platforms-image-specifications} summarizes the image specifications for popular social media platforms.

\textbf{Instagram} supports a variety of image formats, including JPEG, PNG, and GIF. The recommended resolution for Instagram images is 1920x1080 pixels (square), 1080x1350 pixels (vertical), or 1080x608 pixels (horizontal), with an aspect ratio of 1:1, 4:5, or 16:9. The maximum file size for images on Instagram is 30 MB. 

\textbf{Facebook} supports a variety of image formats, including JPEG and PNG. The recommended resolution for Facebook images is 1200x1500 pixels (horizontal) or 1080x1080 pixels (square), with an aspect ratio of 1.91:1 or 1:1. The maximum file size for images on Facebook is 40 MB for JPEG format and 60 MB for PNG format. 

\textbf{Snapchat} supports JPEG and PNG image formats. The recommended resolution for Snapchat images is 1080x1920 pixels (vertical), with an aspect ratio of 9:16. The maximum file size for images on Snapchat is 5 MB. 

\textbf{LinkedIn} supports JPEG, PNG, and GIF image formats. The recommended resolution for LinkedIn images is 1200x627 pixels (horizontal), with an aspect ratio of 1.91:1. The maximum file size for images on LinkedIn is 3 MB. 

\begin{table}[ht]

    \centering
    \scriptsize
    \caption[Image specifications for different social media platforms.]{Image specifications for different social media platforms. TikTok has been omitted due to being a video sharing platform.}
    \label{tab:platforms-image-specifications}
    

        \begin{tabular}{|p{2.25cm}|p{1.5cm}|p{2cm}|p{2.5cm}|p{1.75cm}|p{1.75cm}|p{1.75cm}|}
        \cline{3-7}

        \multicolumn{2}{c|}{}
        & \textbf{Instagram} 
        & \textbf{Facebook} 
        & \textbf{Twitter} 
        & \textbf{SnapChat} 
        & \textbf{LinkedIn} 
        \\ \hline

        \multicolumn{2}{|p{3.75cm}||}{\textbf{File Size}} 
        & \multicolumn{1}{p{2cm}|}{30MB} 
        & \multicolumn{1}{p{2.5cm}|}{45MB for JPEG, 60MB for PNG} 
        & \multicolumn{1}{p{1.75cm}|}{20MB} 
        & \multicolumn{1}{p{1.75cm}|}{5MB} 
        & \multicolumn{1}{p{1.75cm}|}{3MB} 
        \\ \hline

        \multicolumn{1}{|l|}{\multirow{4}{2.25cm}{\textbf{Captions}}}
        & \multicolumn{1}{p{1.5cm}||}{Maximum Length}
        & \multicolumn{1}{p{2cm}|}{2200} 
        & \multicolumn{1}{p{2cm}|}{63206} 
        & \multicolumn{1}{p{1.75cm}|}{280} 
        & \multicolumn{1}{p{1.75cm}|}{250} 
        & \multicolumn{1}{p{1.75cm}|}{3000} 
        \\ \cline{2-7}

        \multicolumn{1}{|l|}{}
        & \multicolumn{1}{p{1.5cm}||}{Truncate Length} 
        & \multicolumn{1}{p{2cm}|}{125} 
        & \multicolumn{1}{p{2cm}|}{80} 
        & \multicolumn{1}{c|}{N/A} 
        & \multicolumn{1}{c|}{N/A} 
        & \multicolumn{1}{p{1.75cm}|}{40} 
        \\ \hline 

        \multicolumn{1}{|l|}{\multirow{5}{2.25cm}{\textbf{File Format}}}
        & \multicolumn{1}{p{1.25cm}||}{\texttt{JPEG}}
        & \multicolumn{1}{c|}{\ding{52}} 
        & \multicolumn{1}{c|}{\ding{52}} 
        & \multicolumn{1}{c|}{\ding{52}} 
        & \multicolumn{1}{c|}{\ding{52}} 
        & \multicolumn{1}{c|}{\ding{52}} 
        \\ \cline{2-7} 
        
        \multicolumn{1}{|l|}{} 
        & \multicolumn{1}{p{1.25cm}||}{\texttt{PNG}}
        & \multicolumn{1}{c|}{\ding{52}} 
        & \multicolumn{1}{c|}{\ding{52}} 
        & \multicolumn{1}{c|}{\ding{52}} 
        & \multicolumn{1}{c|}{\ding{52}} 
        & \multicolumn{1}{c|}{\ding{52}} 
        \\ \cline{2-7} 
        
        \multicolumn{1}{|l|}{} 
        & \multicolumn{1}{p{1.25cm}||}{\texttt{GIF}}
        & \multicolumn{1}{c|}{\ding{52}\tablefootnote{Non-animated.}} 
        & \multicolumn{1}{c|}{\ding{52}} 
        & \multicolumn{1}{c|}{\ding{52}} 
        & \multicolumn{1}{c|}{\ding{54}} 
        & \multicolumn{1}{c|}{\ding{54}} 
        \\ \cline{2-7} 
        
        \multicolumn{1}{|l|}{} 
        & \multicolumn{1}{p{1.25cm}||}{\texttt{BMP}}
        & \multicolumn{1}{c|}{\ding{52}} 
        & \multicolumn{1}{c|}{\ding{52}} 
        & \multicolumn{1}{c|}{\ding{54}} 
        & \multicolumn{1}{c|}{\ding{54}} 
        & \multicolumn{1}{c|}{\ding{54}} 
        \\ \cline{2-7} 
        
        \multicolumn{1}{|l|}{} 
        & \multicolumn{1}{p{1.25cm}||}{\texttt{TIFF}}
        & \multicolumn{1}{c|}{\ding{54}} 
        & \multicolumn{1}{c|}{\ding{52}} 
        & \multicolumn{1}{c|}{\ding{54}} 
        & \multicolumn{1}{c|}{\ding{54}} 
        & \multicolumn{1}{c|}{\ding{54}} 
        \\ \hline 

        \multicolumn{1}{|l|}{\multirow{3}{2.25cm}{\textbf{Supported Orientation 
        }}}
        & \multicolumn{1}{p{1.5cm}||}{Portrait}
        & \multicolumn{1}{c|}{\ding{52}} 
        & \multicolumn{1}{c|}{\ding{52}}  
        & \multicolumn{1}{c|}{\ding{52}} 
        & \multicolumn{1}{c|}{\ding{52}} 
        & \multicolumn{1}{c|}{\ding{52}} 
        \\ \cline{2-7} 
        
        \multicolumn{1}{|l|}{} 
        & \multicolumn{1}{p{1.5cm}||}{Landscape}
        & \multicolumn{1}{c|}{\ding{52}} 
        & \multicolumn{1}{c|}{\ding{52}}  
        & \multicolumn{1}{c|}{\ding{52}} 
        & \multicolumn{1}{c|}{\ding{52}} 
        & \multicolumn{1}{c|}{\ding{52}} 
        \\ \cline{2-7} 
        
        \multicolumn{1}{|l|}{} 
        & \multicolumn{1}{p{1.5cm}||}{Square}
        & \multicolumn{1}{c|}{\ding{54}} 
        & \multicolumn{1}{c|}{\ding{54}}  
        & \multicolumn{1}{c|}{\ding{54}} 
        & \multicolumn{1}{c|}{\ding{54}} 
        & \multicolumn{1}{c|}{\ding{52}} 
        \\ \hline 

        \multicolumn{1}{|l|}{\multirow{3}{2.25cm}{\textbf{Recommended Resolution}}}
        & \multicolumn{1}{p{1.5cm}||}{Portrait}
        & \multicolumn{1}{l|}{\multirow{3}{2cm}{1920 × 1080}} 
        & \multicolumn{1}{l|}{\multirow{3}{2cm}{1200 × 1500}}  
        & \multicolumn{1}{l|}{\multirow{3}{1.75cm}{1600 × 900}}  
        & \multicolumn{1}{l|}{\multirow{3}{1.75cm}{1080 × 1920}} 
        & \multicolumn{1}{|p{1.75cm}|}{1080 x 1350} 
        \\ \hhline{~-~~~~-} 

        \multicolumn{1}{|l|}{} 
        & \multicolumn{1}{p{1.5cm}||}{Landscape}
        & 
        &  
        & 
        & 
        & \multicolumn{1}{|p{1.75cm}|}{1200 x 627} 
        \\ \hhline{~-~~~~-} 
        
        \multicolumn{1}{|l|}{} 
        & \multicolumn{1}{p{1.5cm}||}{Square}
        &  
        & 
        & 
        & 
        & \multicolumn{1}{|p{1.75cm}|}{1200 x 1200} 
        \\ \hline 
        
        \multicolumn{2}{|p{3.75cm}||}{\textbf{Additional Overlay}} 
        & \multicolumn{1}{c|}{\ding{52}} 
        & \multicolumn{1}{c|}{\ding{52}} 
        & \multicolumn{1}{c|}{\ding{54}} 
        & \multicolumn{1}{c|}{\ding{52}} 
        & \multicolumn{1}{c|}{\ding{54}} 
        \\ \hline

        \multicolumn{2}{|p{3.75cm}||}{\textbf{Carousel}} 
        & \multicolumn{1}{c|}{\ding{52}} 
        & \multicolumn{1}{c|}{\ding{52}} 
        & \multicolumn{1}{c|}{\ding{52}\tablefootnote{Up to 4.}} 
        & \multicolumn{1}{c|}{\ding{54}} 
        & \multicolumn{1}{c|}{\ding{52}} 
        \\ \hline 
        
        \end{tabular}


\end{table}
\FloatBarrier

\takeaway{\begin{itemize}
    \item Every social media platform has unique image specifications.
    \item Proper adherence to image specifications ensures professional and optimal image display, neglect can lead to poor visuals or upload failures.
    \item For AI-generated soccer highlights, considering format, resolution, aspect ratio, and file size for each platform can ensure that thrilling moments are showcased effectively, enhancing fan engagement.
\end{itemize}}

\clearpage
\chapter{Social Media Strategies: European Soccer Leagues}\label{section:social-media-strategies-leagues}

In this chapter, we the compare social media strategies of $7$ major European soccer leagues, namely the English Premier League (England), Bundesliga (Germany), La Liga (Spain), Serie A (Italy), Ligue 1 (France), Allsvenskan (Sweden), and Eliteserien (Norway). 

\section{Followers and Engagement}

First, we compare the aforementioned leagues in terms of social media followers and engagement. The goal of the analysis is to determine the relative popularity of each league in terms of their online presence. The data was collected from various sources including TikTok~\cite{PremierLeagueTikTok, BundesligaTikTok, LaLigaTikTok, SerieATikTok, Ligue1TikTok, EliteserienTikTok}, Instagram~\cite{PremierLeagueInstagram, BundesligaInstagram, LaLigaInstagram, SerieAInstagram, Ligue1Instagram, AllsvenskanInstagram, EliteserienInstagram}, Facebook~\cite{PremierLeagueFacebook, BundesligaFacebook, LaLigaFacebook, SerieAFacebook, Ligue1Facebook, AllsvenskanFacebook, EliteserienFacebook}, Twitter~\cite{PremierLeagueTwitter, BundesligaTwitter, LaLigaTwitter, SerieATwitter, Ligue1Twitter, AllsvenskanTwitter, EliteserienTwitter}, and YouTube~\cite{PremierLeagueYouTube, BundesligaYouTube, LaLigaYouTube, SerieAYouTube, Ligue1YouTube, EliteserienYouTube}. 

\textbf{Number of followers:} It is expected that the English Premier League will have the largest number of social media followers, given its global appeal and popularity. Bundesliga is also expected to have a significant following, as it is known for its high-quality soccer and passionate fan culture. La Liga is another league that is expected to perform well in terms of social media followers, given the presence of two of the world's biggest soccer clubs, Real Madrid and Barcelona. Serie A, Ligue 1, Allsvenskan, and Eliteserien are expected to have relatively smaller online followings, but their respective fan bases are still expected to be significant. Overall, the comparison of the number of social media followers of the European soccer leagues will provide valuable insights into the popularity and reach of each league across the world. 

\textbf{Engagement rate:} The engagement rates of the social media accounts of the seven major European soccer leagues are also being analyzed. This analysis is being carried out to determine how effectively each league is engaging with their online followers and fans. Engagement rate refers to the percentage of people who engage with a social media post or account. This includes likes, comments, shares, and any other type of interaction with the content. A higher engagement rate is an indicator that the content is resonating well with the audience and that they are actively participating in the conversation. To calculate the engagement rate, the number of interactions (likes, comments, shares, etc.) is divided by the number of followers, and the result is multiplied by 100 to give a percentage~\cite{EngagementCalculator2023}. A high engagement rate indicates that the content is engaging and is resonating well with the audience. 

By analyzing the engagement rates of each league's social media accounts, it is possible to identify which leagues are creating content that resonates well with their fans and followers. This information can be used to inform future social media strategies and create more engaging content that will increase online followers and improve overall engagement rates. However, it should be noted that calculating the engagement rate for each platform requires the number of views for each post, which unfortunately is not publicly available. Consequently, the total number of followers has been used as the primary source of information to estimate the engagement rate accurately. While this estimation may not be entirely precise, it still provides a good indication of the relative engagement rates of each league's social media accounts. Furthermore, it should be recognized that the engagement rate can vary significantly between different platforms, depending on the type of content being posted and the audience's preferences. Therefore, while the engagement rate is a useful metric for measuring the effectiveness of social media strategies, it should be used in conjunction with other metrics to obtain a more comprehensive understanding of the online presence and engagement of each soccer league. 

\textbf{Other metrics:} There are several other metrics that can be used to measure the online presence and engagement of each soccer league. One important metric is the reach, which refers to the total number of people who have seen a post or piece of content. This metric can provide valuable insights into the size of each league's audience and the effectiveness of their content in reaching a wide audience. 

Another metric is the click-through rate (CTR), which measures the percentage of people who click on a link or call-to-action in a post. This metric can be particularly useful for evaluating the effectiveness of campaigns or promotions that aim to drive traffic to a specific website or landing page. 

Other useful metrics include impressions (the number of times a post or piece of content is displayed), shares (the number of times a post is shared by users), and comments (the number of comments left by users). By analyzing these metrics alongside the engagement rate, it is possible to gain a more comprehensive understanding of each league's social media strategy and effectiveness. 

\begin{figure}[ht]
    
    \centering
    \caption[Comparison of European soccer leagues in terms of social media followers.]{Comparison of European soccer leagues in terms of social media followers.}
    \label{fig:leagues-followers}
    
    \begin{tikzpicture}
        \begin{axis}[
            ybar, area legend, 
            width=15cm,
            height=8cm,
            enlargelimits=0.15,
            legend style={at={(0.5,+1.15)},
              anchor=north,legend columns=-1},
            ylabel={Followers (Million)},
            symbolic x coords={
                Premier League,
                Bundesliga,
                LaLiga,
                Serie A,
                Ligue 1,
                Allsvenskan,
                Eliteserien
            },
            xtick=data,
            xticklabel style={yshift=-1mm}, 
            nodes near coords={}, 
            x tick label style={rotate=20,anchor=east,font=\small}, 
            xmin=Premier League, 
            xmax=Eliteserien, 
            bar width=5.95, 
            ]

            \addplot coordinates {
                (Premier League, 8.9)
                (Bundesliga, 4.5)
                (LaLiga, 14.6)
                (Serie A, 1.9)
                (Ligue 1, 6.1)
                (Allsvenskan, 0.023)
                (Eliteserien, 0.053)
            };
            
            \addplot coordinates {
                (Premier League, 63.3)
                (Bundesliga, 12)
                (LaLiga, 44.7)
                (Serie A, 8.8)
                (Ligue 1, 3.8)
                (Allsvenskan, 0.083)
                (Eliteserien, 0.035)
            };
            
            \addplot coordinates {
                (Premier League, 53)
                (Bundesliga, 9.9)
                (LaLiga, 81.5)
                (Serie A, 4.7)
                (Ligue 1, 9.3)
                (Allsvenskan, 0.06)
                (Eliteserien, 0.056)
            };            

            \addplot coordinates {
                (Premier League, 38.8)
                (Bundesliga, 2.4)
                (LaLiga, 11.2)
                (Serie A, 0.7)
                (Ligue 1, 1.4)
                (Allsvenskan, 0.032)
                (Eliteserien, 0.019)
            };
            
            \addplot coordinates {
                (Premier League, 3.7)
                (Bundesliga, 3.7)
                (LaLiga, 8.8)
                (Serie A, 8.6)
                (Ligue 1, 1.1)
                (Allsvenskan, 0)
                (Eliteserien, 0.003)
            };            
            
            \legend{TikTok, Instagram, Facebook, Twitter, YouTube}
            
        \end{axis}

        \begin{axis}[
            width=5cm,
            height=3cm,
            enlargelimits=0.5,
            ybar,
            ymin=0.01, ymax=0.10,
            ytick={0.01, 0.05, 0.10},
            yticklabels={0.01, 0.05, 0.10},
            ylabel={},
            symbolic x coords={
                Allsvenskan,
                Eliteserien
            },
            xtick=data,
            nodes near coords={},
            x tick label style={rotate=30,anchor=east,font=\tiny},
            ymin=0.01, ymax=0.10,
            bar width=3.95,
            at={(rel axis cs:0.83,0.99)},
            anchor={north west}
            ]

            \addplot coordinates {
                (Allsvenskan, 0.023)
                (Eliteserien, 0.053)
            };   
            
            \addplot coordinates {
                (Allsvenskan, 0.083)
                (Eliteserien, 0.035)
            };
            
            \addplot coordinates {
                (Allsvenskan, 0.06)
                (Eliteserien, 0.056)
            };  
            
            \addplot coordinates {
                (Allsvenskan, 0.032)
                (Eliteserien, 0.019)
            };
            
            \addplot coordinates {
                (Allsvenskan, 0)
                (Eliteserien, 0.003)
            };

        \end{axis}
    \end{tikzpicture}
    
\end{figure}
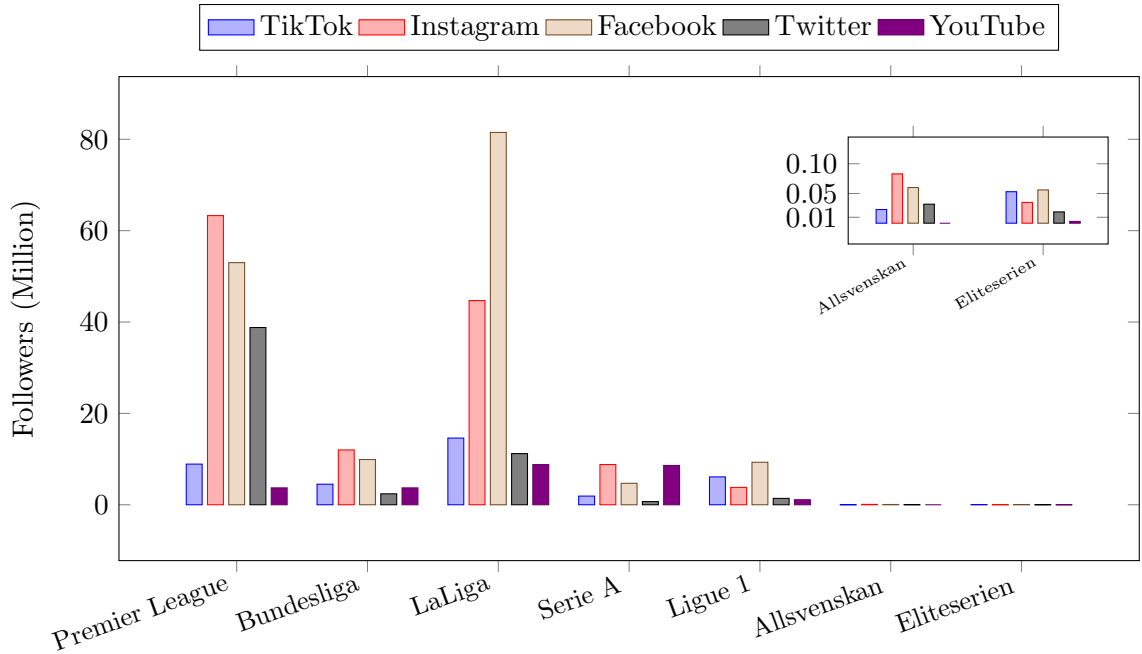

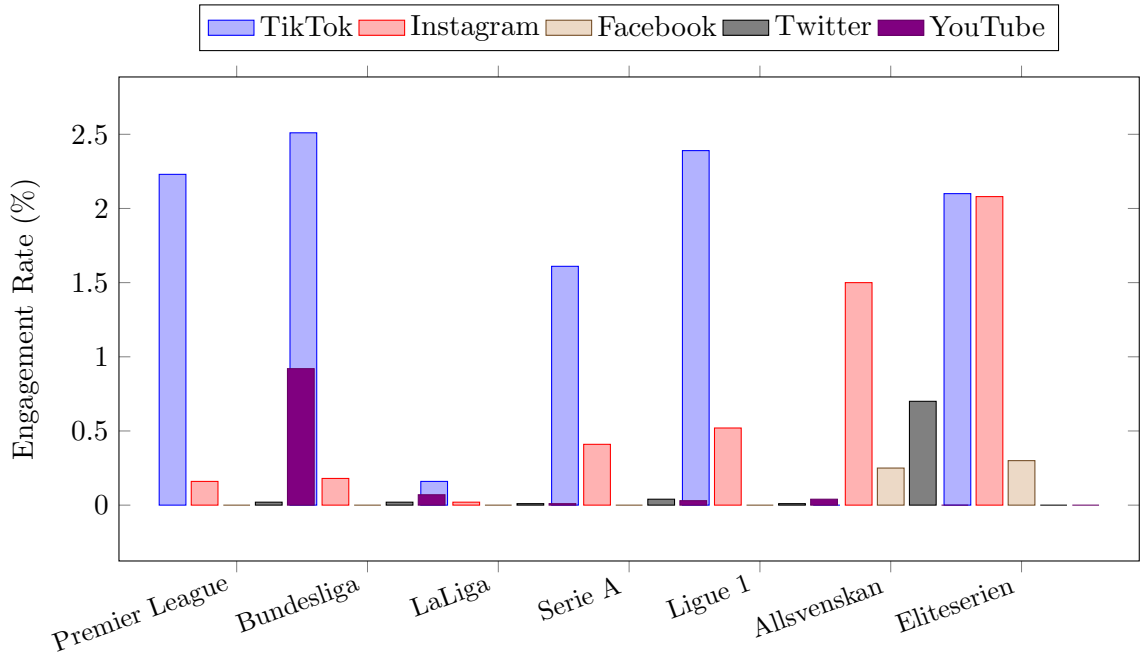
\begin{figure}[ht]

    \centering
    \caption[Comparison of European soccer leagues in terms of social media engagement.]{Comparison of European soccer leagues in terms of social media engagement.}
    \label{fig:leagues-engagement}
    
    \begin{tikzpicture}
        \begin{axis}[
            ybar, area legend, 
            width=15cm,
            height=8cm,
            enlargelimits=0.15,
            legend style={at={(0.5,+1.15)},
              anchor=north,legend columns=-1},
            ylabel={Engagement Rate (\%)},
            symbolic x coords={
                Premier League,
                Bundesliga,
                LaLiga,
                Serie A,
                Ligue 1,
                Allsvenskan,
                Eliteserien
            },
            xtick=data,
            xticklabel style={yshift=-1mm}, 
            nodes near coords={}, 
            x tick label style={rotate=20,anchor=east,font=\small}, 
            xmin=Premier League, 
            xmax=Eliteserien, 
            ]
            
            \addplot coordinates {
                (Premier League, 2.23)
                (Bundesliga, 2.51)
                (LaLiga, 0.16)
                (Serie A, 1.61)
                (Ligue 1, 2.39)
                (Allsvenskan, 0)
                (Eliteserien, 2.1)
            };
            
            \addplot coordinates {
                (Premier League, 0.16)
                (Bundesliga, 0.18)
                (LaLiga, 0.02)
                (Serie A, 0.41)
                (Ligue 1, 0.52)
                (Allsvenskan, 1.50)
                (Eliteserien, 2.08)
            };

            \addplot coordinates {
                (Premier League, 0)
                (Bundesliga, 0)
                (LaLiga, 0)
                (Serie A, 0)
                (Ligue 1, 0)
                (Allsvenskan, 0.25)
                (Eliteserien, 0.30)
            };
            
            \addplot coordinates {
                (Premier League, 0.02)
                (Bundesliga, 0.02)
                (LaLiga, 0.01)
                (Serie A, 0.04)
                (Ligue 1, 0.01)
                (Allsvenskan, 0.7)
                (Eliteserien, 0)
            };
            
            \addplot coordinates {
                (Premier League, 0.92)
                (Bundesliga, 0.07)
                (LaLiga, 0.01)
                (Serie A, 0.03)
                (Ligue 1, 0.04)
                (Allsvenskan, 0)
                (Eliteserien, 0)
            };
                        
            \legend{TikTok, Instagram, Facebook, Twitter, YouTube}
            
        \end{axis}
    \end{tikzpicture}

\end{figure}

Table~\ref{tab:leagues-followers} presents a comparison of selected European soccer leagues in terms of social media followers (the number of followers in millions) and engagement rate (the percentage of followers actively interacting with content)~\cite{EngagementCalculator2023}. Figures~\ref{fig:leagues-followers} and \ref{fig:leagues-engagement} provide visual comparisons for these metrics across different leagues.

\begin{table}[ht]

    \centering
    \scriptsize
    \caption[Comparison of European soccer leagues in terms of social media followers and engagement.]{Comparison of European soccer leagues in terms of social media followers and engagement.}
    \label{tab:leagues-followers}
    

        \begin{tabular}{|c|l||c|c|c|c|c|c|c|}
        \cline{3-9}

        \multicolumn{2}{c|}{} 
        & \multicolumn{1}{|p{1.2cm}|}{\textbf{Premier League}} 
        & \multicolumn{1}{|p{1.7cm}|}{\textbf{Bundesliga}} 
        & \multicolumn{1}{|p{1.1cm}|}{\textbf{LaLiga}} 
        & \multicolumn{1}{|p{1.1cm}|}{\textbf{Serie A}} 
        & \multicolumn{1}{|p{1.1cm}|}{\textbf{Ligue 1}} 
        & \multicolumn{1}{|p{1.7cm}|}{\textbf{Allsvenskan}} 
        & \multicolumn{1}{|p{1.6cm}|}{\textbf{Eliteserien}} 
        \\ \hline

        \multicolumn{1}{|c|}{\multirow{5}{*}{\rotatebox{90}{\textbf{\parbox{1.5cm}{Followers\\(Million)}}}}}
        & TikTok 
        & 8.9 
        & 4.5 
        & 14.6 
        & 1.9 
        & 6.1 
        & 0.023 
        & 0.053 
        \\ \cline{2-9} 
        
        & Instagram 
        & 63.3 
        & 12 
        & 44.7 
        & 8.8 
        & 3.8 
        & 0.083 
        & 0.035 
        \\ \cline{2-9} 
        
        & Facebook 
        & 53 
        & 9.9 
        & 81.5 
        & 4.7 
        & 9.3 
        & 0.06 
        & 0.056 
        \\ \cline{2-9} 
        
        & Twitter 
        & 38.8 
        & 2.4 
        & 11.2 
        & 0.7 
        & 1.4 
        & 0.032 
        & 0.019 
        \\ \cline{2-9}
                
        & YouTube 
        & 3.7 
        & 3.7 
        & 8.8 
        & 8.6 
        & 1.1 
        & 0 
        & 0.003 
        \\ \hline \hline
        
        \multicolumn{1}{|c|}{\multirow{5}{*}{\rotatebox{90}{\textbf{\parbox{1.5cm}{Engagem.\\Rate (\%)}}}}}
        & TikTok 
        & 2.23 
        & 2.51 
        & 0.16 
        & 1.61 
        & 2.39 
        & 0 
        & 2.1 
        \\ \cline{2-9} 
        
        & Instagram 
        & 0.16 
        & 0.18 
        & 0.02 
        & 0.41 
        & 0.52 
        & 1.50 
        & 2.08 
        \\ \cline{2-9} 
                
        & Facebook
        & 0 
        & 0 
        & 0 
        & 0 
        & 0 
        & 0.25 
        & 0.30 
        \\ \cline{2-9} 
        
        & Twitter 
        & 0.02 
        & 0.02 
        & 0.01 
        & 0.04 
        & 0.01 
        & 0.7 
        & 0 
        \\ \cline{2-9} 
        
        & YouTube 
        & 0.92 
        & 0.07 
        & 0.01 
        & 0.03 
        & 0.04 
        & 0 
        & 0 
        \\ \hline 
        
        \end{tabular}
        
    
\end{table}
\FloatBarrier

\FloatBarrier
\FloatBarrier

\takeaway{\begin{itemize}
    \item LaLiga and Premier League dominate in terms of the sheer number of followers. LaLiga leads in terms of popularity on Facebook, where Premier League is most popular on Instagram. 
    \item Serie A, despite having a more subdued presence on other platforms, showcases a strong following on YouTube, nearly rivaling LaLiga.
    \item Ligue 1 has an emerging prominence on TikTok.
    \item Allsvenskan and Eliteserien have modest number of followers in comparison to bigger leagues. However, their engagement rates, especially on Instagram, are notably high.
    \item There is a discrepancy between the relatively low number of followers and the relatively high engagement rate for some platforms (e.g., engagement rate for TikTok is the highest among different platforms for most leagues, but TikTok is never the platform with the highest number of followers).
\end{itemize}}

\NewPageCustom
\section{Content Cadence}

Content cadence refers to the pattern of content sharing, in terms of number and spacing, i.e., frequency. An analysis of publishing frequency in TikTok, Instagram, Facebook, Twitter, and YouTube for different soccer leagues has been conducted to understand social media content cadence. The aim of the analysis was to explore the frequency at which social media content is published on each platform and in various modules. By examining these patterns, it is possible to identify trends and best practices for social media content creation and management. 

The analysis was conducted by collecting data from various social media accounts across different modules. The collected data was then analyzed to identify patterns in the frequency of content publishing. The results of the analysis show that social media content cadence varies across different platforms and modules. 

It was shown by the data that more content is published on Instagram and Facebook by the administration of top European soccer leagues, and that videos and images are more popular than texts. Social media activity is not very high for Allsvenskan and Ellitserien. It was observed that Allsvenskan has no official account on YouTube and has not posted a single video on TikTok since 2023. Ellitserien has not posted anything on Twitter since January 2023 and no new video on YouTube in the last 3 months. 

\begin{figure}[ht]

    \centering
    \caption[Comparison of European soccer leagues in terms of the frequency of their posts on YouTube.]{Comparison of European soccer leagues in terms of the frequency of their posts on YouTube.}
    \label{fig:leagues-frequency-youtube}
    
    \begin{tikzpicture}
    
        \begin{axis}[
            ybar, area legend, 
            width=15cm,
            height=7cm,
            enlargelimits=0.15,
            legend style={at={(0.5,+1.15)},
              anchor=north,legend columns=-1},
            ylabel={Frequency of Posts},
            symbolic x coords={
                Premier League,
                Bundesliga,
                LaLiga,
                Serie A,
                Ligue 1,
                Allsvenskan,
                Eliteserien
            },
            xtick=data,
            xticklabel style={yshift=-1mm}, 
            nodes near coords={}, 
            x tick label style={rotate=20,anchor=east,font=\small}, 
            xmin=Premier League, 
            xmax=Eliteserien, 
            bar width=5.95, 
            ]
            
            \addplot coordinates {
                (Premier League, 1)
                (Bundesliga, 2)
                (LaLiga, 5)
                (Serie A, 4)
                (Ligue 1, 1)
                (Allsvenskan, 0)
                (Eliteserien, 0)
            };

            \addplot coordinates {
                (Premier League, 1)
                (Bundesliga, 1)
                (LaLiga, 2)
                (Serie A, 2)
                (Ligue 1, 0)
                (Allsvenskan, 0)
                (Eliteserien, 0)
            };    
            
            \legend{Video+Text, Shorts+Text}
        \end{axis}
        
    \end{tikzpicture}

\end{figure}
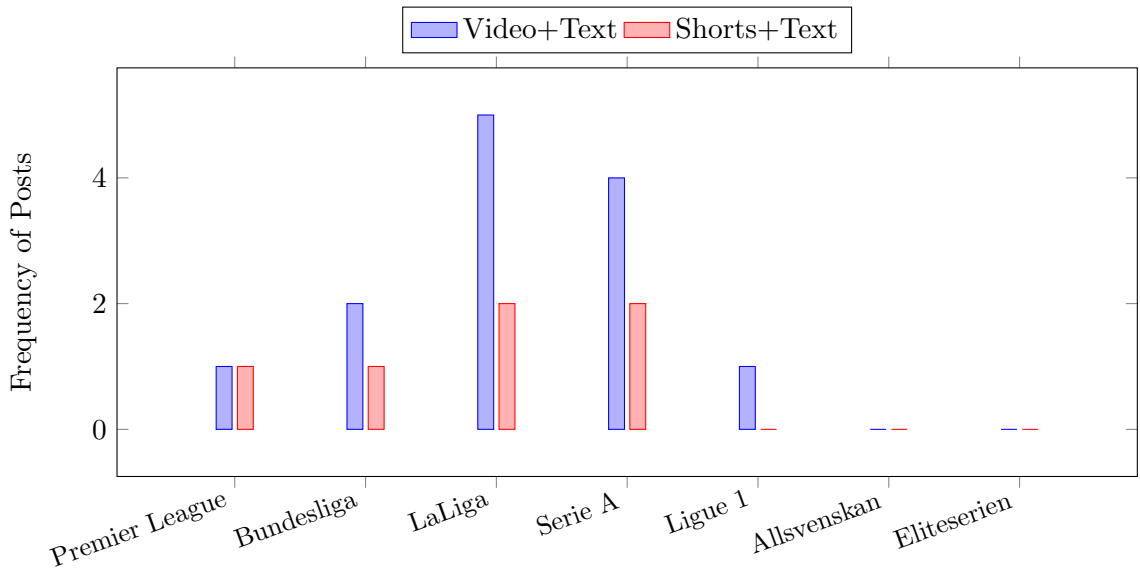

Table~\ref{tab:leagues-frequency} presents a comparison of selected European soccer leagues in terms of social media content cadence. Figures~\ref{fig:leagues-frequency-tiktok}, \ref{fig:leagues-frequency-instagram}, \ref{fig:leagues-frequency-facebook}, \ref{fig:leagues-frequency-twitter}, and \ref{fig:leagues-frequency-youtube} present a visual comparison of the frequency of content posting on various social media platforms for different leagues.

\begin{table}[ht]

    \centering
    \caption[Comparison of European leagues in terms of the frequency of their posts on social media, per platform and modality.]{Comparison of European leagues in terms of the frequency of their posts on social media, per platform and modality.}
    \label{tab:leagues-frequency}
    
    \resizebox{\textwidth}{!}{
    
        \begin{tabular}{|cl||c|c|c|c|c|c|c|}
        \hhline{~~-------}

        \multicolumn{2}{c|}{} 
        & \textbf{Premier League}
        & \textbf{Bundesliga}
        & \textbf{Laliga} 
        & \textbf{Serie A} 
        & \textbf{Ligue 1} 
        & \textbf{Allsvenskan} 
        & \textbf{Elliteserien} 
        \\ \hline
        
        \multicolumn{1}{|c|}{\textbf{TikTok}} 
        & Video 
        & 2 
        & 1 
        & 4 
        & 1 
        & 1 
        & 0 
        & 0.14 
        \\ \hline 

        \multicolumn{1}{|c|}{\multirow{3}{*}{\textbf{Instagram}}} 
        & Video + Text 
        & 7 
        & 8 
        & 30 
        & 2 
        & 2 
        & 0.28 
        & 0.43 
        \\ \cline{2-9} 
        
        \multicolumn{1}{|c|}{} 
        & Image + Text 
        & 18 
        & 3 
        & 7 
        & 6 
        & 1 
        & 0.42 
        & 0.57 
        \\ \cline{2-9} 
        
        \multicolumn{1}{|c|}{} 
        & Story 
        & 0 
        & 3 
        & 12 
        & 6 
        & 4 
        & 0 
        & 0 
        \\ \hline 
  
        \multicolumn{1}{|c|}{\multirow{3}{*}{\textbf{Facebook}}} 
        & Video + Text 
        & 4 
        & 1 
        & 13 
        & 1 
        & 5 
        & 0.21 
        & 0.14 
        \\ \cline{2-9} 

        \multicolumn{1}{|c|}{} 
        & Image + Text 
        & 16 
        & 8 
        & 12 
        & 3 
        & 5 
        & 0.28 
        & 0.57 
        \\ \cline{2-9} 
        
        \multicolumn{1}{|c|}{} 
        & Story 
        & 0 
        & 3 
        & 8 
        & 0 
        & 8 
        & 0 
        & 0 
        \\ \hline 
        
        \multicolumn{1}{|c|}{\multirow{4}{*}{\textbf{Twitter}}} 
        & Video + Text 
        & 3 
        & 3 
        & 5 
        & 3 
        & 1 
        & 0.35 
        & 0 
        \\ \cline{2-9} 
        
        \multicolumn{1}{|c|}{} 
        & Image + Text 
        & 4 
        & 1 
        & 5 
        & 6 
        & 10 
        & 0.5 
        & 0 
        \\ \cline{2-9} 

        \multicolumn{1}{|c|}{} 
        & Text 
        & 1 
        & 0 
        & 0 
        & 1 
        & 0 
        & 0 
        & 0 
        \\ \cline{2-9} 
        
        \multicolumn{1}{|c|}{} 
        & Retweet 
        & 21 
        & 2 
        & 4 
        & 0 
        & 0 
        & 0 
        & 0 
        \\ \hline 
        
        \multicolumn{1}{|c|}{\multirow{2}{*}{\textbf{YouTube}}} 
        & Video + Text 
        & 1 
        & 2 
        & 5 
        & 4 
        & 1 
        & 0 
        & 0 
        \\ \cline{2-9} 

        \multicolumn{1}{|c|}{} 
        & Shorts + Text 
        & 1 
        & 1 
        & 2 
        & 2 
        & 0 
        & 0 
        & 0 
        \\ \hline

        \end{tabular}
        
    }
    
\end{table}
\FloatBarrier

\begin{figure}[ht]

    \centering
    \caption[Comparison of European soccer leagues in terms of the frequency of their posts on TikTok.]{Comparison of European soccer leagues in terms of the frequency of their posts on TikTok.}
    \label{fig:leagues-frequency-tiktok}
    
    \begin{tikzpicture}
        \begin{axis}[
            ybar, area legend, 
            width=15cm,
            height=7cm,
            enlargelimits=0.15,
            legend style={at={(0.5,+1.15)},
              anchor=north,legend columns=-1},
            ylabel={Frequency of Posts},
            symbolic x coords={
                Premier League,
                Bundesliga,
                LaLiga,
                Serie A,
                Ligue 1,
                Allsvenskan,
                Eliteserien
            },
            xtick=data,
            xticklabel style={yshift=-1mm}, 
            nodes near coords={}, 
            x tick label style={rotate=20,anchor=east,font=\small}, 
            xmin=Premier League, 
            xmax=Eliteserien, 
            bar width=5.95, 
            ]
            
            \addplot coordinates {
                (Premier League, 2)
                (Bundesliga, 1)
                (LaLiga, 4)
                (Serie A, 1)
                (Ligue 1, 1)
                (Allsvenskan, 0)
                (Eliteserien, 0.14)
            };
            
            \legend{Video}
            
        \end{axis}

        \begin{axis}[
            width=5cm,
            height=3cm,
            enlargelimits=0.6,
            ybar,
            ylabel={},
            symbolic x coords={
                Allsvenskan,
                Eliteserien
            },
            xtick=data,
            nodes near coords={},
            x tick label style={rotate=30,anchor=east,font=\tiny},
            ymin=0, ymax=0.3,
            bar width=3.95,
            at={(rel axis cs:0.83,0.99)},
            anchor={north west}
            ]

            \addplot coordinates {
                (Allsvenskan, 0)
                (Eliteserien, 0.14)
            };
        \end{axis}
        
    \end{tikzpicture}

\end{figure}
\FloatBarrier
\begin{figure}[ht]

    \centering
    \caption[Comparison of European soccer leagues in terms of the frequency of their posts on Instagram.]{Comparison of European soccer leagues in terms of the frequency of their posts on Instagram.}
    \label{fig:leagues-frequency-instagram}
    
    \begin{tikzpicture}
        \begin{axis}[
            ybar, area legend, 
            width=15cm,
            height=7cm,
            enlargelimits=0.15,
            legend style={at={(0.5,+1.15)},
              anchor=north,legend columns=-1},
            ylabel={Frequency of Posts},
            symbolic x coords={
                Premier League,
                Bundesliga,
                LaLiga,
                Serie A,
                Ligue 1,
                Allsvenskan,
                Eliteserien
            },
            xtick=data,
            xticklabel style={yshift=-1mm}, 
            nodes near coords={}, 
            x tick label style={rotate=20,anchor=east,font=\small}, 
            xmin=Premier League, 
            xmax=Eliteserien, 
            bar width=5.95, 
            ]
            \addplot coordinates {
                (Premier League, 7)
                (Bundesliga, 8)
                (LaLiga, 30)
                (Serie A, 2)
                (Ligue 1, 2)
                (Allsvenskan, 0.28)
                (Eliteserien, 0.43)
            };
            
            \addplot coordinates {
                (Premier League, 18)
                (Bundesliga, 3)
                (LaLiga, 7)
                (Serie A, 6)
                (Ligue 1, 1)
                (Allsvenskan, 0.42)
                (Eliteserien, 0.57)
            };           

            \addplot coordinates {
                (Premier League, 0)
                (Bundesliga, 3)
                (LaLiga, 12)
                (Serie A, 6)
                (Ligue 1, 4)
                (Allsvenskan, 0)
                (Eliteserien, 0)
            };
            
            \legend{Video+Text, Image+Text, Story}
            
        \end{axis}

        \begin{axis}[
            width=5cm,
            height=3cm,
            enlargelimits=0.6,
            ybar,
            ylabel={},
            symbolic x coords={
                Allsvenskan,
                Eliteserien
            },
            xtick=data,
            nodes near coords={},
            x tick label style={rotate=30,anchor=east,font=\tiny},
            ymin=0.1, ymax=0.9,
            bar width=3.95,
            at={(rel axis cs:0.83,0.99)},
            anchor={north west}
            ]

            \addplot coordinates {
                (Allsvenskan, 0.28)
                (Eliteserien, 0.43)
            };
            
            \addplot coordinates {
                (Allsvenskan, 0.42)
                (Eliteserien, 0.57)
            };
            
            \addplot coordinates {
                (Allsvenskan, 0)
                (Eliteserien, 0)
            };
        \end{axis}
        
    \end{tikzpicture}

\end{figure}
\FloatBarrier
\begin{figure}[ht]

    \centering
    \caption[Comparison of European soccer leagues in terms of the frequency of their posts on Facebook.]{Comparison of European soccer leagues in terms of the frequency of their posts on Facebook.}
    \label{fig:leagues-frequency-facebook}
    
    \begin{tikzpicture}
    
        \begin{axis}[
            ybar, area legend, 
            width=15cm,
            height=7cm,
            enlargelimits=0.15,
            legend style={at={(0.5,+1.15)},
              anchor=north,legend columns=-1},
            ylabel={Frequency of Posts},
            symbolic x coords={
                Premier League,
                Bundesliga,
                LaLiga,
                Serie A,
                Ligue 1,
                Allsvenskan,
                Eliteserien
            },
            xtick=data,
            xticklabel style={yshift=-1mm}, 
            nodes near coords={}, 
            x tick label style={rotate=20,anchor=east,font=\small}, 
            xmin=Premier League, 
            xmax=Eliteserien, 
            bar width=5.95, 
            ]
            
            \addplot coordinates {
                (Premier League, 4)
                (Bundesliga, 1)
                (LaLiga, 13)
                (Serie A, 1)
                (Ligue 1, 5)
                (Allsvenskan, 0.21)
                (Eliteserien, 0.14)
            };
            
            \addplot coordinates {
                (Premier League, 16)
                (Bundesliga, 8)
                (LaLiga, 12)
                (Serie A, 3)
                (Ligue 1, 5)
                (Allsvenskan, 0.28)
                (Eliteserien, 0.57)
            };
            
            \addplot coordinates {
                (Premier League, 0)
                (Bundesliga, 3)
                (LaLiga, 8)
                (Serie A, 0)
                (Ligue 1, 8)
                (Allsvenskan, 0)
                (Eliteserien, 0)
            };
            
            \legend{Video+Text, Image+Text, Story}
            
        \end{axis}

        \begin{axis}[
            width=5cm,
            height=3cm,
            enlargelimits=0.6,
            ybar,
            ylabel={},
            symbolic x coords={
                Allsvenskan,
                Eliteserien
            },
            xtick=data,
            nodes near coords={},
            x tick label style={rotate=30,anchor=east,font=\tiny},
            ymin=0.1, ymax=0.7,
            bar width=3.95,
            at={(rel axis cs:0.83,0.99)},
            anchor={north west}
            ]

            \addplot coordinates {
                (Allsvenskan, 0.21)
                (Eliteserien, 0.14)
            };
            
            \addplot coordinates {
                (Allsvenskan, 0.28)
                (Eliteserien, 0.57)
            };
            
            \addplot coordinates {
                (Allsvenskan, 0)
                (Eliteserien, 0)
            };
            
            \addplot coordinates {
                (Allsvenskan, 0)
                (Eliteserien, 0)
            };
        \end{axis}
        
    \end{tikzpicture}

\end{figure}
\FloatBarrier
\begin{figure}[ht]

    \centering
    \caption[Comparison of European soccer leagues in terms of the frequency of their posts on Twitter.]{Comparison of European soccer leagues in terms of the frequency of their posts on Twitter.}
    \label{fig:leagues-frequency-twitter}
    
    \begin{tikzpicture}
    
        \begin{axis}[
            ybar, area legend, 
            width=15cm,
            height=7cm,
            enlargelimits=0.15,
            legend style={at={(0.5,+1.15)},
              anchor=north,legend columns=-1},
            ylabel={Frequency of Posts},
            symbolic x coords={
                Premier League,
                Bundesliga,
                LaLiga,
                Serie A,
                Ligue 1,
                Allsvenskan,
                Eliteserien
            },
            xtick=data,
            xticklabel style={yshift=-1mm}, 
            nodes near coords={}, 
            x tick label style={rotate=20,anchor=east,font=\small}, 
            xmin=Premier League, 
            xmax=Eliteserien, 
            bar width=5.95, 
            ]
            
            \addplot coordinates {
                (Premier League, 3)
                (Bundesliga, 3)
                (LaLiga, 5)
                (Serie A, 3)
                (Ligue 1, 1)
                (Allsvenskan, 0.35)
                (Eliteserien, 0.00)
            };
            
            \addplot coordinates {
                (Premier League, 4)
                (Bundesliga, 1)
                (LaLiga, 5)
                (Serie A, 6)
                (Ligue 1, 10)
                (Allsvenskan, 0.50)
                (Eliteserien, 0.00)
            };
            
            \addplot coordinates {
                (Premier League, 1)
                (Bundesliga, 0)
                (LaLiga, 0)
                (Serie A, 1)
                (Ligue 1, 0)
                (Allsvenskan, 0)
                (Eliteserien, 0)
            };

            \addplot coordinates {
                (Premier League, 21)
                (Bundesliga, 2)
                (LaLiga, 4)
                (Serie A, 0)
                (Ligue 1, 0)
                (Allsvenskan, 0)
                (Eliteserien, 0)
            };
            
            \legend{Video+Text, Image+Text, Text, Retweet}

        \end{axis}

        \begin{axis}[
            width=5cm,
            height=3cm,
            enlargelimits=0.6,
            ybar,
            ylabel={},
            symbolic x coords={
                Allsvenskan,
                Eliteserien
            },
            xtick=data,
            nodes near coords={},
            x tick label style={rotate=30,anchor=east,font=\tiny},
            ymin=0, ymax=0.6,
            bar width=3.95,
            at={(rel axis cs:0.83,0.99)},
            anchor={north west}
            ]

            \addplot coordinates {
                (Allsvenskan, 0.35)
                (Eliteserien, 0.00)
            };
            
            \addplot coordinates {
                (Allsvenskan, 0.50)
                (Eliteserien, 0.00)
            };
            
            \addplot coordinates {
                (Allsvenskan, 0)
                (Eliteserien, 0)
            };
        
            \addplot coordinates {
                (Allsvenskan, 0)
                (Eliteserien, 0)
            };
        \end{axis}
        
    \end{tikzpicture}

\end{figure}
\FloatBarrier
\FloatBarrier

\takeaway{\begin{itemize}
    \item LaLiga exhibits the highest content cadence on TikTok and Instagram, indicating its focus on visual content sharing.
    \item Premier League has the highest number of retweets on Twitter, reflecting a specific type of engagement with its audience and sharing of external content.
    \item Serie A maintains a consistent content-sharing pattern across Facebook, Twitter, and YouTube, highlighting a diversified approach in their social media strategy.
    \item Ligue 1 predominantly uses Instagram for stories, indicating a preference for ephemeral content on this platform.
    \item Allsvenskan shows a limited presence on YouTube and TikTok, with no content being shared on these platforms in the recent past.
    \item Eliteserien has not posted on Twitter since January 2023 and has also been inactive on YouTube for the last months.
    \item Instagram and Facebook are the primary platforms for content sharing for top European soccer leagues, with a greater emphasis on videos and images over text.
    \item Despite being popular platforms, not all soccer leagues actively engage on TikTok and YouTube. For instance, Allsvenskan has not published a single video on TikTok in 2023, and both Allsvenskan and Eliteserien show limited to no activity on YouTube.
    \item While LaLiga and Premier League have a significant presence across most platforms, Allsvenskan and Eliteserien have markedly fewer posts, especially on global platforms such as TikTok, YouTube, and Twitter.
\end{itemize}}

\NewPageCustom
\section{Marketing Channel Spread}

A marketing reach analysis, identifying referral sources, was conducted for the websites of European soccer leagues. The data collected was then compared to provide insights into the sources of visitors to these websites. The results of the analysis are presented in a table that showcases the different sources of traffic, including direct, referrals, search, social, mail, and display. 

In terms of the direct source of traffic, the analysis revealed that visitors to the soccer league websites in England, Spain and Norway mostly came directly to the website, without any external referral. In contrast, visitors to the soccer league websites in Sweden, Germany, Italy and France came mostly from search engines, indicating a stronger marketing approach in those regions. 

The search source of traffic, which includes visitors who come to the website through search engines, was found to be relatively high for all the soccer leagues, which indicate the importance of optimizing the searching algorithms.
Mail, which includes visitors who came to the website through email marketing campaigns, was not a prominent source of traffic for any of the leagues. Display advertising, which includes visitors who came to the website through online display ads, was also not a significant source of traffic for any of the leagues. 

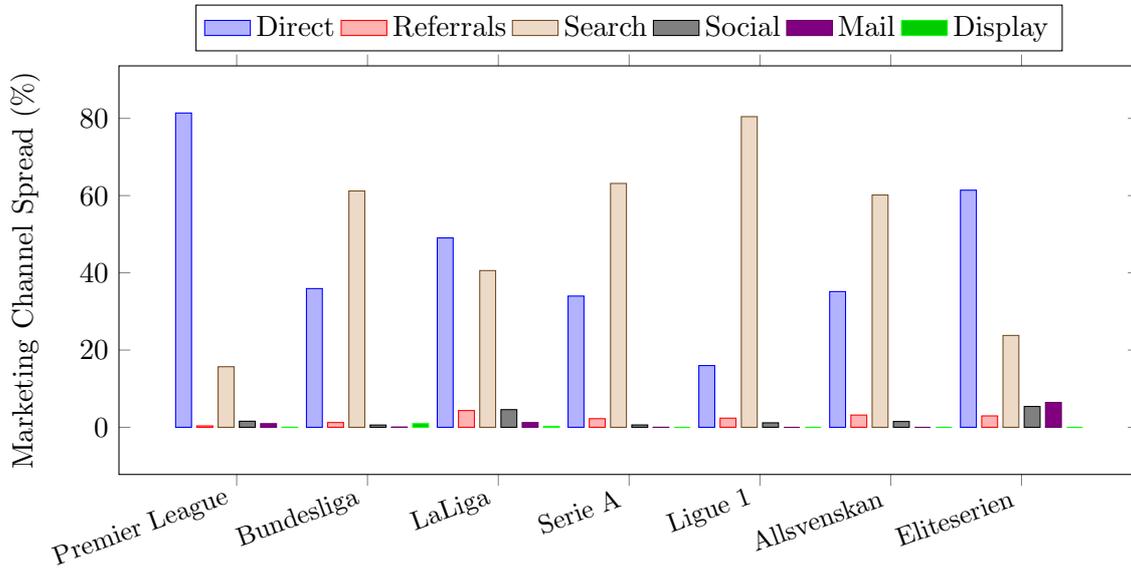
\begin{figure}[ht]

    \centering
    \caption[Comparison of European soccer leagues in terms of the marketing channel spread of their websites.]{Comparison of European soccer leagues in terms of the marketing channel spread of their websites (in percentage).}
    \label{fig:leagues-marketing}
    
    \begin{tikzpicture}
        \begin{axis}[
            ybar, area legend, 
            width=15cm,
            height=7cm,
            enlargelimits=0.15,
            legend style={at={(0.5,+1.15)},
              anchor=north,legend columns=-1},
            ylabel={Marketing Channel Spread (\%)},
            symbolic x coords={
                Premier League,
                Bundesliga,
                LaLiga,
                Serie A,
                Ligue 1,
                Allsvenskan,
                Eliteserien
            },
            xtick=data,
            xticklabel style={yshift=-1mm}, 
            nodes near coords={}, 
            x tick label style={rotate=20,anchor=east,font=\small}, 
            xmin=Premier League, 
            xmax=Eliteserien, 
            bar width=5.95, 
            ]
            
            \addplot coordinates {
                (Premier League, 81.39)
                (Bundesliga, 35.93)
                (LaLiga, 49.05)
                (Serie A, 33.99)
                (Ligue 1, 15.99)
                (Allsvenskan, 35.15)
                (Eliteserien, 61.41)
            };
            
            \addplot coordinates {
                (Premier League, 0.38)
                (Bundesliga, 1.24)
                (LaLiga, 4.35)
                (Serie A, 2.24)
                (Ligue 1, 2.38)
                (Allsvenskan, 3.18)
                (Eliteserien, 2.96)
            };
            
            \addplot coordinates {
                (Premier League, 15.69)
                (Bundesliga, 61.20)
                (LaLiga, 40.59)
                (Serie A, 63.15)
                (Ligue 1, 80.45)
                (Allsvenskan, 60.16)
                (Eliteserien, 23.79)
            };
            
            \addplot coordinates {
                (Premier League, 1.58)
                (Bundesliga, 0.60)
                (LaLiga, 4.58)
                (Serie A, 0.62)
                (Ligue 1, 1.17)
                (Allsvenskan, 1.51)
                (Eliteserien, 5.40)
            };

            \addplot coordinates {
                (Premier League, 0.95)
                (Bundesliga, 0.07)
                (LaLiga, 1.23)
                (Serie A, 0.01)
                (Ligue 1, 0)
                (Allsvenskan, 0)
                (Eliteserien, 6.45)
            };
            
            \addplot coordinates {
                (Premier League, 0.01)
                (Bundesliga, 0.95)
                (LaLiga, 0.20)
                (Serie A, 0)
                (Ligue 1, 0)
                (Allsvenskan, 0)
                (Eliteserien, 0)
            };
            
            \legend{Direct, Referrals, Search, Social, Mail, Display}
            
        \end{axis}
        
    \end{tikzpicture}

\end{figure}

Table~\ref{tab:leagues-marketing} presents a comparison of European soccer leagues in terms of the marketing channel spread of their websites~\cite{SimilarWeb2023}. Figure~\ref{fig:leagues-marketing} presents a visual comparison of the marketing channel spread of the websites of different leagues. 

\begin{table}[ht]

    \centering
    \caption[Comparison of European soccer leagues in terms of the marketing channel spread of their websites.]{Comparison of European soccer leagues in terms of the marketing channel spread of their websites (in percentage).}
    \label{tab:leagues-marketing}

    \resizebox{\textwidth}{!}{
    
        \begin{tabular}{|l||c|c|c|c|c|c|c|}
        \hline

        \textbf{Referral Source} 
        & \textbf{Premier League} 
        & \textbf{Bundesliga} 
        & \textbf{Laliga} 
        & \textbf{Serie A} 
        & \textbf{Ligue 1} 
        & \textbf{Allsvenskan} 
        & \textbf{Eliteserien} 
        \\ \hline
  
        Direct 
        & 81.39 
        & 35.93 
        & 49.05 
        & 33.99 
        & 15.99 
        & 35.15 
        & 61.41 
        \\ \hline 
          
        Referrals 
        & 0.38 
        & 1.24 
        & 4.35 
        & 2.24 
        & 2.38 
        & 3.18 
        & 2.96 
        \\ \hline 
        
        Search 
        & 15.69 
        & 61.20 
        & 40.59 
        & 63.15 
        & 80.45 
        & 60.16 
        & 23.79 
        \\ \hline 
        
        Social 
        & 1.58 
        & 0.60 
        & 4.58 
        & 0.62 
        & 1.17 
        & 1.51 
        & 5.40 
        \\ \hline 
        
        Mail 
        & 0.95 
        & 0.07 
        & 1.23 
        & 0.01 
        & 0 
        & 0 
        & 6.45 
        \\ \hline 
        
        Display 
        & 0.01 
        & 0.95 
        & 0.20 
        & 0 
        & 0 
        & 0 
        & 0 
        \\ \hline

        \end{tabular}

    }
    
\end{table}
\FloatBarrier

\FloatBarrier

\takeaway{\begin{itemize}
    \item Majority of website visitors arrive directly without external referral, implying a strong brand presence or awareness.
    \item Search engine driven website traffic plays a significant role for all soccer leagues, underscoring the universal importance of search engine visibility.
    \item Despite its ubiquity in the digital age, email marketing seems to be a less effective channel for most leagues, with the exception of Eliteserien that has a notable 6.45\% traffic from this source.
    \item Display advertising is not a significant driver of traffic for the analyzed leagues.
    \item Social media posting frequency varies across platforms and leagues. For instance, LaLiga is quite active on platforms such as TikTok and Instagram, while leagues such as Allsvenskan and Eliteserien have a more subdued online activity.
\end{itemize}}

\NewPageCustom
\section{Website Referrals}

An analysis was conducted on the social media website referrals for the top European soccer leagues, including England, Germany, Spain, Italy, France, Sweden, and Norway. The objective of the analysis was to gain insights into the social media platforms driving traffic to the league websites. 

The data provided in the table highlights the monthly visits and monthly unique visitors for the top soccer leagues' websites in Europe, along with the social traffic referring sources. The data indicates that the English Premier League website receives the highest number of monthly visits with 64,000,000, followed by the German Bundesliga with 3,260,000 visits and the Spanish LaLiga with 1,820,000 visits. The Italian Serie A and French Ligue 1 websites have 1,300,000 and 378,072 visits, respectively, while the Swedish Allsvenskan and Norwegian Eliteserien have relatively fewer visits with 75,525 and 36,299, respectively. 

In terms of social traffic referring sources, YouTube is the most significant source of traffic for the German Bundesliga with 48.21\% of the social traffic, while Twitter generates the most significant percentage of social traffic for the Spanish LaLiga with 52.29\%. Facebook is the primary source of social traffic for the Italian Serie A and French Ligue 1, generating 43.87\% and 100\%, respectively. The Swedish Allsvenskan and Norwegian Eliteserien websites receive the most significant percentage of social traffic from Facebook. The analysis further revealed that social media website referrals accounted for a considerable portion of the overall traffic to the league websites. 

\begin{table}[ht]

    \centering
    \caption[Comparison of European soccer leagues in terms of their social media website referrals.]{Comparison of European soccer leagues in terms of their social media website referrals (in percentage).}
    \label{tab:leagues-website}

    \resizebox{\textwidth}{!}{
    
        \begin{tabular}{|c|l||c|c|c|c|c|c|c|}
        \cline{3-9}
        
        \multicolumn{2}{c|}{} 
        & \textbf{Premier League} 
        & \textbf{Bundesliga} 
        & \textbf{Laliga} 
        & \textbf{Serie A} 
        & \textbf{Ligue 1}
        & \textbf{Allsvenskan} 
        & \textbf{Elliteserien} 
        \\ \hline
        
        \multicolumn{2}{|c||}{\textbf{Monthly Visits}} 
        & 64,000,000 
        & 3,260,000 
        & 1,820,000 
        & 1,300,000 
        & 378,072 
        & 75,525 
        & 36,299 
        \\ \hline 

        \multicolumn{1}{|c|}{\multirow{8}{*}
        {\rotatebox{90}{\textbf{Referral Source}}}}
        & \textbf{Instagram} 
        & 0 
        & 5.23 
        & 20.27 
        & 21.86 
        & 0 
        & 0 
        & 0 
        \\ \cline{2-9}

        & \textbf{Facebook} 
        & 14.37 
        & 7.80 
        & 0 
        & 43.87 
        & 0 
        & 100 
        & 100 
        \\ \cline{2-9}
        
        & \textbf{Twitter} 
        & 24.21 
        & 5.44 
        & 52.29 
        & 0 
        & 100 
        & 0 
        & 0 
        \\ \cline{2-9} 
        
        & \textbf{YouTube} 
        & 38.78 
        & 48.21 
        & 8.26 
        & 21.46 
        & 0 
        & 0 
        & 0 
        \\ \cline{2-9} 
        
        & \textbf{Reddit} 
        & 7.97 
        & 33.30 
        & 0 
        & 12.81 
        & 0 
        & 0 
        & 0 
        \\ \cline{2-9}
        
        & \textbf{WhatsApp} 
        & 4.97 
        & 0 
        & 4.81 
        & 0 
        & 0 
        & 0 
        & 0 
        \\ \cline{2-9}
        
        & \textbf{Quora} 
        & 0 
        & 0 
        & 5.09 
        & 0 
        & 0 
        & 0 
        & 0 
        \\ \cline{2-9}
        
        & \textbf{Other} 
        & 9.71 
        & 0 
        & 9.28 
        & 0 
        & 0 
        & 0 
        & 0 
        \\ \hline 

        \end{tabular}
        
    }
    
\end{table}
\FloatBarrier

Table~\ref{tab:leagues-website} presents an analysis of marketing reach for top European soccer league websites. 

\takeaway{\begin{itemize}
    \item Premier League has an unmatched number of monthly visits, standing out distinctly.
    \item Bundesliga and LaLiga show specific reliance: Bundesliga leans heavily on YouTube, whereas LaLiga primarily uses Twitter.
    \item Serie A effectively leverages Facebook for referrals, this trend is also seen in smaller leagues like Allsvenskan and Eliteserien.
    \item Bundesliga uniquely boasts a significant presence on Reddit.
    \item Premier League's referral sources are diversified: almost 10\% comes from "Other" sources, suggesting a broad outreach.
    \item Social media referrals have an undeniable influence on consistently driving audience engagement, this emphasizes the quality of the referrals over the mere volume of traffic.
\end{itemize}}

\NewPageCustom
\section{Visual Content}

We conducted an in-depth analysis of the visual content posted by top European soccer leagues from England, Germany, Spain, Italy, and France across social media platforms such as TikTok, Instagram, Facebook, Twitter, and YouTube. Our focus was on evaluating the length and type of content shared, including the duration of videos and stories (measured in seconds), as well as the length of captions (counted in characters) accompanying goal, highlight, and event multimedia. Table~\ref{tab:leagues-content} presents an analysis of visual content for top European soccer leagues on social media.

\begin{table}[ht]

    \centering
    \caption[Comparison of European soccer leagues in terms of their social media content.]{Comparison of European soccer leagues in terms of their social media content.}
    \label{tab:leagues-content}

    \resizebox{\textwidth}{!}{
    
        \begin{tabular}{|cll||c|c|c|c|c|}
        \cline{4-8}

        \multicolumn{3}{c|}{} 
        & \textbf{Premier League} 
        & \textbf{Bundesliga} 
        & \textbf{LaLiga} 
        & \textbf{Serie A} 
        & \textbf{Ligue 1} 
        \\ \hline
        
        \multicolumn{1}{|c|}{\multirow{6}{*}{\textbf{TikTok}}} 
        & \multicolumn{1}{l|}{\multirow{2}{*}{Goals}} 
        & Video 
        & \textless{}35s 
        & {N/A}
        & $\sim$30s 
        & $\sim$10s 
        & $\sim$10 
        \\ \cline{3-8} 
        
        \multicolumn{1}{|c|}{} 
        & \multicolumn{1}{c|}{} 
        & Caption 
        & $\sim$45 
        & {N/A}
        & 100 
        & \textless{}100 
        & $\sim$70 
        \\ \cline{2-8} 

        \multicolumn{1}{|c|}{} 
        & \multicolumn{1}{l|}{\multirow{2}{*}{Highlights}} 
        & Video 
        & {N/A}
        & \textless{}20 
        & $\sim$30s 
        & {N/A}
        & {N/A}
        \\ \cline{3-8} 
        
        \multicolumn{1}{|c|}{} 
        & \multicolumn{1}{c|}{} 
        & Caption 
        & {N/A}
        & 92 
        & 120 
        & {N/A}
        & {N/A}
        \\ \cline{2-8} 
        
        \multicolumn{1}{|c|}{} 
        & \multicolumn{1}{l|}{\multirow{2}{*}{Events}} 
        & Video 
        & \textless{}10s 
        & {N/A}
        & \textless{}10s 
        & $\sim$15s 
        & $\sim$15 
        \\ \cline{3-8} 
        
        \multicolumn{1}{|c|}{} 
        & \multicolumn{1}{c|}{} 
        & Caption 
        & $\sim$35 
        & {N/A}
        & 140 
        & \textless{}100 
        & $\sim$70 
        \\ \hline 
        
        \multicolumn{1}{|c|}{\multirow{9}{*}{\textbf{Instagram}}} 
        & \multicolumn{1}{l|}{\multirow{3}{*}{Goals}} 
        & Video 
        & \textless{}30 
        & \textless{}20 
        & $\sim$30s 
        & $\sim$20s 
        & $\sim$15 
        \\ \cline{3-8} 
        
        \multicolumn{1}{|c|}{} 
        & \multicolumn{1}{c|}{} 
        & Story 
        & {N/A}
        & {N/A}
        & {N/A}
        & {N/A}
        & {N/A}
        \\ \cline{3-8} 

        \multicolumn{1}{|c|}{} 
        & \multicolumn{1}{c|}{} 
        & Caption 
        & $\sim$30 
        & 108 
        & $\sim$40 
        & $\sim$70 
        & $\sim$140 
        \\ \cline{2-8} 
        
        \multicolumn{1}{|c|}{} 
        & \multicolumn{1}{l|}{\multirow{3}{*}{Highlights}} 
        & Video 
        & $\sim$90s 
        & \textless{}60 
        & {N/A}
        & {N/A}
        & $\sim$60 
        \\ \cline{3-8} 
        
        \multicolumn{1}{|c|}{} 
        & \multicolumn{1}{c|}{} 
        & Story 
        & {N/A}
        & {N/A}
        & {N/A}
        & {N/A}
        & {N/A}
        \\ \cline{3-8} 

        \multicolumn{1}{|c|}{} 
        & \multicolumn{1}{c|}{} 
        & Caption 
        & $\sim$30 
        & 150 
        & {N/A}
        & {N/A}
        & $\sim$140 
        \\ \cline{2-8} 
        
        \multicolumn{1}{|c|}{} 
        & \multicolumn{1}{l|}{\multirow{3}{*}{Events}} 
        & Video 
        & \textless{}15s 
        & \textless{}50 
        & $\sim$15s 
        & $\sim$15s 
        & $\sim$30 
        \\ \cline{3-8} 
        
        \multicolumn{1}{|c|}{} 
        & \multicolumn{1}{c|}{} 
        & Story 
        & {N/A}
        & {N/A}
        & {N/A}
        & {N/A}
        & $\sim$7s 
        \\ \cline{3-8} 

        \multicolumn{1}{|c|}{} 
        & \multicolumn{1}{c|}{} 
        & Caption 
        & $\sim$30 
        & 71 
        & $\sim$40 
        & $\sim$60 
        & $\sim$140 
        \\ \hline 

        \multicolumn{1}{|c|}{\multirow{9}{*}{\textbf{Facebook}}} 
        & \multicolumn{1}{l|}{\multirow{3}{*}{Goals}} 
        & Videos 
        & \textless{}60s 
        & \textless{}20s 
        & \textless{}30s 
        & {N/A}
        & $\sim$30s 
        \\ \cline{3-8} 
        
        \multicolumn{1}{|c|}{} 
        & \multicolumn{1}{c|}{} 
        & Story 
        & {N/A}
        & {N/A}
        & \textless{}30s 
        & {N/A}
        & {N/A}
        \\ \cline{3-8} 

        \multicolumn{1}{|c|}{} 
        & \multicolumn{1}{c|}{} 
        & Caption 
        & $\sim$70 char. 
        & 79 
        & {N/A}
        & {N/A}
        & $\sim$80 
        \\ \cline{2-8} 
        
        \multicolumn{1}{|c|}{} 
        & \multicolumn{1}{l|}{\multirow{3}{*}{Highlights}}  
        & Video 
        & \textless{}90s 
        & \textless{}60s 
        & \textless{}60s 
        & \textless{}60s 
        & $\sim$120s 
        \\ \cline{3-8} 
        
        \multicolumn{1}{|c|}{} 
        & \multicolumn{1}{c|}{} 
        & Story 
        & {N/A}
        & {N/A}
        & {N/A}
        & {N/A}
        & {N/A}
        \\ \cline{3-8} 

        \multicolumn{1}{|c|}{} 
        & \multicolumn{1}{c|}{} 
        & Caption 
        & $\sim$30 char. 
        & 80 
        & {N/A}
        & 70 char. 
        & $\sim$150 
        \\ \cline{2-8} 
        
        \multicolumn{1}{|c|}{} 
        & \multicolumn{1}{l|}{\multirow{3}{*}{Events}} 
        & Video 
        & \textless{}120 
        & \textless{}50s 
        & $\sim$15s 
        & {N/A}
        & $\sim$30s 
        \\ \cline{3-8} 
        
        \multicolumn{1}{|c|}{} 
        & \multicolumn{1}{c|}{} 
        & Story 
        & {N/A}
        & {N/A}
        & {N/A}
        & {N/A}
        & {N/A}
        \\ \cline{3-8} 

        \multicolumn{1}{|c|}{} 
        & \multicolumn{1}{c|}{} 
        & Caption 
        & $\sim$30 
        & 105 
        & {N/A}
        & {N/A}
        & $\sim$70 
        \\ \hline 

        \multicolumn{1}{|c|}{\multirow{6}{*}{\textbf{Twitter}}} 
        & \multicolumn{1}{l|}{\multirow{2}{*}{Goals}} 
        & Video 
        & \textless{}60 
        & \textless{}60 
        & \textless{}60s 
        & $\sim$20s 
        & $\sim$30 
        \\ \cline{3-8} 
        
        \multicolumn{1}{|c|}{} 
        & \multicolumn{1}{c|}{} 
        & Caption 
        & $\sim$30 
        & 95 
        & 90 
        & $\sim$60 
        & $\sim$115 
        \\ \cline{2-8} 
        
        \multicolumn{1}{|c|}{} 
        & \multicolumn{1}{l|}{\multirow{2}{*}{Highlights}} 
        & Videos 
        & {N/A}
        & \textless{}300 
        & $\sim$120 
        & {N/A}
        & $\sim$200 
        \\ \cline{3-8} 
        
        \multicolumn{1}{|c|}{} 
        & \multicolumn{1}{c|}{} 
        & Caption 
        & {N/A}
        & 95 
        & $\sim$60 
        & {N/A}
        & $\sim$190 
        \\ \cline{2-8} 

        \multicolumn{1}{|c|}{} 
        & \multicolumn{1}{l|}{\multirow{2}{*}{Events}} 
        & Video 
        & \textless{}120s 
        & \textless{}40 
        & \textless{}10 
        & $\sim$15s 
        & $\sim$30 
        \\ \cline{3-8} 
        
        \multicolumn{1}{|c|}{} 
        & \multicolumn{1}{c|}{} 
        & Caption 
        & $\sim$30 
        & 80 
        & $\sim$60 
        & $\sim$60 
        & $\sim$100 
        \\ \hline

        \multicolumn{1}{|c|}{\multirow{6}{*}{\textbf{YouTube}}} 
        & \multicolumn{1}{l|}{\multirow{2}{*}{Goals}} 
        & Video + link 
        & {N/A}
        & \textless{}15s 
        & $\sim$120s 
        & $\sim$150s 
        & $\sim$90 
        \\ \cline{3-8} 
        
        \multicolumn{1}{|c|}{} 
        & \multicolumn{1}{c|}{} 
        & Caption + link 
        & {N/A}
        & 115 
        & $\sim$800 
        & $\sim$700 
        & $\sim$730 
        \\ \cline{2-8} 
        
        \multicolumn{1}{|c|}{} 
        & \multicolumn{1}{l|}{\multirow{2}{*}{Highlights}} 
        & Video + link 
        & {N/A}
        & \textless{}200 
        & 180s 
        & $\sim$200s 
        & $\sim$200 
        \\ \cline{3-8} 

        \multicolumn{1}{|c|}{} 
        & \multicolumn{1}{c|}{} 
        & Caption + link 
        & {N/A}
        & 800 
        & $\sim$900 
        & $\sim$700 
        & $\sim$1200 
        \\ \cline{2-8} 
        
        \multicolumn{1}{|c|}{} 
        & \multicolumn{1}{l|}{\multirow{2}{*}{Events}} 
        & Video + link 
        & {N/A}
        & \textless{}10 
        & $\sim$120s 
        & $\sim$300s 
        & $\sim$90 
        \\ \cline{3-8} 
        
        \multicolumn{1}{|c|}{} 
        & \multicolumn{1}{c|}{} 
        & Caption + link 
        & {N/A}
        & 110 
        & $\sim$800 
        & $\sim$700 
        & $\sim$950 
        \\ \cline{1-8} 

        \end{tabular}
        
    }
    
\end{table}
\FloatBarrier

\takeaway{\begin{itemize}
    \item Leagues tailor their content uniquely for platforms. Not all leagues post uniformly across platforms, suggesting strategic content prioritization.    
    \item Facebook and Twitter have platform-specific content length, with Bundesliga's lengthy highlights on Twitter being notable. 
    \item Platforms such as TikTok have shorter captions, while YouTube and Instagram opt for detailed descriptions. 
    \item Most leagues prioritize content below 35 seconds on TikTok, emphasizing quick engagement. 
    \item Premier League, in particular, uses longer highlight reels on Instagram, possibly to maximize viewer time. 
    \item LaLiga and Ligue 1 utilize extensive captions (detailed descriptions) on YouTube for added context. 
    \item Event highlights, especially on YouTube, are given extended durations, underscoring their importance.   
\end{itemize}}

\clearpage
\chapter{Social Media Strategies: European Soccer Teams}\label{section:social-media-strategies-teams}

In this section, we compare the social media strategies of two major soccer teams, Liverpool (England) and Real Madrid (Spain), and three Norwegian soccer teams, VIF, RBK, and TIL. 

This analysis was carried out as part of the broader study that looked into the social media strategies of major soccer leagues (see Section~\ref{section:social-media-strategies-leagues}). By comparing the social media presence of soccer teams, we can gain a better understanding of how different teams utilize social media to engage with their fans and promote their brand. 

\section{Followers and Engagement}

First, we compare the aforementioned teams in terms of their social media followers and engagement. The goal of the analysis is to determine the relative popularity of each team in terms of their online presence. The data was collected from various sources including TikTok~\cite{Liverpool_TikTok, Realmadrid_TikTok, VIF_TikTok, TIL_TikTok}, Instagram~\cite{Liverpool_Instagram, Realmadrid_Instagram, VIF_Instagram, RBK_Instagram, TIL_Instagram}, 
Facebook~\cite{Liverpool_Facebook, Realmadrid_Facebook, VIF_Facebook, RBK_Facebook, TIL_Facebook}, 
Twitter~\cite{Liverpool_Twitter, Realmadrid_Twitter, VIF_Twitter, RBK_Twitter, TIL_Twitter}, and 
YouTube~\cite{Liverpool_YouTube, Realmadrid_YouTube, VIF_YouTube, RBK_YouTube, TIL_YouTube}. 

Table~\ref{tab:teams-followers} presents a comparison of selected European soccer teams in terms of social media followers and engagement, with the number of followers presented in millions, and the engagement rate~\cite{EngagementCalculator2023} presented as a percentage.\footnote{Engagement rates exceeding 100\% denote that, every follower interacted more than once on average, or that a significant portion of the engagement came from viewers who aren't followers, emphasizing the content's broad appeal.}    

\begin{table}[ht]

    \centering
    \caption[Comparison of European soccer teams in terms of social media followers and engagement.]{Comparison of European soccer teams in terms of social media followers and engagement.}
    \label{tab:teams-followers}
    

        \begin{tabular}{|c|c||c|c|c|c|c|}
        \cline{3-7}
        
        \multicolumn{2}{c|}{} 
        & \textbf{Liverpool} 
        & \textbf{Real Madrid}
        & \textbf{VIF} 
        & \textbf{RBK} 
        & \textbf{TIL} 
        \\ \hline
        
        \multicolumn{1}{|c|}{\multirow{5}{*}{\rotatebox{90}{\textbf{\parbox{2cm}{Followers\\(Million)}}}}} 
        
        & TikTok            
        & 12,400,000 
        & 26,100,000 
        & 0 
        & 33,300 
        & 0 
        \\ \cline{2-7} 
        
        & Instagram 
        & 41,400,000 
        & 132,000,000 
        & 28,200 
        & 76,200 
        & 13,800 
        \\ \cline{2-7} 
        
        & Facebook 
        & 44,253,409 
        & 116,363,808 
        & 73,705 
        & 218,092 
        & 39,524 
        \\ \cline{2-7} 
        
        & Twitter 
        & 23,300,000 
        & 46,700,000 
        & 20,000 
        & 61,100 
        & 9,966 
        \\ \cline{2-7}
        
        & YouTube          
        & 8,030,000 
        & 9,360,000 
        & 2,340 
        & 15,400 
        & 0 
        \\ \hline \hline 
        
        \multicolumn{1}{|c|}{\multirow{5}{*}{\rotatebox{90}{\textbf{\parbox{2cm}{Engagem.\\Rate (\%)}}}}}
        
        & TikTok 
        & 0.67 
        & 0.83 
        & 0 
        & 4.06 
        & 0
        \\ \cline{2-7} 
        
        & Instagram 
        & 0.70 
        & 0.35 
        & 3.90 
        & 1.75 
        & 1.26 
        \\ \cline{2-7} 
        
        & Facebook
        & 0.02 
        & 0.07 
        & 0.73 
        & 0.18 
        & 0.10 
        \\ \cline{2-7} 
        
        & Twitter 
        & 1.01 
        & 1.90 
        & 0.38 
        & 0.69 
        & 0.49 
        \\ \cline{2-7} 
        
        & YouTube 
        & 1.70 
        & 3.76 
        & 106.90 
        & 61.44 
        & 0 
        \\ \hline 
        
        \end{tabular}
        

\end{table}
\FloatBarrier

\takeaway{\begin{itemize}    
    \item Liverpool and Real Madrid lead in terms of the total number of followers, far surpassing Norwegian teams VIF, RBK, and TIL.
    \item Real Madrid boasts over 132 million followers on Instagram.
    \item Liverpool presents a balanced presence across major social media platforms.
    \item Despite having fewer followers overall, Norwegian teams, and VIF in particular, exhibit high engagement rates.  
    \item Both Liverpool and Real Madrid have a substantial number of followers on TikTok. 
    \item Norwegian teams have yet to fully tap into TikTok as a platform.
    \item Overall, our findings highlight the balance between the sheer number of followers and the quality of engagement, as well as stress the significance of platform-specific strategies for optimal outreach and engagement.
\end{itemize}}

\NewPageCustom
\section{Content Cadence}

An analysis of publishing frequency in TikTok, Instagram, Facebook, Twitter, and YouTube for different soccer teams has been conducted to understand social media content cadence. The aim of the analysis was to explore the frequency at which social media content is published on each platform and in various modules. By examining these patterns, it is possible to identify trends and best practices for social media content creation and management. 

The analysis was conducted by collecting data from various social media accounts across different modules. The collected data was then analyzed to identify patterns in the frequency of content publishing. The results of the analysis show that social media content cadence varies across different platforms and modules. 

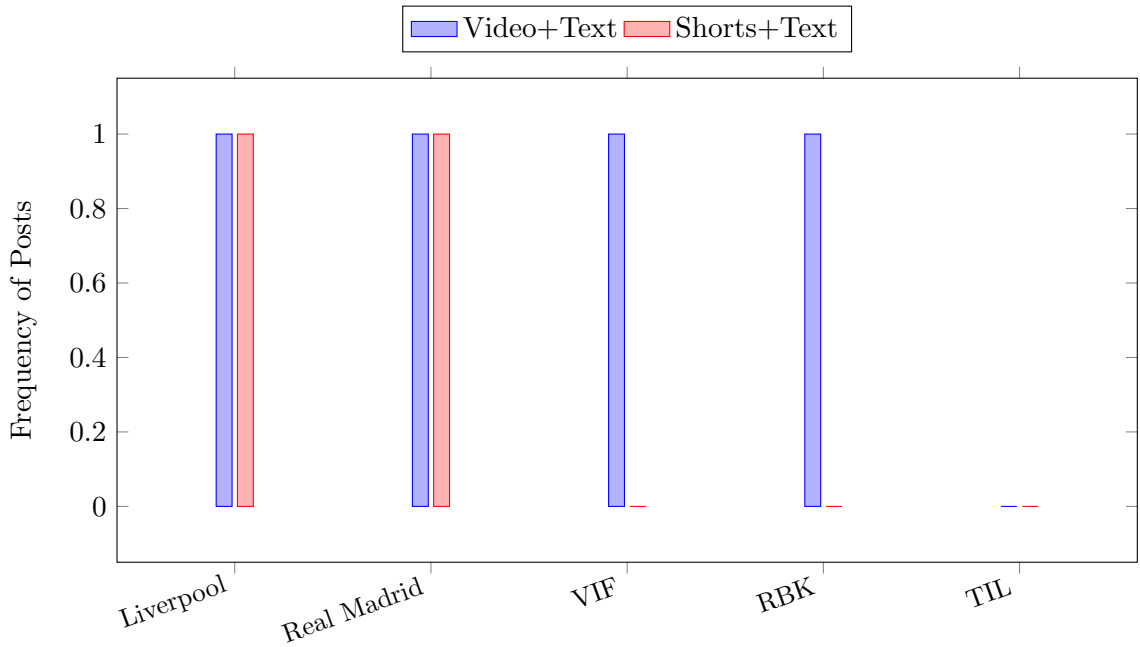
\begin{figure}[ht]

    \centering
    \caption[Comparison of European soccer teams in terms of the frequency of their posts on YouTube.]{Comparison of European soccer teams in terms of the frequency of their posts on YouTube.}
    \label{fig:teams-frequency-youtube}
    
    \begin{tikzpicture}
    
        \begin{axis}[
            ybar, area legend, 
            width=15cm,
            height=8cm,
            enlargelimits=0.15,
            legend style={at={(0.5,+1.15)},
              anchor=north,legend columns=-1},
            ylabel={Frequency of Posts},
            symbolic x coords={
                Liverpool,
                Real Madrid,
                VIF,
                RBK,
                TIL
            },
            xtick=data,
            xticklabel style={yshift=-1mm}, 
            nodes near coords={}, 
            x tick label style={rotate=20,anchor=east,font=\small}, 
            xmin=Liverpool, 
            xmax=TIL, 
            bar width=5.95, 
            ]
            
            \addplot coordinates {
                (Liverpool, 1)
                (Real Madrid, 1)
                (VIF, 1)
                (RBK, 1)
                (TIL, 0)
            };
            
            \addplot coordinates {
                (Liverpool, 1)
                (Real Madrid, 1)
                (VIF, 0)
                (RBK, 0)
                (TIL, 0)
            }; 
            
            \legend{Video+Text, Shorts+Text}
            
        \end{axis}
        
    \end{tikzpicture}

\end{figure}

Table~\ref{tab:teams-frequency} presents a comparison of selected European soccer teams in terms of social media content cadence. 
Figures~\ref{fig:teams-frequency-tiktok}, \ref{fig:teams-frequency-instagram}, \ref{fig:teams-frequency-facebook}, \ref{fig:teams-frequency-twitter}, and \ref{fig:teams-frequency-youtube} present a visual comparison of the frequency of content posting on various social media platforms for different teams.\footnote{(Table~\ref{tab:teams-frequency} and Figures~\ref{fig:teams-frequency-tiktok}, \ref{fig:teams-frequency-instagram}, \ref{fig:teams-frequency-facebook}, \ref{fig:teams-frequency-twitter}, \ref{fig:teams-frequency-youtube}) The numbers represent the average daily frequency of posts, which were derived from observing posts over a one-week period, and then calculating the daily average from the weekly observation.}

\begin{table}[ht]

    \centering
    \caption{Comparison of European teams in terms of the frequency of their posts on social media, per modality.}
    \label{tab:teams-frequency}

    \resizebox{\textwidth}{!}{
    
        \begin{tabular}{|cl||c|c|c|c|c|}
        \cline{3-7}

        \multicolumn{2}{c|}{} 
        & \textbf{Liverpool} 
        & \textbf{Real Madrid} 
        & \textbf{VIF} 
        & \textbf{RBK} 
        & \textbf{TIL} 
        \\ \hline
  
        \multicolumn{1}{|c|}{\multirow{4}{*}{\textbf{Facebook}}} 
        & Video + Text 
        & 4 
        & 3 
        & 1 
        & 1 
        & 1 
        \\ \cline{2-7} 
        
        \multicolumn{1}{|c|}{} 
        & Image + Text 
        & 8 
        & 14 
        & 1 
        & 2 
        & 1 
        \\ \cline{2-7} 
        
        \multicolumn{1}{|c|}{} 
        & Text 
        & 0 
        & 0 
        & 0 
        & 0 
        & 0 
        \\ \cline{2-7} 
        
        \multicolumn{1}{|c|}{} 
        & Story 
        & 0 
        & 0 
        & 0 
        & 0 
        & 3 
        \\ \hline 
        
        \multicolumn{1}{|c|}{\multirow{3}{*}{\textbf{Instagram}}} 
        & Video + Text 
        & 3 
        & 2 
        & 1 
        & 1 
        & 1 
        \\ \cline{2-7} 
        
        \multicolumn{1}{|c|}{} 
        & Image + Text 
        & 6 
        & 10 
        & 1 
        & 1 
        & 1 
        \\ \cline{2-7} 
        
        \multicolumn{1}{|c|}{} 
        & Story 
        & 20 
        & 10 
        & 5 
        & 0 
        & 3 
        \\ \hline 
        
        \multicolumn{1}{|c|}{\textbf{TikTok}} 
        & Video + Music 
        & 2 
        & 2 
        & 0 
        & 1 
        & 0.2 
        \\ \hline 
        
        \multicolumn{1}{|c|}{\multirow{4}{*}{\textbf{Twitter}}} 
        & Video + Text 
        & 6 
        & 6 
        & 3 
        & 3 
        & 1 
        \\ \cline{2-7} 
        
        \multicolumn{1}{|c|}{} 
        & Image + Text 
        & 12 
        & 18 
        & 1 
        & 1 
        & 1 
        \\ \cline{2-7} 
        
        \multicolumn{1}{|c|}{} 
        & Text 
        & 0 
        & 0 
        & 0 
        & 1 
        & 0 
        \\ \cline{2-7} 
        
        \multicolumn{1}{|c|}{} 
        & Retweet 
        & 0 
        & 0 
        & 0 
        & 0 
        & 0 
        \\ \hline 
        
        \multicolumn{1}{|c|}{\multirow{2}{*}{\textbf{YouTube}}} 
        & Video + Text 
        & 1 
        & 1 
        & 1 
        & 1 
        & 0 
        \\ \cline{2-7} 
        
        \multicolumn{1}{|c|}{} 
        & Shorts + Text 
        & 1 
        & 1 
        & 0 
        & 0 
        & 0 
        \\ \hline 

        \end{tabular}
        
    }
    
\end{table}
\FloatBarrier

\begin{figure}[ht]

    \centering
    \caption[Comparison of European soccer teams in terms of the frequency of their posts on TikTok.]{Comparison of European soccer teams in terms of the frequency of their posts on TikTok.}
    \label{fig:teams-frequency-tiktok} 
    
    \begin{tikzpicture}
  
        \begin{axis}[
            ybar, area legend, 
            width=15cm,
            height=8cm,
            enlargelimits=0.15,
            legend style={at={(0.5,+1.15)},
              anchor=north,legend columns=-1},
            ylabel={Frequency of Posts},
            symbolic x coords={
                Liverpool,
                Real Madrid,
                VIF,
                RBK,
                TIL
            },
            xtick=data,
            xticklabel style={yshift=-1mm}, 
            nodes near coords={}, 
            x tick label style={rotate=20,anchor=east,font=\small}, 
            xmin=Liverpool, 
            xmax=TIL, 
            bar width=5.95, 
            ]
            
            \addplot coordinates {
                (Liverpool, 2)
                (Real Madrid, 2)
                (VIF, 0)
                (RBK, 1)
                (TIL, 0.2)
            };
            
            \legend{Video+Text}
            
        \end{axis}
        
    \end{tikzpicture}

\end{figure}
\FloatBarrier
\begin{figure}[ht]

    \centering
    \caption[Comparison of European soccer teams in terms of the frequency of their posts on Instagram.]{Comparison of European soccer teams in terms of the frequency of their posts on Instagram.}
    \label{fig:teams-frequency-instagram}
    
    \begin{tikzpicture}
        \begin{axis}[
            ybar, area legend, 
            width=15cm,
            height=8cm,
            enlargelimits=0.15,
            legend style={at={(0.5,+1.15)},
              anchor=north,legend columns=-1},
            ylabel={Frequency of Posts},
            symbolic x coords={
                Liverpool,
                Real Madrid,
                VIF,
                RBK,
                TIL
            },
            xtick=data,
            xticklabel style={yshift=-1mm}, 
            nodes near coords={}, 
            x tick label style={rotate=20,anchor=east,font=\small}, 
            xmin=Liverpool, 
            xmax=TIL, 
            bar width=5.95, 
            ]
            
            \addplot coordinates {
                (Liverpool, 3)
                (Real Madrid, 2)
                (VIF, 1)
                (RBK, 1)
                (TIL, 1)
            };
            
            \addplot coordinates {
                (Liverpool, 6)
                (Real Madrid, 10)
                (VIF, 1)
                (RBK, 1)
                (TIL, 1)
            };
            
            \addplot coordinates {
                (Liverpool, 20)
                (Real Madrid, 10)
                (VIF, 5)
                (RBK, 0)
                (TIL, 3)
            };
            
            \legend{Video+Text, Image+Text, Story}
            
        \end{axis}
        
    \end{tikzpicture}

\end{figure}
\FloatBarrier
\begin{figure}[ht]

    \centering
    \caption[Comparison of European soccer teams in terms of the frequency of their posts on Facebook.]{Comparison of European soccer teams in terms of the frequency of their posts on Facebook.}
    \label{fig:teams-frequency-facebook}
    
    \begin{tikzpicture}
    
        \begin{axis}[
            ybar, area legend, 
            width=15cm,
            height=8cm,
            enlargelimits=0.15,
            legend style={at={(0.5,+1.15)},
              anchor=north,legend columns=-1},
            ylabel={Frequency of Posts},
            symbolic x coords={
                Liverpool,
                Real Madrid,
                VIF,
                RBK,
                TIL
            },
            xtick=data,
            xticklabel style={yshift=-1mm}, 
            nodes near coords={}, 
            x tick label style={rotate=20,anchor=east,font=\small}, 
            xmin=Liverpool, 
            xmax=TIL, 
            bar width=5.95, 
            ]
            
            \addplot coordinates {
                (Liverpool, 4)
                (Real Madrid, 3)
                (VIF, 1)
                (RBK, 1)
                (TIL, 1)
            };
            
            \addplot coordinates {
                (Liverpool, 8)
                (Real Madrid, 14)
                (VIF, 1)
                (RBK, 2)
                (TIL, 1)
            };
            
            \addplot coordinates {
                (Liverpool, 0)
                (Real Madrid, 0)
                (VIF, 0)
                (RBK, 0)
                (TIL, 0)
            };
            
            \addplot coordinates {
                (Liverpool, 0)
                (Real Madrid, 0)
                (VIF, 0)
                (RBK, 0)
                (TIL, 3)
            };
            
            \legend{Video+Text, Image+Text, Text, Story}
            
        \end{axis}
        
    \end{tikzpicture}

\end{figure}
\FloatBarrier
\begin{figure}[ht]

    \centering
    \caption[Comparison of European soccer teams in terms of the frequency of their posts on Twitter.]{Comparison of European soccer teams in terms of the frequency of their posts on Twitter.}
    \label{fig:teams-frequency-twitter}
    
    \begin{tikzpicture}
    
        \begin{axis}[
            ybar, area legend, 
            width=15cm,
            height=8cm,
            enlargelimits=0.15,
            legend style={at={(0.5,+1.15)},
              anchor=north,legend columns=-1},
            ylabel={Frequency of Posts},
            symbolic x coords={
                Liverpool,
                Real Madrid,
                VIF,
                RBK,
                TIL
            },
            xtick=data,
            xticklabel style={yshift=-1mm}, 
            nodes near coords={}, 
            x tick label style={rotate=20,anchor=east,font=\small}, 
            xmin=Liverpool, 
            xmax=TIL, 
            bar width=5.95, 
            ]
            
            \addplot coordinates {
                (Liverpool, 6)
                (Real Madrid, 6)
                (VIF, 3)
                (RBK, 3)
                (TIL, 1)
            };
            
            \addplot coordinates {
                (Liverpool, 12)
                (Real Madrid, 18)
                (VIF, 1)
                (RBK, 1)
                (TIL, 1)
            };
            
            \addplot coordinates {
                (Liverpool, 0)
                (Real Madrid, 0)
                (VIF, 0)
                (RBK, 1)
                (TIL, 0)
            };
            
            \addplot coordinates {
                (Liverpool, 0)
                (Real Madrid, 0)
                (VIF, 0)
                (RBK, 0)
                (TIL, 0)
            };
            
            \legend{Video+Text, Image+Text, Text, Retweet}
            
        \end{axis}

        \begin{axis}[
            width=5cm,
            height=3cm,
            enlargelimits=0.2,
            ybar,
            ylabel={},
            symbolic x coords={
                VIF,
                RBK,
                TIL
            },
            xtick=data,
            nodes near coords={},
            x tick label style={rotate=30,anchor=east,font=\tiny},
            ymin=0.1, ymax=3.9, 
            bar width=3.95,
            at={(rel axis cs:0.80,0.99)}, 
            anchor={north west}
            ]

            \addplot coordinates {
                (VIF, 3)
                (RBK, 3)
                (TIL, 1)
            };
            
            \addplot coordinates {
                (VIF, 1)
                (RBK, 1)
                (TIL, 1)
            };
            
            \addplot coordinates {
                (VIF, 0)
                (RBK, 1)
                (TIL, 0)
            };
            
            \addplot coordinates {
                (VIF, 0)
                (RBK, 0)
                (TIL, 0)
            };
        \end{axis}
        
    \end{tikzpicture}

\end{figure}
\FloatBarrier
\FloatBarrier

\takeaway{\begin{itemize}
    \item Liverpool and Real Madrid maintain active content schedules across platforms such as Facebook and Instagram.   
    \item Real Madrid prefers image-based posts on Facebook, with 14 posts overall.
    \item Liverpool posts less frequently on Facebook with 8 posts overall.
    \item Liverpool is more active on Instagram with stories, doubling Real Madrid's frequency.
    \item Among Norwegian teams, TIL stands out by leveraging Facebook stories.
    \item Other content modalities see less consistent posting among Norwegian teams.
    \item Both Liverpool and Real Madrid post at similar rates on TikTok and YouTube.
    \item Overall, our findings emphasize the role of strategic content scheduling on targeting and engaging different audience segments across multiple platforms.
\end{itemize}}

\NewPageCustom
\section{Marketing Channel Spread}

A marketing research analysis, identifying referral sources, was conducted for the websites of European soccer teams. The data collected was then compared to provide insights into the sources of visitors to these websites. The results of the analysis are presented in a table that showcases the different sources of traffic, including direct, referrals, search, social, mail, and display. 

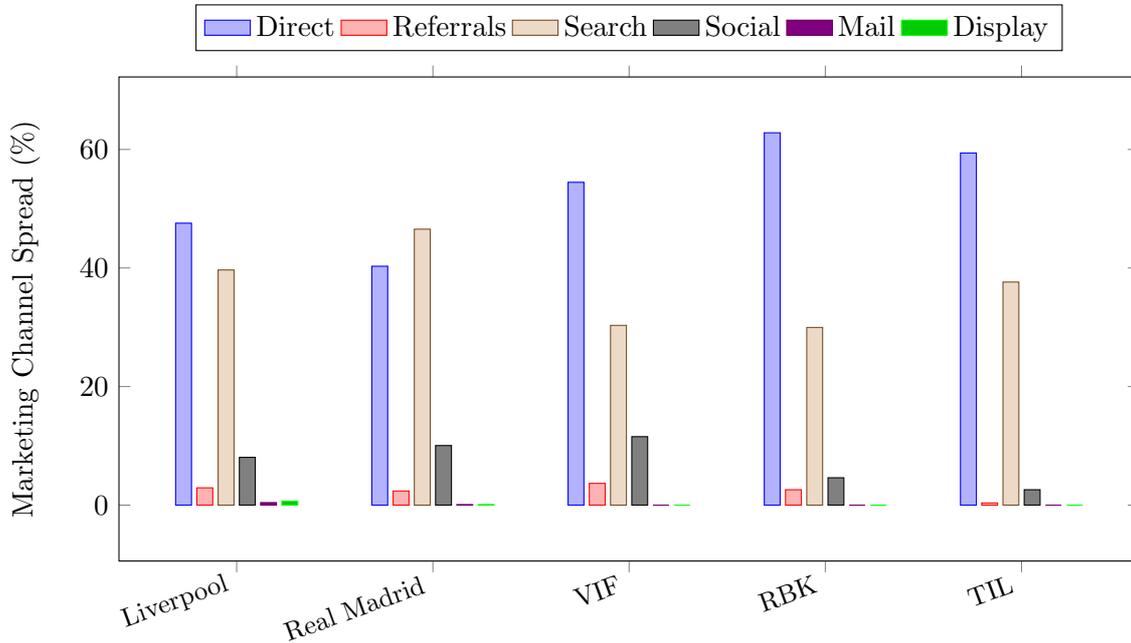
\begin{figure}[ht]

    \centering
    \caption[Comparison of European soccer teams in terms of the marketing channel spread of their websites.]{Comparison of European soccer teams in terms of the marketing channel spread of their websites (in percentage).}
    \label{fig:teams-marketing}
    
    \begin{tikzpicture}
    
        \begin{axis}[
            ybar, area legend, 
            width=15cm,
            height=8cm,
            enlargelimits=0.15,
            legend style={at={(0.5,+1.15)},
              anchor=north,legend columns=-1},
            ylabel={Marketing Channel Spread (\%)},
            symbolic x coords={
                Liverpool,
                Real Madrid,
                VIF,
                RBK,
                TIL
            },
            xtick=data,
            xticklabel style={yshift=-1mm}, 
            nodes near coords={}, 
            x tick label style={rotate=20,anchor=east,font=\small}, 
            xmin=Liverpool, 
            xmax=TIL, 
            bar width=5.95, 
            ]
            
            \addplot coordinates {
                (Liverpool, 47.56)
                (Real Madrid, 40.30)
                (VIF, 54.46)
                (RBK, 62.81)
                (TIL, 59.40)
            };
            
            \addplot coordinates {
                (Liverpool, 2.92)
                (Real Madrid, 2.39)
                (VIF, 3.67)
                (RBK, 2.60)
                (TIL, 0.37)
            };
            
            \addplot coordinates {
                (Liverpool, 39.67)
                (Real Madrid, 46.55)
                (VIF, 30.32)
                (RBK, 29.97)
                (TIL, 37.63)
            };
            
            \addplot coordinates {
                (Liverpool, 8.06)
                (Real Madrid, 10.05)
                (VIF, 11.55)
                (RBK, 4.62)
                (TIL, 2.60)
            };

            \addplot coordinates {
                (Liverpool, 0.43)
                (Real Madrid, 0.09)
                (VIF, 0)
                (RBK, 0)
                (TIL, 0)
            };
            
            \addplot coordinates {
                (Liverpool, 0.67)
                (Real Madrid, 0.09)
                (VIF, 0)
                (RBK, 0)
                (TIL, 0)
            };
            
            \legend{Direct, Referrals, Search, Social, Mail, Display}
            
        \end{axis}
        
    \end{tikzpicture}

\end{figure}

Table~\ref{tab:teams-marketing} presents a comparison of European soccer teams in terms of the marketing channel spread of their websites~\cite{SimilarWeb2023}. Figure~\ref{fig:teams-marketing} presents a visual comparison of the marketing channel spread of the websites of different teams. 

\begin{table}[ht]

    \centering
    \caption[Comparison of European soccer teams in terms of the marketing channel spread of their websites.]{Comparison of European soccer teams in terms of the marketing channel spread of their websites (in percentage).}
    \label{tab:teams-marketing}
    
    \resizebox{\textwidth}{!}{
    
        \begin{tabular}{|l||c|c|c|c|c|}
        \hline
        
        \textbf{Referral Source} 
        & \textbf{Liverpool} 
        & \textbf{Real Madrid} 
        & \textbf{VIF} 
        & \textbf{RBK} 
        & \textbf{TIL} 
        \\ \hline
  
        Direct 
        & 47.56 
        & 40.30 
        & 54.46 
        & 62.81 
        & 59.40 
        \\ \hline 
        
        Referrals 
        & 2.92 
        & 2.39 
        & 3.67 
        & 2.60 
        & 0.37 
        \\ \hline 
        
        Search 
        & 39.67 
        & 46.55 
        & 30.32 
        & 29.97 
        & 37.63 
        \\ \hline 
        
        Social 
        & 8.06 
        & 10.05 
        & 11.55 
        & 4.62 
        & 2.60 
        \\ \hline 
        
        Mail 
        & 0.43 
        & 0.09 
        & 0 
        & 0 
        & 0 
        \\ \hline 
        
        Display 
        & 0.67 
        & 0.09 
        & 0 
        & 0 
        & 0 
        \\ \hline
        
        \end{tabular}
    }
    
\end{table}
\FloatBarrier
 
\FloatBarrier 

\takeaway{\begin{itemize}
    \item Direct traffic dominates the marketing channel spread for most European soccer team websites, RBK leads with 62.81\% of its traffic being direct.
    \item Search engines are crucial for Real Madrid, comprising 46.55\% of its website traffic.
    \item While social media referrals are generally modest, VIF stands out with 11.55\% of its traffic coming from social platforms.
    \item Despite the common use of email and display advertising in digital marketing, their impact on website traffic for selected teams is minimal. 
    \item Overall, our findings indicate that direct traffic and search engines play a pivotal role in driving online engagement for selected soccer teams. 
\end{itemize}}

\NewPageCustom
\section{Website Referrals}

An analysis was conducted in the field of social media referrals to the websites of two major soccer teams, Liverpool and Real Madrid, as well as three Norwegian soccer teams, namely VIF, RBK, and TIl. The purpose of this analysis was to compare the social media referral patterns of these teams and to identify any potential differences in their respective online presence. Various metrics were collected and analyzed, and the results are presented in the following tables. The data presented in these tables will be used to inform further discussions on the effectiveness of social media referrals for soccer teams and the potential impact on their online presence. The monthly visits to each of their websites, as well as their country ranking, have also been analyzed. Upon reviewing the table, it is evident that the Norwegian teams do not have a strong ranking within the country when compared to Liverpool and Real Madrid. It appears that there is ample opportunity for these teams to improve their online presence and attract more visitors through social media channels~\cite{SimilarWeb2023}. 

\begin{table}[ht]

    \centering
    \caption[Comparison of European soccer teams in terms of their social media website referrals.]{Comparison of European soccer teams in terms of their social media website referrals (in percentage).}
    \label{tab:teams-website}

    \resizebox{\textwidth}{!}{
    
        \begin{tabular}{|c|l||c|c|c|c|c|}
        \cline{3-7}

        \multicolumn{2}{c|}{} 
        & \textbf{Liverpool} 
        & \textbf{Real Madrid}
        & \textbf{VIF} 
        & \textbf{RBK} 
        & \textbf{TIL} 
        \\ \hline
        
        \multicolumn{2}{|c||}{\textbf{Monthly Visits}} 
        & 5,886,000 
        & 7,704,000 
        & 53,141 
        & 58,715 
        & 7,860 
        \\ \hline 
        
        \multicolumn{2}{|c||}{\textbf{Country Ranking}} 
        & 857 
        & 404 
        & 5,383 
        & 5,504 
        & 421,140 
        \\ \hline 
        
        \multicolumn{1}{|c|}{\multirow{8}{*}{\rotatebox{90}{\textbf{Referral Source}}}}

        & \textbf{Instagram} 
        & 5.6 
        & 3.6 
        & 0 
        & 0 
        & 0 
        \\ \cline{2-7}
        
        & \textbf{Facebook} 
        & 46.8 
        & 57.2 
        & 59.1 
        & 26.9 
        & 0 
        \\ \cline{2-7}
        
        & \textbf{Twitter} 
        & 5.3 
        & 5.0 
        & 40.9 
        & 0 
        & 0 
        \\ \cline{2-7} 

        & \textbf{YouTube} 
        & 4.0 
        & 7.8 
        & 0 
        & 0 
        & 0 
        \\ \cline{2-7}
        
        & \textbf{LinkedIn} 
        & 0 
        & 0 
        & 0 
        & 73.9 
        & 0 
        \\ \cline{2-7}
        
        & \textbf{Reddit} 
        & 33.2 
        & 25.8 
        & 0 
        & 0 
        & 0 
        \\ \cline{2-7}
        
        & \textbf{WhatsApp} 
        & 0 
        & 0 
        & 0 
        & 0 
        & 0 
        \\ \cline{2-7}
        
        & \textbf{Quora} 
        & 0 
        & 0 
        & 0 
        & 0 
        & 0 
        \\ \cline{2-7}
        
        & \textbf{Other} 
        & 5.1 
        & 0.5 
        & 0 
        & 0 
        & 0 
        \\ \hline 

        \end{tabular} 
        
    }
    
\end{table}
\FloatBarrier

\takeaway{\begin{itemize}
  \item Liverpool and Real Madrid have significantly higher website traffic (5,886,000 and 7,704,000 monthly visits, respectively) compared to Norwegian teams (VIF, RBK, TIL).
  \item Liverpool and Real Madrid have strong country rankings. Norwegian teams, especially TIL, rank lower.  
  \item Liverpool and Real Madrid benefit from varied referral sources, including Facebook and Reddit. VIF mainly gets traffic from Facebook and Twitter. RBK has a notable referral pattern from LinkedIn.
  \item Norwegian teams have a potential for better utilization of channels such as Instagram, YouTube, Reddit, WhatsApp, Quora, and more.
  \item RBK's dominant LinkedIn presence is distinctive. Liverpool shows significant traction from Reddit.
  \item Norwegian teams, particularly VIF and TIL, could diversify their online presence to boost website traffic. Adopting strategies from giants such as Liverpool and Real Madrid might be advantageous.
\end{itemize}}

\NewPageCustom
\section{Visual Content}

In this analysis, the average video duration and caption length for different types of posts on social media platforms of two major soccer teams, Liverpool and Real Madrid, are compared to those of three Norwegian soccer teams, VIF, RBK, and Til. By comparing the average video duration and caption length of posts on social media platforms, it is possible to identify trends and patterns in the social media strategies of these teams. The analysis can reveal which teams prioritize longer or shorter video content, or use more or fewer characters in their posts. Table~\ref{tab:teams-content} presents an analysis of visual content for the selected European soccer teams on social media.

Analyzing this data, it appears that Liverpool and Real Madrid have a consistent pattern across all social media platforms, with slightly longer average video duration and longer captions compared to the Norwegian soccer teams VIF, RBK, and TIL. 

On TikTok, Liverpool and Real Madrid have slightly longer average video duration and longer captions compared to the Norwegian soccer teams for goal and event videos. All teams post videos with a duration of less than 60 seconds. 

On Facebook, Liverpool and Real Madrid post videos with an average duration of less than 10 minutes for events, and less than 3 minutes for highlights. For captions, Liverpool averages around 71 characters, while Real Madrid averages around 150 characters. In contrast, VIF, RBK, and TIL all post videos with shorter duration captions. 

On Instagram, Liverpool and Real Madrid post goal and event videos with average duration of less than 30 seconds, while VIF, RBK, and TIL have slightly longer average duration. For captions, Liverpool and Real Madrid have slightly longer captions compared to the Norwegian soccer teams. 

On Twitter, Liverpool and Real Madrid once again have slightly longer average video duration and longer captions compared to the Norwegian soccer teams for event videos. All teams post videos with a duration of less than 6 minutes. 

On YouTube, Liverpool and Real Madrid have longer average video duration for highlights and event videos compared to the Norwegian soccer teams. Liverpool and Real Madrid also have slightly longer captions. 

Overall, the data suggests that Liverpool and Real Madrid have a consistent pattern of longer video duration and captions compared to the Norwegian soccer teams across various social media platforms. However, it is important to note that the analysis only examines a limited number of factors, and there may be other variables at play that could affect social media strategies. Further analysis and research would be needed to fully understand the social media strategies of these soccer teams.

\begin{table}[ht]

    \centering
    \caption[Comparison of European soccer teams in terms of their social media content.]{Comparison of European soccer teams in terms of their social media content.}
    \label{tab:teams-content}

    \resizebox{\textwidth}{!}{ 
    
        \begin{tabular}{|ccc||c|c|c|c|c|}
        \cline{4-8}

        \multicolumn{3}{c|}{} 
        & \textbf{Liverpool} 
        & \textbf{Real Madrid}
        & \textbf{VIF} 
        & \textbf{RBK} 
        & \textbf{TIL} 
        \\ \hline
        
        \multicolumn{1}{|c|}{\multirow{9}{*}{\textbf{Facebook}}} 
        & \multicolumn{1}{c|}{\multirow{3}{*}{Goals}} 
        & Videos 
        & \textless{}60s 
        & {N/A} 
        & {N/A}
        & {N/A}
        & \textless{}20 s 
        \\ \cline{3-8} 
        
        \multicolumn{1}{|c|}{} 
        & \multicolumn{1}{c|}{} 
        & Story 
        & {N/A}
        & {N/A}
        & {N/A}
        & {N/A}
        & {N/A}
        \\ \cline{3-8} 
        
        \multicolumn{1}{|c|}{} 
        & \multicolumn{1}{c|}{} 
        & Caption 
        & $\sim$71 char. 
        & {N/A} 
        & {N/A}
        & {N/A}
        & $\sim$65 char. 
        \\ \cline{2-8} 
        
        \multicolumn{1}{|c|}{} 
        & \multicolumn{1}{c|}{\multirow{3}{*}{Highlights}} 
        & Video 
        & \textless{}120s 
        & \textless 3min 
        & {N/A}
        & {N/A}
        & {N/A}
        \\ \cline{3-8} 
        
        \multicolumn{1}{|c|}{} 
        & \multicolumn{1}{c|}{} 
        & Story 
        & {N/A}
        & {N/A}
        & {N/A}
        & {N/A}
        & {N/A}
        \\ \cline{3-8} 
        
        \multicolumn{1}{|c|}{} 
        & \multicolumn{1}{c|}{} 
        & Caption 
        & $\sim$85 char. 
        & $\sim$150 char. 
        & {N/A}
        & {N/A} 
        & {N/A}
        \\ \cline{2-8} 
        
        \multicolumn{1}{|c|}{} 
        & \multicolumn{1}{c|}{\multirow{3}{*}{Events}} 
        & Video 
        & \textless{}10min 
        & \textless{}15min 
        & \textless{}3min 
        & \textless{}3min 
        & \textless{}20s 
        \\ \cline{3-8} 
        
        \multicolumn{1}{|c|}{} 
        & \multicolumn{1}{c|}{} 
        & Story 
        & {N/A}
        & {N/A}
        & {N/A}
        & {N/A}
        & {N/A}
        \\ \cline{3-8} 
        
        \multicolumn{1}{|c|}{} 
        & \multicolumn{1}{c|}{} 
        & Caption 
        & $\sim$20 char. 
        & $\sim$120 
        & $\sim$207 
        & $\sim$120 
        & $\sim$103 char. 
        \\ \hline 
        
        \multicolumn{1}{|c|}{\multirow{9}{*}{\textbf{Instagram}}} 
        & \multicolumn{1}{c|}{\multirow{3}{*}{Goals}} 
        & Video 
        & \textless{}30s (D angles) 
        & \textless{}10s 
        & {N/A}
        & {N/A}
        & {N/A}
        \\ \cline{3-8} 
        
        \multicolumn{1}{|c|}{} 
        & \multicolumn{1}{c|}{} 
        & Story 
        & {N/A}
        & {N/A}
        & {N/A}
        & {N/A}
        & {N/A}
        \\ \cline{3-8} 
        
        \multicolumn{1}{|c|}{} 
        & \multicolumn{1}{c|}{} 
        & Caption 
        & 87 char. 
        & $\sim$70 
        & {N/A}
        & {N/A} 
        & {N/A} 
        \\ \cline{2-8} 
        
        \multicolumn{1}{|c|}{} 
        & \multicolumn{1}{c|}{\multirow{3}{*}{Highlights}} 
        & Video 
        & {N/A}
        & \textless{}100s 
        & {N/A}
        & {N/A}
        & {N/A}
        \\ \cline{3-8} 
        
        \multicolumn{1}{|c|}{} 
        & \multicolumn{1}{c|}{} 
        & Story 
        & {N/A}
        & {N/A}
        & {N/A}
        & {N/A}
        & {N/A}
        \\ \cline{3-8} 
        
        \multicolumn{1}{|c|}{} 
        & \multicolumn{1}{c|}{} 
        & Caption 
        & {N/A} 
        & - $\sim$60 char 
        & {N/A}
        & {N/A} 
        & {N/A}
        \\ \cline{2-8} 
        
        \multicolumn{1}{|c|}{} 
        & \multicolumn{1}{c|}{\multirow{3}{*}{Events}} 
        & Video 
        & \textless{}30 
        & \textless{}30s 
        & \textless{}60s 
        & \textless{}60s 
        & \textless{}75s 
        \\ \cline{3-8} 
        
        \multicolumn{1}{|c|}{} 
        & \multicolumn{1}{c|}{} 
        & Story 
        & {N/A}
        & {N/A}
        & {N/A}
        & {N/A}
        & $\sim$7s 
        \\ \cline{3-8} 
        
        \multicolumn{1}{|c|}{} 
        & \multicolumn{1}{c|}{} 
        & Caption 
        & $\sim$45 char. 
        & $\sim$75 char. 
        & $\sim$30 char. 
        & $\sim$150 char. 
        & \textless{}70 char. 
        \\ \hline 
        
        \multicolumn{1}{|c|}{\multirow{6}{*}{\textbf{TikTok}}} 
        & \multicolumn{1}{c|}{\multirow{2}{*}{Goals}} 
        & Video + music
        & \textless{}30s 
        & \textless{}30s 
        & {N/A}
        & \textless{}30s 
        & \textless{}60s 
        \\ \cline{3-8} 
        
        \multicolumn{1}{|c|}{} 
        & \multicolumn{1}{c|}{} 
        & Caption 
        & $\sim$90 char. 
        & $\sim$75 
        & {N/A} 
        & $\sim$90 char. 
        & $\sim$89 char. 
        \\ \cline{2-8} 
        
        \multicolumn{1}{|c|}{} 
        & \multicolumn{1}{c|}{\multirow{2}{*}{Highlights}} 
        & Video 
        & {N/A}
        & {N/A}
        & {N/A}
        & {N/A}
        & {N/A}
        \\ \cline{3-8} 
        
        \multicolumn{1}{|c|}{} 
        & \multicolumn{1}{c|}{} 
        & Caption 
        & {N/A}
        & {N/A}
        & {N/A}
        & {N/A}
        & {N/A}
        \\ \cline{2-8} 
        
        \multicolumn{1}{|c|}{} 
        & \multicolumn{1}{c|}{\multirow{2}{*}{Events}} 
        & Video + music
        & \textless{}60s 
        & \textless{}30s 
        & {N/A}
        & \textless{}60s 
        & \textless{}75s 
        \\ \cline{3-8} 
        
        \multicolumn{1}{|c|}{} 
        & \multicolumn{1}{c|}{} 
        & Caption 
        & $\sim$100 char. 
        & $\sim$85 
        & {N/A}
        & $\sim$67 char. 
        & $\sim$124 char. 
        \\ \hline 
        
        \multicolumn{1}{|c|}{\multirow{6}{*}{\textbf{YouTube}}} 
        & \multicolumn{1}{c|}{\multirow{2}{*}{Goals}} 
        & Video + link 
        & \textless{}60s 
        & \textless{}60s 
        & {N/A}
        & {N/A}
        & {N/A}
        \\ \cline{3-8} 
        
        \multicolumn{1}{|c|}{} 
        & \multicolumn{1}{c|}{} 
        & Caption + link 
        & $\sim$46 char. 
        & $\sim$55 char. 
        & {N/A}
        & {N/A}
        & {N/A}
        \\ \cline{2-8} 
        
        \multicolumn{1}{|c|}{} 
        & \multicolumn{1}{c|}{\multirow{2}{*}{Highlights}} 
        & Video + link 
        & \textless{}10min 
        & \textless{}5min 
        & \textless{}4min 
        & \textless{}5min 
        & {N/A}
        \\ \cline{3-8} 
        
        \multicolumn{1}{|c|}{} 
        & \multicolumn{1}{c|}{} 
        & Caption +link 
        & $\sim$660 char. 
        & $\sim$970 
        & $\sim$72 char. 
        & $\sim$327 char. 
        & {N/A}
        \\ \cline{2-8} 
        
        \multicolumn{1}{|c|}{} 
        & \multicolumn{1}{c|}{\multirow{2}{*}{Events}} 
        & Video + link 
        & \textless{}60min 
        & \textless{}10min 
        & \textless{}10min 
        & \textless{}60min 
        & {N/A}
        \\ \cline{3-8} 
        
        \multicolumn{1}{|c|}{} 
        & \multicolumn{1}{c|}{} 
        & Caption + link 
        & $\sim$660 char. 
        & $\sim$670 char. 
        & $\sim$495 char. 
        & $\sim$500 char. 
        & {N/A}
        \\ \cline{1-8} 
        
        \multicolumn{1}{|c|}{\multirow{6}{*}{\textbf{Twitter}}} 
        & \multicolumn{1}{c|}{\multirow{2}{*}{Goals}} 
        & Video 
        & \textless{}60s 
        & {N/A}
        & {N/A}
        & {N/A}
        & {N/A}
        \\ \cline{3-8} 
        
        \multicolumn{1}{|c|}{} 
        & \multicolumn{1}{c|}{} 
        & Caption 
        & 84 char. 
        & {N/A}
        & {N/A}
        & {N/A}
        & {N/A}
        \\ \cline{2-8} 
        
        \multicolumn{1}{|c|}{} 
        & \multicolumn{1}{c|}{\multirow{2}{*}{Highlights}} 
        & Videos 
        & {N/A}
        & {N/A}
        & {N/A}
        & {N/A}
        & {N/A}
        \\ \cline{3-8}
        
        \multicolumn{1}{|c|}{} 
        & \multicolumn{1}{c|}{} 
        & Caption 
        & {N/A}
        & {N/A}
        & {N/A}
        & {N/A}
        & {N/A}
        \\ \cline{2-8} 
        
        \multicolumn{1}{|c|}{} 
        & \multicolumn{1}{c|}{\multirow{2}{*}{Events}} 
        & Video 
        & \textless{}6min 
        & \textless{}60s 
        & \textless{}60s 
        & \textless{}3min 
        & \textless{}90s 
        \\ \cline{3-8} 
        
        \multicolumn{1}{|c|}{} 
        & \multicolumn{1}{c|}{} 
        & Caption 
        & $\sim$97 char. 
        & $\sim$116 char. 
        & $\sim$117 char. 
        & $\sim$110 char. 
        & $\sim$75 char. 
        \\ \hline
        
        \end{tabular} 
        
    }
    
\end{table}
\FloatBarrier

\takeaway{\begin{itemize}
  \item Liverpool and Real Madrid substantially lead in social media metrics compared to Norwegian teams VIF, RBK, and TIL.
  \item Liverpool and Real Madrid showcase longer video duration and captions across platforms. This indicates a broader and more intensive content strategy for these major teams.
  \item All teams are active on major platforms: TikTok, Instagram, Facebook, Twitter, and YouTube. However, Norwegian teams exhibit areas for growth in optimizing content for increased engagement.
  \item Overall, our findings highlight the need for bespoke social media strategies to enhance fan engagement and online visibility.
\end{itemize}}

\clearpage
\chapter{Conclusion}\label{section:conclusion}

In the era of digitalization, social media has become an integral part of our lives, serving as a significant hub for individuals and businesses to share information, communicate, and engage. This is also the case for professional sports, where leagues, clubs and players are using social media to reach out to their fans. In this respect, a huge amount of time is spent curating multimedia content for various social media platforms and their target users. With the emergence of Artificial Intelligence (AI), AI-based tools for automating content generation and enhancing user experiences on social media have become widely popular. However, to effectively utilize such tools, it is imperative to comprehend the demographics and preferences of users on different platforms, understand how content providers post information in these channels, and how different types of multimedia are consumed by audiences. This report presents an analysis of social media platforms, in terms of demographics, supported multimedia modalities, and distinct features and specifications for different modalities, followed by a comparative case study of select European soccer leagues and teams, in terms of their social media practices. Through this analysis, we demonstrate that social media, while being very important for and widely used by supporters from all ages, also requires a fine-tuned effort on the part of soccer professionals, in order to elevate fan experiences and foster engagement.

In the first part of this report, we provide an analysis of $7$ social media platforms, namely TikTok~\cite{TikTokHelp2023}, Instagram~\cite{InstagramHelp2023}, Facebook~\cite{FacebookHelp2023}, Twitter~\cite{TwitterHelp2023}, Snapchat~\cite{SnapchatHelp2023}, LinkedIn~\cite{LinkedInHelp2023}, and YouTube~\cite{YouTubeHelp2023}, in terms of demographics (e.g., age and gender distribution), as well as supported modalities and multimedia specifications. Our analysis will help identify the most popular social media platforms for different age groups and genders, and understand the unique features of each platform. This knowledge can be used to curate highlights including video, image, and text, which are optimized for each platform, develop personalized content, and enhance user engagement, ultimately leading to better social media experiences. 

In the second part of this report, we present specific case studies. First, we explore the social media presence of $5$ prominent European soccer leagues, namely the English Premier League (England)~\cite{PremierLeague2023}, Bundesliga (Germany)~\cite{Bundesliga2023}, La Liga (Spain)~\cite{LaLiga2023}, Serie A (Italy)~\cite{SerieA2023}, and Ligue 1 (France)~\cite{Ligue1UberEats2023}, and compare their activities to the social media presence of two smaller leagues from Scandinavia, namely Allsvenskan (Sweden)~\cite{Allsvenskan2023} and Eliteserien (Norway)~\cite{Eliteserien2023}. Next, we compare the social media activities of some of the most prominent teams in the aforementioned leagues with those of Norwegian soccer teams. Through this analysis, we hope to provide insights into the differences and similarities in the social media practices of different soccer leagues and teams, and how they utilize social media platforms to engage with fans and promote their respective organizations.

European soccer leagues exhibit diverse online presences across social media platforms. The English Premier League, the German Bundesliga, and the Spanish La Liga dominate in terms of the number of followers, with Premier League expectedly leading due to its global appeal. Engagement rates serve as key indicators of content resonance, with direct and search traffic being predominant sources for website visits. While YouTube, Twitter, and Facebook emerge as significant referrers, the type of visual content varies across leagues, emphasizing the importance of platform-specific content strategies for optimal fan engagement.

The social media presence of major European soccer teams, Liverpool and Real Madrid, significantly outpace that of Norwegian teams VIF, RBK, and TIL. Liverpool and Real Madrid consistently exhibit longer video duration and captions across platforms, suggesting a more extensive content strategy. While all teams utilize platforms such as TikTok, Facebook, Instagram, Twitter, and YouTube, the Norwegian teams have opportunities for growth, especially in terms of optimizing their content for engagement. The data underscores the importance of tailored social media strategies to maximize fan engagement and online visibility.

\clearpage
\scriptsize
\bibliography{references}

\begin{thebibliography}{100}

\bibitem{sproutsocial2021}
How different generations use social media{\ifmmode---\else\textemdash\fi}and
  what this means for your business.
\newblock
  \url{https://sproutsocial.com/insights/guides/social-media-use-by-generation},
  July 2021.
\newblock [Online; accessed 28. Sep. 2023].

\bibitem{Allsvenskan2023}
Allsvenskan.
\newblock \url{https://www.allsvenskan.se/}, 2023.
\newblock Accessed: 24/05/2022.

\bibitem{AllsvenskanFacebook}
{Allsvenskan} official {Facebook}.
\newblock \url{https://www.facebook.com/Allsvenskan}, 2023.
\newblock Accessed: 24/05/2023.

\bibitem{EliteserienFacebook}
{Allsvenskan} official {Facebook}.
\newblock \url{https://www.facebook.com/eliteserien/}, 2023.
\newblock Accessed: 24/05/2023.

\bibitem{AllsvenskanInstagram}
{Allsvenskan} official {Instagram}.
\newblock \url{https://www.instagram.com/allsvenskan/}, 2023.
\newblock Accessed: 24/05/2023.

\bibitem{AllsvenskanTwitter}
{Allsvenskan Official Twitter}.
\newblock \url{https://twitter.com/AllsvenskanSE}, 2023.
\newblock Accessed: 24/05/2023.

\bibitem{Bundesliga2023}
Bundesliga.
\newblock \url{https://www.bundesliga.com/en/bundesliga}, 2023.
\newblock Accessed: 24/05/2022.

\bibitem{BundesligaFacebook}
Bundesliga official {Facebook}.
\newblock \url{https://www.facebook.com/BundesligaOfficial}, 2023.
\newblock Accessed: 24/05/2023.

\bibitem{BundesligaInstagram}
Bundesliga official {Instagram}.
\newblock \url{https://instagram.com/bundesliga/}, 2023.
\newblock Accessed: 24/05/2023.

\bibitem{BundesligaTikTok}
Bundesliga official {TikTok}.
\newblock \url{https://www.tiktok.com/channel/germany-football?lang=en}, 2023.
\newblock Accessed: 24/05/2023.

\bibitem{BundesligaTwitter}
Bundesliga official {Twitter}.
\newblock \url{https://twitter.com/Bundesliga_EN}, 2023.
\newblock Accessed: 24/05/2023.

\bibitem{BundesligaYouTube}
Bundesliga official {YouTube}.
\newblock \url{https://www.youtube.com/user/bundesliga}, 2023.
\newblock Accessed: 24/05/2023.

\bibitem{Eliteserien2023}
Eliteserien.
\newblock \url{https://www.eliteserien.no/}, 2023.
\newblock Accessed: 24/05/2022.

\bibitem{EliteserienInstagram}
{Eliteserien} official {Instagram}.
\newblock \url{https://www.instagram.com/eliteserien/?hl=am-et}, 2023.
\newblock Accessed: 24/05/2023.

\bibitem{EliteserienTikTok}
{Eliteserien} official {TikTok}.
\newblock \url{https://www.tiktok.com/@eliteserien}, 2023.
\newblock Accessed: 24/05/2023.

\bibitem{EliteserienTwitter}
{Eliteserien} official {Twitter}.
\newblock \url{https://twitter.com/eliteserien}, 2023.
\newblock Accessed: 24/05/2023.

\bibitem{EliteserienYouTube}
{Eliteserien} official {YouTube}.
\newblock \url{https://www.youtube.com/channel/UCKB1WCVnk1j3UeM-82MTScA}, 2023.
\newblock Accessed: 24/05/2023.

\bibitem{EngagementCalculator2023}
Engagement calculator.
\newblock \url{https://phlanx.com/engagement-calculator}, 2023.
\newblock Accessed: 24/05/2022.

\bibitem{FacebookHelp2023}
{Facebook} help center.
\newblock \url{https://www.facebook.com/help}, 2023.
\newblock Accessed: 24/05/2022.

\bibitem{InstagramHelp2023}
{Instagram} help center.
\newblock \url{https://help.instagram.com/}, 2023.
\newblock Accessed: 24/05/2022.

\bibitem{LaLiga2023}
{LaLiga}.
\newblock \url{https://www.laliga.com/en-GB}, 2023.
\newblock Accessed: 24/05/2022.

\bibitem{LaLigaFacebook}
{LaLiga} official {Facebook}.
\newblock \url{https://www.facebook.com/LaLiga}, 2023.
\newblock Accessed: 24/05/2023.

\bibitem{LaLigaInstagram}
{LaLiga} official {Instagram}.
\newblock \url{https://www.instagram.com/laliga/}, 2023.
\newblock Accessed: 24/05/2023.

\bibitem{LaLigaTikTok}
{LaLiga} official {TikTok}.
\newblock \url{https://www.tiktok.com/@laliga}, 2023.
\newblock Accessed: 24/05/2023.

\bibitem{LaLigaTwitter}
{LaLiga} official {Twitter}.
\newblock \url{https://twitter.com/LaLigaEN}, 2023.
\newblock Accessed: 24/05/2023.

\bibitem{LaLigaYouTube}
{LaLiga} official {YouTube}.
\newblock \url{https://www.youtube.com/user/laliga}, 2023.
\newblock Accessed: 24/05/2023.

\bibitem{Ligue1UberEats2023}
Ligue 1 - {Uber Eats France}.
\newblock \url{https://www.ligue1.com/}, 2023.
\newblock Accessed: 24/05/2022.

\bibitem{Ligue1Facebook}
{Ligue 1 - Uber Eats France Official Facebook}.
\newblock \url{https://www.facebook.com/Ligue1UberEats/}, 2023.
\newblock Accessed: 24/05/2023.

\bibitem{Ligue1Instagram}
{Ligue 1 - {Uber Eats France}} official {Instagram}.
\newblock \url{https://www.instagram.com/ligue1ubereats/}, 2023.
\newblock Accessed: 24/05/2023.

\bibitem{Ligue1TikTok}
Ligue 1 - {Uber Eats France} official {TikTok}.
\newblock \url{https://www.tiktok.com/channel/french-league}, 2023.
\newblock Accessed: 24/05/2023.

\bibitem{Ligue1Twitter}
{Ligue 1 - Uber Eats France Official Twitter}.
\newblock \url{https://twitter.com/Ligue1UberEats}, 2023.
\newblock Accessed: 24/05/2023.

\bibitem{Ligue1YouTube}
{Ligue 1 - Uber Eats France Official YouTube}.
\newblock \url{https://www.youtube.com/c/Ligue1official}, 2023.
\newblock Accessed: 24/05/2023.

\bibitem{LinkedInHelp2023}
{LinkedIn} help center.
\newblock \url{https://www.linkedin.com/help/linkedin?lang=en}, 2023.
\newblock Accessed: 24/05/2022.

\bibitem{Liverpool_Facebook}
Liverpool official {Facebook}.
\newblock \url{https://www.facebook.com/LiverpoolFC}, Jan 2023.

\bibitem{Liverpool_Instagram}
Liverpool official {Instagram}.
\newblock \url{https://www.instagram.com/liverpoolfc}, Jan 2023.

\bibitem{Liverpool_TikTok}
Liverpool official {TikTok}.
\newblock \url{https://www.tiktok.com/@liverpoolfc?lang=en}, Jan 2023.

\bibitem{Liverpool_Twitter}
Liverpool official {Twitter}.
\newblock \url{https://twitter.com/LFC}, Jan 2023.

\bibitem{Liverpool_YouTube}
Liverpool official {YouTube}.
\newblock \url{https://www.youtube.com/liverpoolfc}, Jan 2023.

\bibitem{statistaFacebook2023}
Number of monthly active facebook users worldwide.
\newblock
  \url{https://www.statista.com/statistics/264810/number-of-monthly-active-facebook-users-worldwide/?ref=buffer.com},
  2023.
\newblock Accessed: 24/05/2023.

\bibitem{PremierLeague2023}
{Premier League England}.
\newblock \url{https://www.premierleague.com/}, 2023.
\newblock Accessed: 24/05/2022.

\bibitem{PremierLeagueFacebook}
{Premier League} official {Facebook}.
\newblock \url{https://www.facebook.com/premierleague}, 2023.
\newblock Accessed: 24/05/2023.

\bibitem{PremierLeagueInstagram}
{Premier League} official {Instagram}.
\newblock \url{https://www.instagram.com/premierleague/}, 2023.
\newblock Accessed: 24/05/2023.

\bibitem{PremierLeagueTikTok}
{Premier League} official {TikTok}.
\newblock \url{https://www.tiktok.com/@premierleague?lang=en}, 2023.
\newblock Accessed: 24/05/2023.

\bibitem{PremierLeagueTwitter}
{Premier League} official {Twitter}.
\newblock \url{https://twitter.com/premierleague}, 2023.
\newblock Accessed: 24/05/2023.

\bibitem{PremierLeagueYouTube}
{Premier League} official {YouTube}.
\newblock \url{https://www.youtube.com/premierleague}, 2023.
\newblock Accessed: 24/05/2023.

\bibitem{RBK_Facebook}
{RBK} official {Facebook}.
\newblock \url{https://www.facebook.com/rosenborg/}, Jan 2023.

\bibitem{RBK_Instagram}
{RBK} official {Instagram}.
\newblock \url{https://www.instagram.com/rosenborgballklub/}, Jan 2023.

\bibitem{RBK_Twitter}
{RBK} official {Twitter}.
\newblock \url{https://twitter.com/RBKfotball}, Jan 2023.

\bibitem{RBK_YouTube}
{RBK} official {YouTube}.
\newblock \url{https://www.youtube.com/channel/UCMzBr0zvB2uhUTRvnSFzwIw}, Jan
  2023.

\bibitem{Realmadrid_Facebook}
{Real Madrid} official {Facebook}.
\newblock \url{https://www.facebook.com/RealMadrid}, Jan 2023.

\bibitem{Realmadrid_Instagram}
{Real Madrid} official {Instagram}.
\newblock \url{https://www.instagram.com/realmadrid/}, Jan 2023.

\bibitem{Realmadrid_TikTok}
{Real Madrid} official {TikTok}.
\newblock \url{https://www.tiktok.com/@realmadrid?lang=en}, Jan 2023.

\bibitem{Realmadrid_Twitter}
{Real Madrid} official {Twitter}.
\newblock \url{https://twitter.com/realmadriden}, Jan 2023.

\bibitem{Realmadrid_YouTube}
{Real Madrid} official {YouTube}.
\newblock \url{https://www.youtube.com/channel/UCWV3obpZVGgJ3j9FVhEjF2Q}, Jan
  2023.

\bibitem{SerieA2023}
{Serie A}.
\newblock \url{https://www.legaseriea.it/en}, 2023.
\newblock Accessed: 24/05/2022.

\bibitem{SerieAFacebook}
{Serie A} official {Facebook}.
\newblock \url{https://www.facebook.com/SerieA}, 2023.
\newblock Accessed: 24/05/2023.

\bibitem{SerieAInstagram}
{Serie A} official {Instagram}.
\newblock \url{https://www.instagram.com/seriea/}, 2023.
\newblock Accessed: 24/05/2023.

\bibitem{SerieATikTok}
{Serie A} official {TikTok}.
\newblock \url{https://www.tiktok.com/discover/italian-serie-a-soccer}, 2023.
\newblock Accessed: 24/05/2023.

\bibitem{SerieATwitter}
{Serie A} official {Twitter}.
\newblock \url{https://twitter.com/SerieA}, 2023.
\newblock Accessed: 24/05/2023.

\bibitem{SerieAYouTube}
{Serie A} official {YouTube}.
\newblock \url{https://www.youtube.com/SerieA}, 2023.
\newblock Accessed: 24/05/2023.

\bibitem{SnapchatHelp2023}
{Snapchat} help center.
\newblock \url{https://help.snapchat.com/hc/en-us}, 2023.
\newblock Accessed: 24/05/2022.

\bibitem{Statista2023}
Social media demographic world.
\newblock
  \url{https://www.statista.com/search/?q=social+media+demographic+world&qKat=search&newSearch=true&p=1},
  2023.
\newblock Accessed: 24/05/2022.

\bibitem{TikTokHelp2023}
{TikTok} help center.
\newblock \url{https://support.tiktok.com/en/}, 2023.
\newblock Accessed: 24/05/2022.

\bibitem{TIL_Facebook}
{TIL} official {Facebook}.
\newblock \url{https://www.facebook.com/TromsoIL/}, Jan 2023.

\bibitem{TIL_Instagram}
{TIL} official {Instagram}.
\newblock \url{https://www.instagram.com/tromsoil/}, Jan 2023.

\bibitem{TIL_TikTok}
{TIL} official {TikTok}.
\newblock \url{https://www.tiktok.com/@tromsoil?lang=en}, Jan 2023.

\bibitem{TIL_Twitter}
{TIL} official {Twitter}.
\newblock \url{https://twitter.com/TromsoIL}, Jan 2023.

\bibitem{TIL_YouTube}
{TIL} official {YouTube}.
\newblock \url{https://www.youtube.com/channel/UCz0e2RRUfC07SKaYTJAi1DQ}, Jan
  2023.

\bibitem{TwitterHelp2023}
{Twitter} help center.
\newblock \url{https://help.twitter.com/en}, 2023.
\newblock Accessed: 24/05/2022.

\bibitem{VIF_Facebook}
{VIF} official {Facebook}.
\newblock \url{https://www.facebook.com/VaalerengaFotballElite/}, Jan 2023.

\bibitem{VIF_Instagram}
{VIF} official {Instagram}.
\newblock \url{https://www.instagram.com/valerengaoslo/}, Jan 2023.

\bibitem{VIF_TikTok}
{VIF} official {TikTok}.
\newblock \url{https://www.tiktok.com/@vaalerengaoslo}, Jan 2023.

\bibitem{VIF_Twitter}
{VIF} official {Twitter}.
\newblock \url{https://twitter.com/ValerengaOslo}, Jan 2023.

\bibitem{VIF_YouTube}
{VIF} official {YouTube}.
\newblock \url{https://www.youtube.com/c/V%C3%A5lerengaFotballElite}, Jan 2023.

\bibitem{SimilarWeb2023}
Website analysis - {Uncover} the entire digital landscape of any website.
\newblock \url{https://www.similarweb.com/}, 2023.
\newblock Accessed: 24/05/2022.

\bibitem{YouTubeHelp2023}
{YouTube} help center.
\newblock \url{https://support.google.com/youtube/}, 2023.
\newblock Accessed: 24/05/2023.

\bibitem{ACCC2023}
ACCC.
\newblock {ACCC} to investigate how big tech platforms use smart device data.
\newblock
  \url{https://www.afr.com/politics/federal/accc-probes-data-flows-from-smart-devices-20230310-p5cr5w},
  2023.
\newblock Accessed: 24/05/2022.

\bibitem{Alduayji2019}
Mohammed~Nasser Alduayji.
\newblock {Online co-creation behaviour in a sports context}.
\newblock {\em the UWA Profiles and Research Repository}, 2019.

\bibitem{apostolidis2021}
Konstantinos Apostolidis and Vasileios Mezaris.
\newblock A fast smart-cropping method and dataset for video retargeting.
\newblock In {\em 2021 IEEE International Conference on Image Processing
  (ICIP)}, pages 2618--2622. IEEE, 2021.

\bibitem{Belsky2023}
Ira Belsky.
\newblock {How Generative AI Is Changing Creative Work}.
\newblock {\em Forbes}, May 2023.

\bibitem{Darbinyan2023}
Rem Darbinyan.
\newblock {How AI Transforms Social Media}.
\newblock {\em Forbes}, March 2023.

\bibitem{Eddy2021}
Terry Eddy, B.~Colin Cork, Katie Lebel, and Erin~Howie Hickey.
\newblock {Examining Engagement With Sport Sponsor Activations on Twitter}.
\newblock {\em International Journal of Sport Communication}, 14(1):79--108,
  January 2021.

\bibitem{gautam2022}
Sushant Gautam.
\newblock {AI}-based soccer game summarization: From video highlights to
  dynamic text summaries.
\newblock
  \url{https://www.researchgate.net/publication/363857936_AI-based_Soccer_Game_Summarization_From_Video_Highlights_to_Dynamic_Text_Summaries},
  2022.

\bibitem{gautam2023}
Sushant Gautam.
\newblock {Bridging Multimedia Modalities: Enhanced Multimodal AI Understanding
  and Intelligent Agents}.
\newblock In {\em {ICMI '23: Proceedings of the 25th International Conference
  on Multimodal Interaction}}, pages 695--699. Association for Computing
  Machinery, New York, NY, USA, October 2023.

\bibitem{gautam2022_summarization}
Sushant Gautam, Cise Midoglu, Saeed Shafiee~Sabet, Dinesh~Baniya Kshatri, and
  P{\aa}l Halvorsen.
\newblock Assisting soccer game summarization via audio intensity analysis of
  game highlights.
\newblock In {\em Proceedings of 12th IOE Graduate Conference}, volume~12,
  pages 25 -- 32. Institute of Engineering, Tribhuvan University, Nepal,
  October 2022.

\bibitem{gautam2022_soccer}
Sushant Gautam, Cise Midoglu, Saeed Shafiee~Sabet, Dinesh~Baniya Kshatri, and
  P{\aa}l Halvorsen.
\newblock Soccer game summarization using audio commentary, metadata, and
  captions.
\newblock In {\em Proceedings of the 1st Workshop on User-centric Narrative
  Summarization of Long Videos}, pages 13--22, 2022.

\bibitem{Hootsuite2022}
Hootsuite.
\newblock Social media image sizes: A quick reference guide for each network.
\newblock \url{https://blog.hootsuite.com/social-media-image-sizes-guide/},
  2022.

\bibitem{Hou2022}
Sujuan Hou, Jiacheng Li, Weiqing Min, Qiang Hou, Yanna Zhao, Yuanjie Zheng, and
  Shuqiang Jiang.
\newblock {Deep Learning for Logo Detection: A Survey}.
\newblock {\em ACM Trans. Multimedia Comput. Commun. Appl.}, 2022.

\bibitem{Husa2022-HOST-ATS}
Andreas Husa, Cise Midoglu, Malek Hammou, P\r{a}l Halvorsen, and Michael~A.
  Riegler.
\newblock Host-ats: Automatic thumbnail selection with dashboard-controlled ml
  pipeline and dynamic user survey.
\newblock In {\em Proceedings of the ACM Multimedia Systems Conference}, MMSys
  '22, page 334–340, 2022.

\bibitem{Husa2022-Automatic}
Andreas Husa, Cise Midoglu, Malek Hammou, Steven~A. Hicks, Dag Johansen, Tomas
  Kupka, Michael~A. Riegler, and P\r{a}l Halvorsen.
\newblock Automatic thumbnail selection for soccer videos using machine
  learning.
\newblock In {\em Proceedings of the ACM Multimedia Systems Conference}, MMSys
  '22, page 73–85, 2022.

\bibitem{Jin_2023}
Jun-gyu Jin, Jaehyun Bae, Han-gyul Baek, and Sang-hyo Park.
\newblock Object-ratio-preserving video retargeting framework based on
  segmentation and inpainting.
\newblock In {\em Proceedings of the IEEE/CVF Winter Conference on Applications
  of Computer Vision (WACV) Workshops}, pages 497--503, January 2023.

\bibitem{Kapoor2018}
Kawaljeet~Kaur Kapoor, Kuttimani Tamilmani, Nripendra~P. Rana, Pushp Patil,
  Yogesh~K. Dwivedi, and Sridhar Nerur.
\newblock {Advances in Social Media Research: Past, Present and Future}.
\newblock {\em Inf. Syst. Front.}, 20(3):531--558, June 2018.

\bibitem{DIGITAL_2023_STATSHOT_REPORT}
Simon Kemp.
\newblock Digital 2023 april global statshot report — datareportal – global
  digital insights.
\newblock
  \url{https://datareportal.com/reports/digital-2023-april-global-statshot},
  Apr 2023.

\bibitem{Lu2021}
Heng-yang Lu, Chenyou Fan, Xiaoning Song, and Wei Fang.
\newblock {A novel few-shot learning based multi-modality fusion model for
  COVID-19 rumor detection from online social media}.
\newblock {\em PeerJ Comput. Sci.}, 7:e688, August 2021.

\bibitem{mallmann2020}
Jackson Mallmann, Altair~Olivo Santin, Eduardo~Kugler Viegas, Roger~Robson dos
  Santos, and Jhonatan Geremias.
\newblock Ppcensor: Architecture for real-time pornography detection in video
  streaming.
\newblock {\em Future Generation Computer Systems}, 112:945--955, 2020.

\bibitem{McCarthy2022}
Jeff McCarthy, Jenny Rowley, and Brendan~J. Keegan.
\newblock {Social media marketing strategy in English football clubs}.
\newblock {\em Soccer {\&} Society}, pages 513--528, April 2022.

\bibitem{midoglu2022}
Cise Midoglu, Steven~A Hicks, Vajira Thambawita, Tomas Kupka, and P{\aa}l
  Halvorsen.
\newblock {MMSys' 22 Grand Challenge on AI-based Video Production for Soccer}.
\newblock {\em arXiv preprint arXiv:2202.01031}, 2022.

\bibitem{Naraine2019}
Michael~L. Naraine, Henry~T. Wear, and Damien~J. Whitburn.
\newblock {User engagement from within the Twitter community of professional
  sport organizations}.
\newblock {\em Managing Sport and Leisure}, pages 275--293, June 2019.

\bibitem{Neoreach2022}
NeoReach.
\newblock The ultimate guide to social media post sizes for 2022.
\newblock \url{https://neoreach.com/social-media-post/}, 2022.

\bibitem{Nergård2020}
Olav~A. Norgård~Rongved, Steven~A. Hicks, Vajira Thambawita, Håkon~K.
  Stensland, Evi Zouganeli, Dag Johansen, Michael~A. Riegler, and Pål
  Halvorsen.
\newblock Real-time detection of events in soccer videos using 3d convolutional
  neural networks.
\newblock In {\em Proc. of IEEE International Symposium on Multimedia (ISM)},
  pages 135--144, 2020.

\bibitem{Nergård2021-Automated}
Olav~Andre Norgård~Rongved, Markus Stige, Steven~Alexander Hicks,
  Vajira~Lasantha Thambawita, Cise Midoglu, Evi Zouganeli, Dag Johansen,
  Michael~Alexander Riegler, and Pål Halvorsen.
\newblock Automated event detection and classification in soccer: The potential
  of using multiple modalities.
\newblock {\em Machine Learning and Knowledge Extraction}, 3(4):1030--1054,
  2021.

\bibitem{Obradovic2019}
Maja Obradovi{\ifmmode\acute{c}\else\'{c}\fi}, Slavko
  Al{\ifmmode\check{c}\else\v{c}\fi}akovi{\ifmmode\acute{c}\else\'{c}\fi},
  Daria Vyugina, and Sandra Tasevski.
\newblock {Use of Social Media in Communication Strategies of {Premier League}
  Football Clubs}.
\newblock {\em Proceedings of the International Scientific Conference - Sinteza
  2019}, 2019.

\bibitem{Nergård2021-3D}
Olav A.~Nergård Rongved, Steven~A. Hicks, Vajira Thambawita, Håkon~K.
  Stensland, Evi Zouganeli, Dag Johansen, Cise Midoglu, Michael~Alexander
  Riegler, and Pål Halvorsen.
\newblock Using 3d convolutional neural networks for real-time detection of
  soccer events.
\newblock {\em International Journal of Semantic Computing}, 15(2):161–187,
  2021.

\bibitem{Sosnovik2023}
Ivan Sosnovik, A.~Moskalev, Cees Kaandorp, and A.~Smeulders.
\newblock {Learning to Summarize Videos by Contrasting Clips}.
\newblock {\em ArXiv}, 2023.

\bibitem{Trail2017}
Galen~T. Trail, Dean~F. Anderson, and Don Lee.
\newblock {A longitudinal study of team-fan role identity on self-reported
  attendance behavior and future intentions}.
\newblock {\em JAS}, 3(1), March 2017.

\bibitem{Valand2021-Automated}
Joakim~O. Valand, Haris Kadragic, Steven~A. Hicks, Vajira Thambawita, Cise
  Midoglu, Tomas Kupka, Dag Johansen, Michael~A. Riegler, and Pål Halvorsen.
\newblock Automated clipping of soccer events using machine learning.
\newblock In {\em 2021 IEEE International Symposium on Multimedia (ISM)}, pages
  210--214, 2021.

\bibitem{Valand2021-AI-Based}
Joakim~Olav Valand, Haris Kadragic, Steven~Alexander Hicks, Vajira~Lasantha
  Thambawita, Cise Midoglu, Tomas Kupka, Dag Johansen, Michael~Alexander
  Riegler, and Pål Halvorsen.
\newblock Ai-based video clipping of soccer events.
\newblock {\em Machine Learning and Knowledge Extraction}, 3(4):990--1008,
  2021.

\bibitem{Wang2022}
Hang Wang, Pengcheng Zhou, Chong Zhou, Zhao Zhang, and Xing Sun.
\newblock {PAC-Net: Highlight Your Video via History Preference Modeling}.
\newblock
  \url{https://www.semanticscholar.org/paper/PAC-Net%3A-Highlight-Your-Video-via-History-Modeling-Wang-Zhou/c02abc84e50041b5de7af1a78e86535eb1a73d07},
  2022.
\newblock [Online; accessed 28. Sep. 2023].

\bibitem{Winand2019}
Mathieu Winand, Matthew Belot, Sebastian Merten, and Dimitrios Kolyperas.
\newblock {International sport federations' social media communication: a
  content analysis of FIFA's Twitter account}.
\newblock {\em International Journal of Sport Communication}, 12(2):209--233,
  June 2019.

\bibitem{Xiong2023}
Ziwei Xiong and Han Wang.
\newblock {Dual-Stream Multimodal Learning for Topic-Adaptive Video Highlight
  Detection}.
\newblock {\em Proceedings of the 2023 ACM International Conference on
  Multimedia Retrieval}, 2023.

\bibitem{Yun2020}
Jin~Ho Yun, Philip~J. Rosenberger, and Kristi Sweeney.
\newblock {Drivers of soccer fan loyalty: Australian evidence on the influence
  of team brand image, fan engagement, satisfaction and enduring involvement}.
\newblock {\em Asia Pacific Journal of Marketing and Logistics},
  33(3):755--782, July 2020.

\bibitem{Zogan2021}
Hamad Zogan, Imran Razzak, Shoaib Jameel, and Guandong Xu.
\newblock {DepressionNet: Learning Multi-modalities with User Post
  Summarization for Depression Detection on Social Media}.
\newblock In {\em {SIGIR '21: Proceedings of the 44th International ACM SIGIR
  Conference on Research and Development in Information Retrieval}}, pages
  133--142. Association for Computing Machinery, New York, NY, USA, July 2021.

\end{thebibliography}
\bibliographystyle{plain}

\end{document}